\documentclass[10pt,journal,compsoc]{IEEEtran}
\usepackage[utf8]{inputenc}
\usepackage[T1]{fontenc}

\usepackage{stmaryrd}

\usepackage{xcolor}
\usepackage{soul}
\usepackage{amsmath}
\usepackage{amssymb}
\usepackage[binary-units,per-mode=symbol]{siunitx}
\usepackage{relsize}
\usepackage{soul}

\usepackage{balance}
\usepackage{ctable}

\usepackage{dblfloatfix}

\usepackage{subcaption}
\ifCLASSOPTIONcompsoc
  \usepackage[nocompress]{cite}
\else
  \usepackage{cite}
\fi

\begin{document}
\title{ Security Properties of Gait for \\ Mobile Device Pairing}

\author{Arne~Brüsch,
        Ngu~Nguyen,~\IEEEmembership{Member,~IEEE,}
        Dominik Schürmann,~\IEEEmembership{Member,~IEEE,}
        Stephan Sigg,~\IEEEmembership{Member,~IEEE,}
        and Lars~Wolf,~\IEEEmembership{Member,~IEEE}
\IEEEcompsocitemizethanks{\IEEEcompsocthanksitem A. Brüsch, D. Schürmann and L. Wolf are with TU Braunschweig, Germany.
\IEEEcompsocthanksitem N. Nguyen and S. Sigg are with Aalto University, Finnland.}
\thanks{\textbf{All authors contributed equally to this work and are listed in alphabetic order.} Manuscript received October 20, 2017, revised June 24, 2018.}}

\markboth{Preprint accepted in IEEE Transactions on Mobile Computing, 2019}%
{Bruesch \MakeLowercase{\textit{et al.}}: Security Properties of Gait for Mobile Device Pairing}

\IEEEtitleabstractindextext{%
\begin{abstract}
Gait has been proposed as a feature for mobile device pairing across arbitrary positions on the human body. 
Results indicate that the correlation in gait-based features across different body locations is sufficient to establish secure device pairing. 
However, the population size of the studies is limited and powerful attackers with e.g. capability of video recording are not considered. 
We present a concise discussion of security properties of gait-based pairing schemes including quantization, classification and analysis of attack surfaces,  
of statistical properties of generated sequences, an entropy analysis, as well as possible threats and security weaknesses. 
For one of the schemes considered, we present modifications to fix an identified security flaw.
As a general limitation of gait-based authentication or pairing systems, we further demonstrate that an adversary with video support can create key sequences that are sufficiently close to on-body generated acceleration sequences to breach gait-based security mechanisms. 
\end{abstract}

}

\maketitle

 \IEEEdisplaynontitleabstractindextext
 \IEEEpeerreviewmaketitle

\section{Introduction}
\IEEEPARstart{W}{ith} the proliferation of mobile devices and the upcoming Internet of Things, interaction between these devices will drastically increase~\cite{guo2013opportunistic}. 
In particular, smart textile and digital assistants are to generate dense body area networks~\cite{dawy2017toward}. 
This is extended by spontaneous pairings {to other} devices in the context of use~\cite{Cryptography_Sigg_2011-5}. 
In such environment where device pairings raise by $n$ with each $n+1$st additional device, and device count and type changes on a sub-day schedule, manual pairing is impractical.
Implicit pairing has been proposed e.g. based on acceleration~\cite{findling2016shakeunlock}, audio~\cite{Cryptography_Schuerman_2011}, magnetometer~\cite{jin2016magpairing} and RF features~\cite{mathur2011proximate}. 
Especially gait~\cite{muaaz2017smartphone} is well suited in wearable settings as it is confined to a single person and can be read out at arbitrary body location~\cite{kunze2011compensating}. 

Gait-based pairing schemes~\cite{schuermann2017bandana} (extended in~\cite{schuermann2018jagger}), \cite{sun2017secure},\cite{groza2012saphe},\cite{Xu_2016_WalkieTalkie} (extended in~\cite{xu2017gaitkey} ), \cite{revadigar2017smartwearables} share the common goal to protect against Man-in-the-Middle~(MitM) attacks, where an attacker actively places herself between devices to modify intercepted communication.
In contrast to other schemes such as Bluetooth Secure Simple Pairing~(SSP) that typically requires the comparison of PINs, gait is leveraged for automatic MitM protection.
However, no concise study of the security properties of quantization approaches for gait-based pairing has been presented to-date.

We close this gap by providing a comprehensive classification of attack surfaces for gait-based pairing and authentication.  
We analyse four recent quantization schemes covering protocol-specific attacks and potential security weaknesses, as well as distribution, statistical and entropy analysis of key sequences.  
Finally, we show that a sophisticated adversary using video can break gait-based pairing if executed in real-time.
Our contributions are 
\begin{itemize}
\item a concise investigation and comparison of popular quantization schemes for gait-based device-pairing, 
\item a comprehensive discussion of attack surfaces, 
\item an entropy, pattern and statistical analysis, 
\item an improved quantization for one of the approaches to mitigate an identified security weakness, 
\item the first ever empirical demonstration that video poses a significant threat to gait-based security.  
\end{itemize}

We first introduce technical details of the four quantization schemes and their performance on a common dataset (Section~\ref{sec:quantization}).
Then, to spot conceptual flaws, we analyse properties of generated keys in terms of bit distribution and statistical tests in Section~\ref{sec:randomness}, before identifying scheme-specific security weaknesses (Section~\ref{sec:SecurityAnalysis}).
In particular, we consider the one-shot success probability, quantization-specific attacks, effects of error correction on security properties, gait mimicry, as well as, impersonation via video recording. 
In Section~\ref{sec:improving}, we suggest improvements for specific schemes.
Our work concludes in Section~\ref{sec:conclusion} with the main quantization differences and discusses the most promising scheme.

\section{Related Work}\label{sectionRelatedWork} 
We first discuss gait recognition approaches, before summariz{ing} recent progress in gait-based authentication and gait-based pairing. 
In the remainder of the discussion, we then focus on using acceleration sequences from natural gait for device pairing. 

\subsection{Gait Recognition}
Traditionally, {g}ait recognition has been applied exploiting machine vision~\cite{nixon1996automatic,han2006individual,liu2006improved,Sakar2005Gait}. 
Systems then comprise one or multiple cameras to capture natural gait and contain image recognition steps including background subtraction, feature extraction and recognition~\cite{boulgouris2005gait}. 
First work goes back to perception experiments on light point displays conducted in~\cite{Cutting1977}. 
This work was further developed in~\cite{niyogi1994analyzing} with computer vision approaches to recognize people from gait. 
In preceding years drastic improvements have been made in gait recognition algorithms~\cite{boyd2004synchronization,nixon2004advances}.
Gait recognition approaches can be grouped into (1) temporal alignment-based, (2) static parameter-based and (3) silhouette shape-based approaches~\cite{liu2006improved}. 
From these,~\cite{veeraraghavan2005matching} found that shape is more significant for person identification than kinematics. 

Temporal alignment-based approaches emphasize both shape and dynamics and first extract silhouette features before aligning sequences of these e.g. with temporal correlation, dynamic time warping or hidden Markov models.  

Static parameter-based approaches exploit gait dynamics such as stride length, cadence and stride speed~\cite{johnson2001multi}. However, they are least successful for gait-based identification due to their need for 3D calibration information.

Finally, silhouette shape-based approaches use silhouette shape similarity and disregard temporal information, often considering averaged silhouettes or treating silhouette shapes as collection without specific order~\cite{liu2006improved}.
For all above methods, gait recognition can be improved by combining statistical gait features from real and synthetic templates~\cite{han2006individual}

Due to the increasing availability of wearable sensors such as gyroscopes (rotation), accelerometers (acceleration) or force sensors (force during walking), gait recognition via such wearable sensors is increasingly investigated~\cite{gafurov2008performance,morris2002shoe,huang2007gait,ailisto2005identifying,rong2007identification,rong2007wearable,vildjiounaite2006unobtrusive}.
In these approaches, acceleration sequences are recorded from various body locations, most prominently at the waist. 
The acceleration signal is then denoised e.g. by applying wavelet transformation~\cite{rong2007wearable} and changes in walking speed are mitigated utilizing dynamic time warping~\cite{kale2003gait} or similar approaches. 
Individual steps are identified from the resulting signal by searching for minima and by applying pattern or template matching~\cite{rong2007wearable}. 
Similarity can be estimated by the computation of cross-correlation~\cite{ailisto2005identifying}.
Alternatively, machine learning classifiers are trained and applied~\cite{huang2007gait}.

Finally, a recent technique employed to acquire human gait is to monitor phase changes of an electromagnetic signal reflected from a subject walking towards a transceiver~\cite{Wang:2016:GRU:2971648.2971670,Zeng2016Gait}.
The authors exploit changes in channel state information (CSI) from WiFi devices for the detection of gait. 
After generating spectograms from CSI measurements, similar to Doppler radars, and applying autocorrelation on the torso reflection to remove imperfection in these spectrograms, gait patterns are extracted.

Note that frequently, sensors installed in the floor such as pressure sensing mats are also mentioned as modalities for gait recognition~\cite{Jenkins2007,nakajima2000footprint}.
However, in these cases, not gait itself is extracted but other features such as footprints~\cite{nakajima2000footprint}, ground reaction force~\cite{orr2000smart} or heel-to-toe ratio~\cite{middleton2005floor}.

\subsection{Gait as a Biometric Pattern for Authentication}
\label{sec:RelWork:Attacks}
{Biometric} authentication systems comprise sensors converting analog stimuli to digital input that can then be quantized and compared to a database of previously stored {biometric} features.
Gait as a discriminating feature was first studied in~\cite{Johansson1973,Cutting1977}. 
It has been realized that characteristic features in gait enable identification of subjects also in larger gait databases~\cite{sarkar2005humanid,wang2005gait,lam2005new,nixon2010human}.
In addition, multiple studies have demonstrated that the success probability of an imposter trying to mimic a subjects gait are low~\cite{gafurov2007spoof} even when trained professionals with similar physical characteristics are employed~\cite{muaaz2017smartphone}.
For instance, Hoang et al.~\cite{hoang2015gaitauthentication} generated a key fingerprint from the difference of a mean gait spanning the complete population to the individual's mean gait. 
In this way, the authors assured that the resulting sequence is well balanced and uniformly distributed. A good overview on gait-based user authentication is provided in~\cite{derawi2010unobtrusive,derawi2012smartphones}.

\begin{figure*}
	\begin{footnotesize}
		\fbox{
			\begin{minipage}[c]{0.98\textwidth}
				ShakeUnlock  protocol\\
				\begin{minipage}[c]{0.4\columnwidth}
					\begin{enumerate}
						\item Record acceleration sequences
						\item Remove gravity per axis, calculate magnitude and normalize to $[-1,1]$
						\item Share magnitude via secure channel
					\end{enumerate}
				\end{minipage}
				\begin{minipage}[c]{0.58\columnwidth}
					\begin{enumerate}
						\setcounter{enumi}{3}
						\item Slice magnitude segments; transform to frequency domain
						\item Compute pairwise coherence via cross spectral- \& power spectral density
						\item Calculate the mean over all coherence values
						\item Unlock IFF mean coherence exceeds threshold
					\end{enumerate}
				\end{minipage}
		\end{minipage}}
		\vspace{.1cm}
		
		\fbox{
			\begin{minipage}[c]{0.25\textwidth}
				Candidate Key protocol (SAPHE)\\[-.5cm]
				\begin{enumerate}
					\item Extract features on devices
					\item Hash feature values
					\item Exchange hashes to identify matching values
					\item When sufficient entropy collected (matching values), concatenate matching values to give secure key.
				\end{enumerate}
		\end{minipage}}
		\hfill
		\fbox{
			\begin{minipage}[c]{0.705\textwidth}
				Walkie-Talkie  protocol\\[-.6cm]
				\begin{enumerate}
					\item Agree on heel-strike count. Then, record acceleration.
					\item Use ICA for source separation; apply FFT on independent components 
					\item Low-pass filter (3Hz) in gravity direction (reduce noise and detect local maxima (heel-strikes))
					\item Rotate acceleration data using gyroscope to same body coordinate system 
					\item Low pass filter (10Hz); normalize 3D acceleration to zero mean, unit length 
					\item Samples $\frac{\geq}{\leq}\mu+\alpha\sigma$ are interpreted as 1/0 where $\mu$ and $\sigma$ are computed per window
					\item Matching samples chosen define key.
					IFF $\leq0.5+\varepsilon$ overlap, abort (counter impersonation)
					\item XOR sequences between consecutive windows to obtain keys, axes are interleaved.
				\end{enumerate}
		\end{minipage}}
		\vspace{.1cm}
		
		\fbox{
			\begin{minipage}[c]{0.98\textwidth}
				BANDANA protocol\\
				\begin{minipage}[c]{0.49\columnwidth}
					\begin{enumerate}
						\item Collect acceleration readings from the z-axis
						\item Correct rotation wrt gravity (gyroscope)
						\item Bandpass between 0.5Hz and 12Hz
						\item Resampling (40 samples/gait) and gait detection
						\item Compute mean gait
					\end{enumerate}
				\end{minipage}
				\begin{minipage}[c]{0.49\columnwidth}
					\begin{enumerate}\setcounter{enumi}{5}
						\item Difference between mean and instantaneous gait translates to binary sequence
						\item Calculate reliability of bits, disregard least reliable
						\item Share reliability ordering \& create fingerprint
						\item Fuzzy cryptography: Get key from fingerprint
					\end{enumerate}
				\end{minipage}
		\end{minipage}}
		\vspace{.1cm}
		
		\fbox{
			\begin{minipage}[c]{0.98\textwidth}
				Inter-Pulse-Interval (IPI) protocol\\[-.1cm]
				\begin{minipage}[c]{0.53\columnwidth}
					\begin{enumerate}
						\item[(1-4)] Analog to the BANDANA Protocol\setcounter{enumi}{4}
						\item Detect left/right-foot-flat peaks from acceleration
					\end{enumerate}
				\end{minipage}
				\begin{minipage}[c]{0.45\columnwidth}
					\begin{enumerate}\setcounter{enumi}{5}
						\item $\overline{IPI}_{gray}=\mbox{Graycode}\left(\left\lfloor\frac{IPI}{m\cdot 1000/f_s}\mod 2^q\right\rfloor\right)$ 
						\item Obtain key as first $k$ bits in $\overline{IPI}_{gray}$
					\end{enumerate}
				\end{minipage}
		\end{minipage}}
		\caption{Description of acceleration-based device-pairing protocols}
		\label{fig:protocols}
	\end{footnotesize}
\end{figure*}

However, despite studies asserting that gait can be used as biometric feature~\cite{rong2007identification,gafurov2007survey,jain2007handbook}, we remark that there is a lack of studies investigating the security features and entropy of gait as an authentication mechanism.

Several attacks though have been considered (cf. Table~\ref{tab:attacks}).
\setlength{\tabcolsep}{2pt}
\ctable[
    caption = {Attacks on gait-based wearable authentication systems},
    label = tab:attacks,
    pos = tbp,
    width=1\columnwidth,
    doinside=\footnotesize
]{@{}p{2.3cm}p{2.1cm}p{4.2cm}@{}}{%
}{                          \FL
\textbf{Paper}               & \textbf{Applications}    & \textbf{Attacking} \ML
Muaaz et al.~\cite{muaaz2017smartphone}    & Gait recognition & Active imposter (imitation),  \NN
                                                      && 20\% EER \NN
Xu et al.~\cite{Xu_2016_WalkieTalkie}      & Device pairing   & Active imposter (imitation), \NN
                                                      && passive imposter, MitM \NN
Kumar et al.~\cite{kumar_2015_treadmill}   & Gait recognition & Treadmill attack        \NN
Trippel et al.~\cite{kwongyou}             & Injection of false acceleration & Poisoning acoustic injection attack \NN
Derawi et al.~\cite{derawi2010unobtrusive} &           & Active imposter, 20\% EER, significant random success probability \NN
Mjaland et al.~\cite{mjaaland2010walk}     & Gait biometrics  & Active long-term trained impostors \NN
Stang~\cite{stang2007gait}                 & Gait biometrics  & Training impostors with continuous visual feedback \LL
}
For instance, Mjaaland et al~\cite{mjaaland2010walk} trained seven individuals to imitate one specific victim. 
Even after intensive training over two weeks (5 hours every day), it was not possible for the subjects to accurately imitate the walking pattern of the victim. 
Also, the provision of continuous visual feedback did not suffice to assist imitators in~\cite{stang2007gait}. 
Furthermore, the authors of~\cite{gafurov2007spoof} investigated the success probability of an attacker towards a particular subject on a database of 100 subjects and concluded that it is unlikely for an adversary to mimic the subjects gait with sufficient accuracy. 
This result has been confirmed by~\cite{muaaz2017smartphone} who employed professional actors to mimic the gait of 15 subjects with close physical properties. 
Indeed, the attempt to mimic gait incorporates the risk of asymmetric gait cycles and thus even lowers the chance of success. 
However, as indicated in~\cite{gafurov2007spoof}, the probability of random matches significantly exceeds the expected probability in the birthday {paradox}. 

This means that an attacker with knowledge of the template database can select persons that are close to him in terms of similarity as suitable victims.
This poses a serious threat to gait-based authentication in general. 
This is confirmed in~\cite{muaaz2013analysis,derawi2010unobtrusive} who report an equal error rate (EER)\footnote{Equal rates for false acceptance and false rejection} of 20\%  for gait authentication.
In addition, given the gait features of the victim and exploiting a treadmill to control speed, length of steps, thigh lift, hip movement and width of steps, the authors in~\cite{kumar_2015_treadmill} could reach a false acceptance rate (FAR) of 46.66\%. 

In addition, the high accuracy of video-based gait recognition systems also empowers an adversary to generate a database of gait information on multiple subjects unnoticed. 
Video-based attacks on gait-authentication systems are insufficiently investigated in the literature.
In Section~\ref{sec:videoattacks}, we demonstrate that a sophisticated adversary with video support can estimate gait sufficiently {accurate} in order to break gait-based authentication and pairing schemes. 

We conclude that gait-based authentication faces serious security threats and gait appears not feasible as sole basis for authentication, especially in systems where the adversary is targeting not a specific but any subject in the system. 
Furthermore, gait changes over time~\cite{boulgouris2005gait} and is affected by clothing, footwear, walking surface~\cite{nixon1996automatic}, walking speed~\cite{boulgouris2005gait} and emotion~\cite{sloman1982gait}.
These effects are insufficiently studied and render gait-based authentication a challenging undertaking.

\subsection{Acceleration-Based Pairing of Devices}
Device pairing protocols execute quantization on one or more devices at the same time to generate similar bit sequences.
In contrast to user authentication, these sequences are not matched against a template database.
Instead they are used to authenticate a key agreement between all participanting parties.
Recently, several authors have considered acceleration or gait for the pairing of devices co-present on the same body~\cite{srivastava2015step,heinz2003experimental,muaaz2013analysis}. 
In particular, these approaches exploit correlation in acceleration signals when devices are worn on the same body~\cite{lester2004you,cornelius2011recognizing} or shaken together~\cite{findling2014shakeunlock,findling2016shakeunlock}.
Note that, in contrast to exploiting gait for authentication, the existence of a unique and reproducible biometric gait sequence is not required {for these approaches}. 
Instead, the protocols exploit instantaneous, correlated acceleration sequences that can not be re-used at different time as the system can be restricted to single attempts~\cite{schuermann2017bandana}. 
The above {described} weaknesses for gait as biometric pattern therefore do not apply. 
{Instead,} the strength of the pairing approach is conditioned on the quantization used, what entropy that approach can guarantee and whether or not it leaks information to a powerful (realistic) attacker. 

In~\cite{findling2014shakeunlock,findling2016shakeunlock} the ShakeUnlock protocol is presented to unlock a mobile device when it is shaken simultaneously with a smartwatch. 
The individual steps of this protocol are briefly described in Figure~\ref{fig:protocols}. 
This approach requires the direct comparison of acceleration sequences in order to compute correlation and therefore needs an established secure channel to exchange this information.

However, other approaches that do not require {an} already established secure connection have been proposed recently. 
For authentication based on arbitrary co-aligned sensor data, Mayrhofer~\cite{mayrhofer2007candidate} proposes the candidate key protocol.
An advanced variant {that solves the known issue of low-entropy input data} is implemented in SAPHE~\cite{groza2012saphe}.
It interactively exchanges hashes from feature sequences as short secrets and concatenates the key from the secrets with matching hashes (cf. Figure~\ref{fig:protocols}).

Walkie-Talkie, an alternative approach conditioned on correlated acceleration sequences from a person's gait, is presented in~\cite{Xu_2016_WalkieTalkie}.
The authors achieve a high bitrate by using individual samples for key bits if they deviate by at least $\alpha$ standard deviations from the mean (cf. Figure~\ref{fig:protocols}).
\begin{figure*}
    \begin{subfigure}[t]{0.48\textwidth}
        \includegraphics[width=\textwidth]{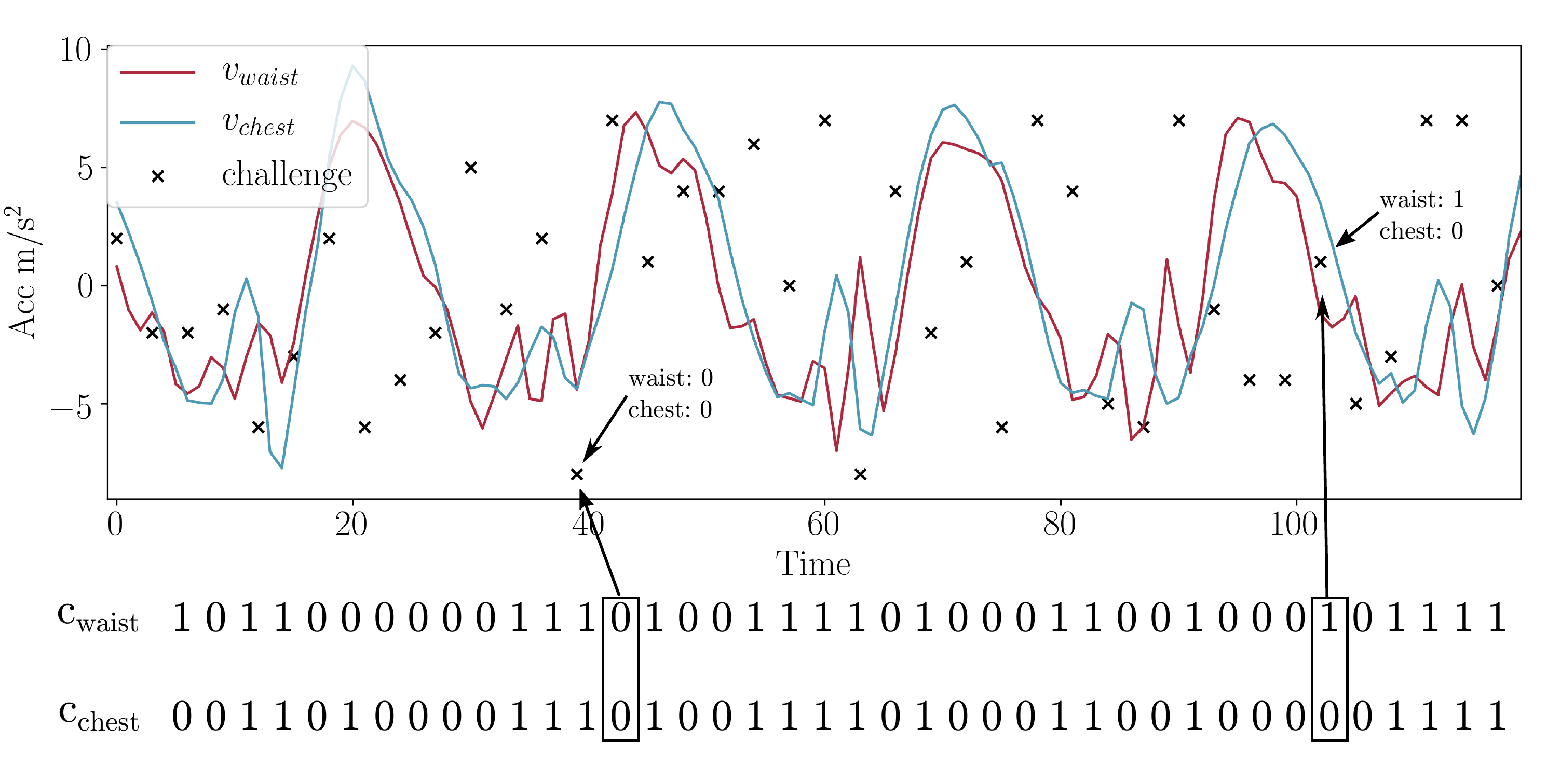}
        \caption{SAPHE}
        \label{fig:quantizationSAPHE}
    \end{subfigure}
    \quad
    \begin{subfigure}[t]{0.48\textwidth}
        \includegraphics[width=\textwidth]{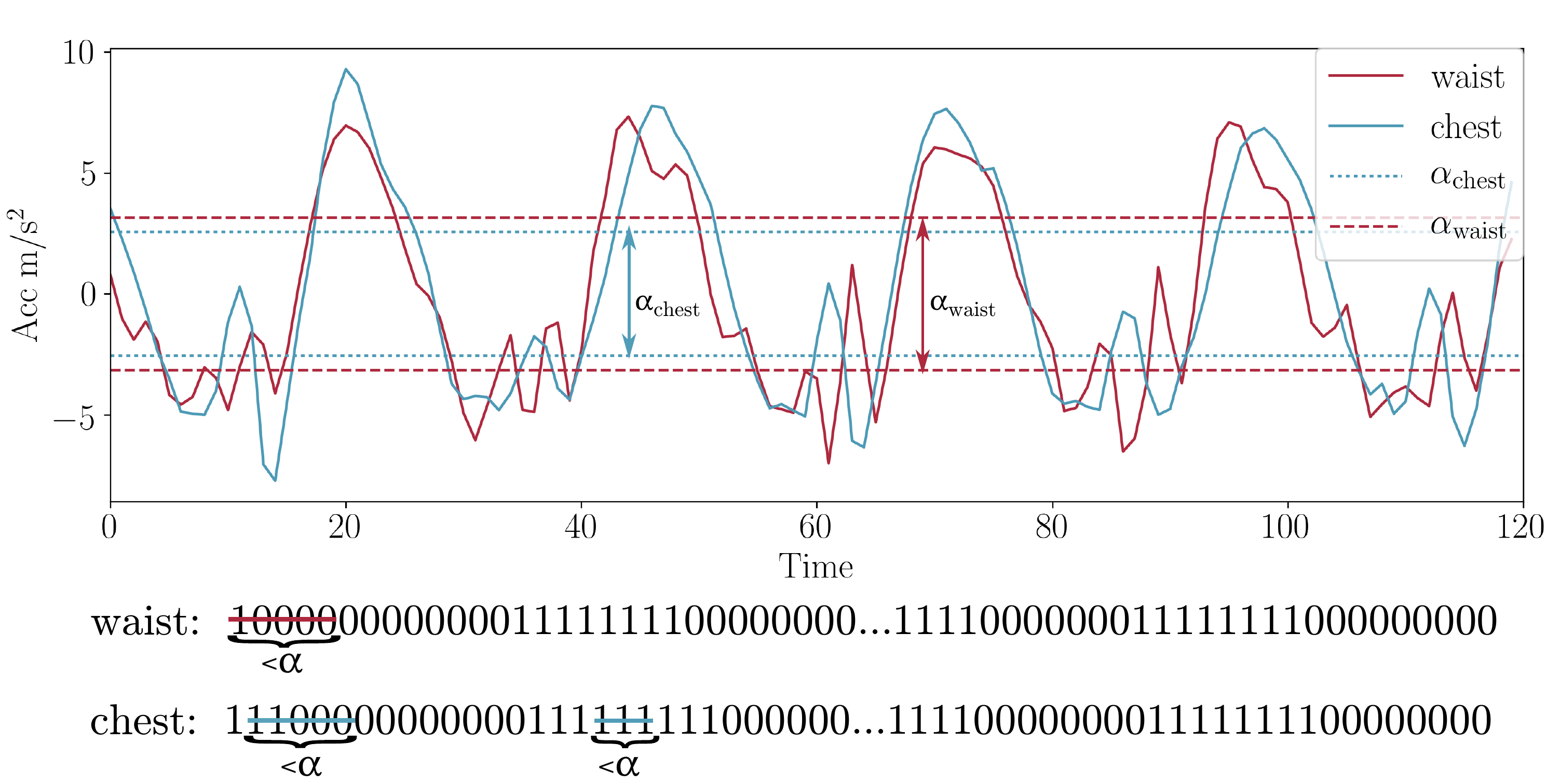}
        \caption{Walkie-Talkie}
        \label{fig:quantizationWalkieTalkie}
    \end{subfigure}
    
    \begin{subfigure}[t]{0.48\textwidth}
        \includegraphics[width=\textwidth]{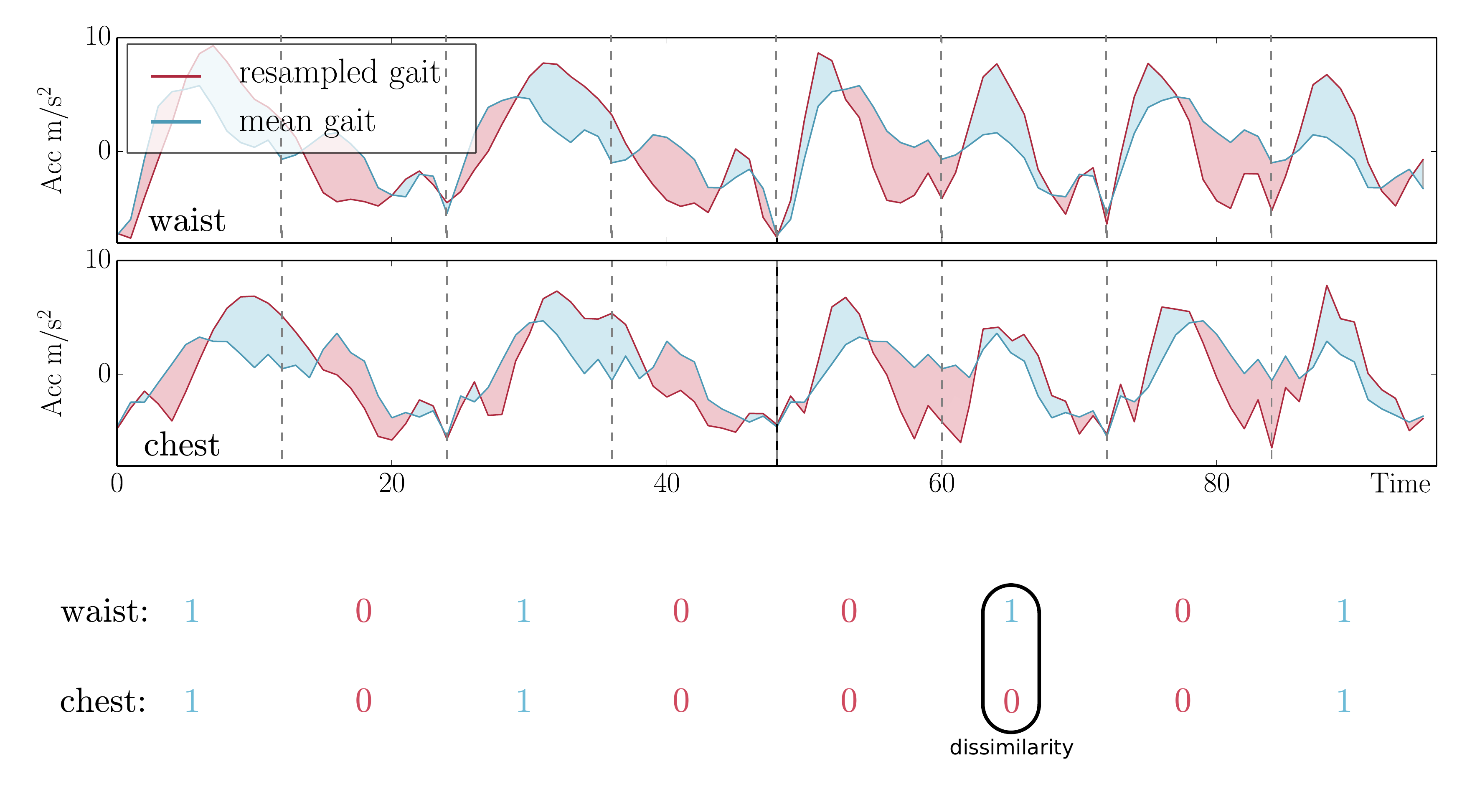}
        \caption{BANDANA}
        \label{fig:quantizationBandana}
    \end{subfigure}
    \quad
    \begin{subfigure}[t]{0.48\textwidth}
        \includegraphics[width=\textwidth]{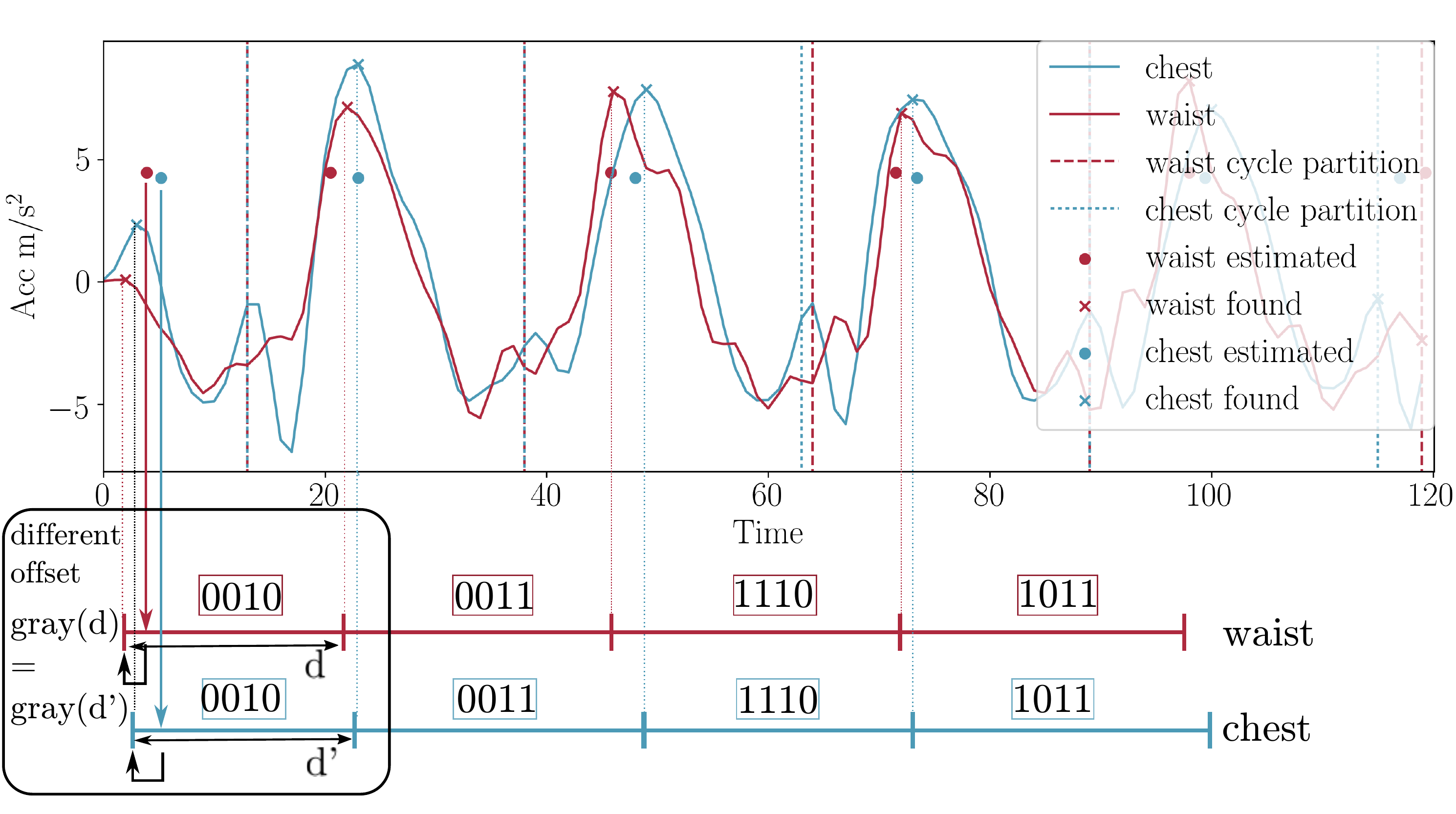}
        \caption{IPI}
        \label{fig:quantizationIPI}
    \end{subfigure}

    \caption{Descriptive examples for the evaluated quantization schemes}
    \label{fig:quantization}
\end{figure*}

An extended version has been published as Gait-Key~\cite{xu2017gaitkey} providing a higher bitrate by applying multiple thresholds.
Walkie-Talkie and Gait-Key use the acceleration values along gravity, walking and sideways direction.
In another scheme by the same authors, movement on all three axis is used as a random source to generate a group key~\cite{revadigar2017smartwearables}.
This group key is locked in a fuzzy vault using a secret set based on the acceleration along gravity only (cf. Walkie-Talkie).
Other wearables can unlock the vault using a secret set, sufficiently similar to the original one, to retrieve the group key.
When an attacker intercepts the locked fuzzy vault the security of this scheme solely depends on the secret set.

The BANDANA protocol~\cite{schuermann2017bandana} exploits acceleration along the z-axis only and conditions the gait fingerprint on the difference between instantaneous gait and mean gait at that body location. 
It thereby achieves normalization among acceleration sequences across body locations. 
Remaining dissimilarities in fingerprints are corrected with fuzzy cryptography exploiting BCH codes (cf. Figure~\ref{fig:protocols}).

In an extended version~\cite{schuermann2018jagger}, the required key length has been reduced to 16 bit by using a Password Authenticated Key Exchange~(PAKE).

Recently, the Inter-Pulse-Interval (IPI) between consecutive steps has been exploited for secure key generation from gait~\cite{sun2017secure}.   
The protocol exploits the acceleration along the z-axis and concatenates the key sequence as gray-coded, scaled and rounded IPIs.
As reported in~\cite{sun2017secure} (cf. Table~\ref{tab:entropy} in Section~\ref{sec:attackTypes}) the security and inter-class similarity depends on the speed of consecutive steps and steplength. 
The protocol was verified on gait captured from devices on the torso of subjects (lower back, upper right arm and right ear). 

The quantization methods in these approaches diverge and result in different properties of the generated binary fingerprints, as described in Section~\ref{sec:quantization}.

An attack on acceleration-based pairing is described in~\cite{kwongyou}. 
An active adversary emitting modulated acoustic interference at the resonant frequency of materials in MEMS sensors can control or modify measured acceleration, and thus inject changes to acceleration sequences.

\section{Comparison of Quantization Schemes}
\label{sec:quantization}
A crucial part in gait-based pairing is the quantization used.
It has to preserve a \emph{high similarity} between generated keys on different body parts, and generate \emph{sufficiently unpredictable} bit sequences for the use as cryptographic keys that withstand a computationally unconstrained adversary. 

In this section, we analyze the quantization of SAPHE, Walkie-Talkie, BANDANA and IPI and describe their working principles along Figure~\ref{fig:quantization}.
In particular, we study the similarity of keys generated for pairs of devices on different body locations.
Additionally, we evaluate how they fulfill the first requirement, i.e. to generate keys with \emph{high similarity} between different locations on the same body (intra-body) and no similarity between different bodies (inter-body).

All quantization schemes have been analyzed using walking data recorded in~\cite{sztyler2016onbody}\footnote{ The dataset includes 15 subjects, 10 minutes walking each, acceleration sensors at 7 different body locations (50Hz) and is available at http://sensor.informatik.uni-mannheim.de}, pre-processed by Madgwick's algorithm to correct accelerometer orientation.
Each quantization scheme generated keys from same length walking data.
Due to the protocols' different efficiency, key length may vary across schemes. 
The performance of the schemes to withstand adversaries is discussed in Section~\ref{sec:randomness}.

\begin{figure*}
    \begin{subfigure}[t]{0.5\textwidth}
        \includegraphics[width=\textwidth, trim={.35cm 0 2.3cm 1cm},clip]{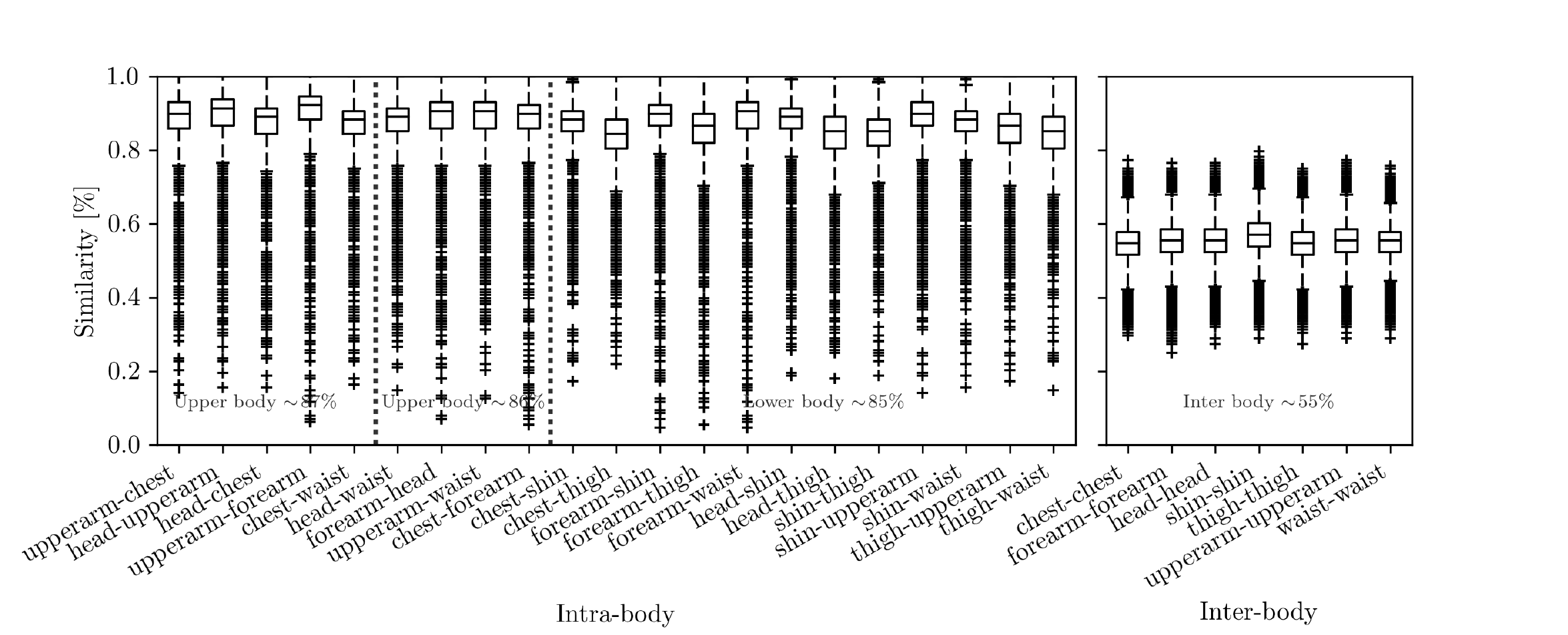}
        \caption{SAPHE}
        \label{fig:similaritySAPHE}
    \end{subfigure}
    \begin{subfigure}[t]{0.5\textwidth}
        \includegraphics[width=\textwidth, trim={.35cm 0 2.3cm 1cm},clip]{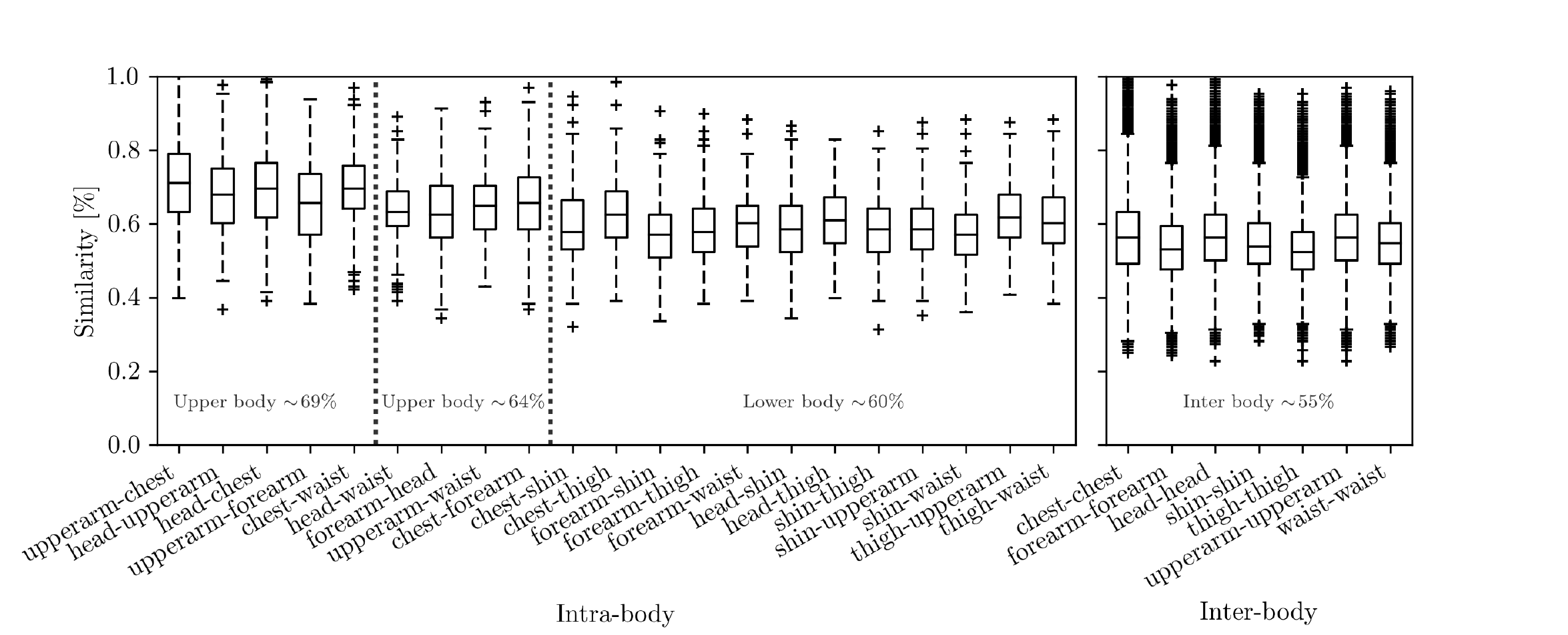}
        \caption{Walkie-Talkie}
        \label{fig:similarityWalkieTalkie}
    \end{subfigure}
    
    \vspace{.2cm}
    
    \begin{subfigure}[t]{0.5\textwidth}
        \includegraphics[width=\textwidth, trim={.35cm 0 2.3cm 1cm},clip]{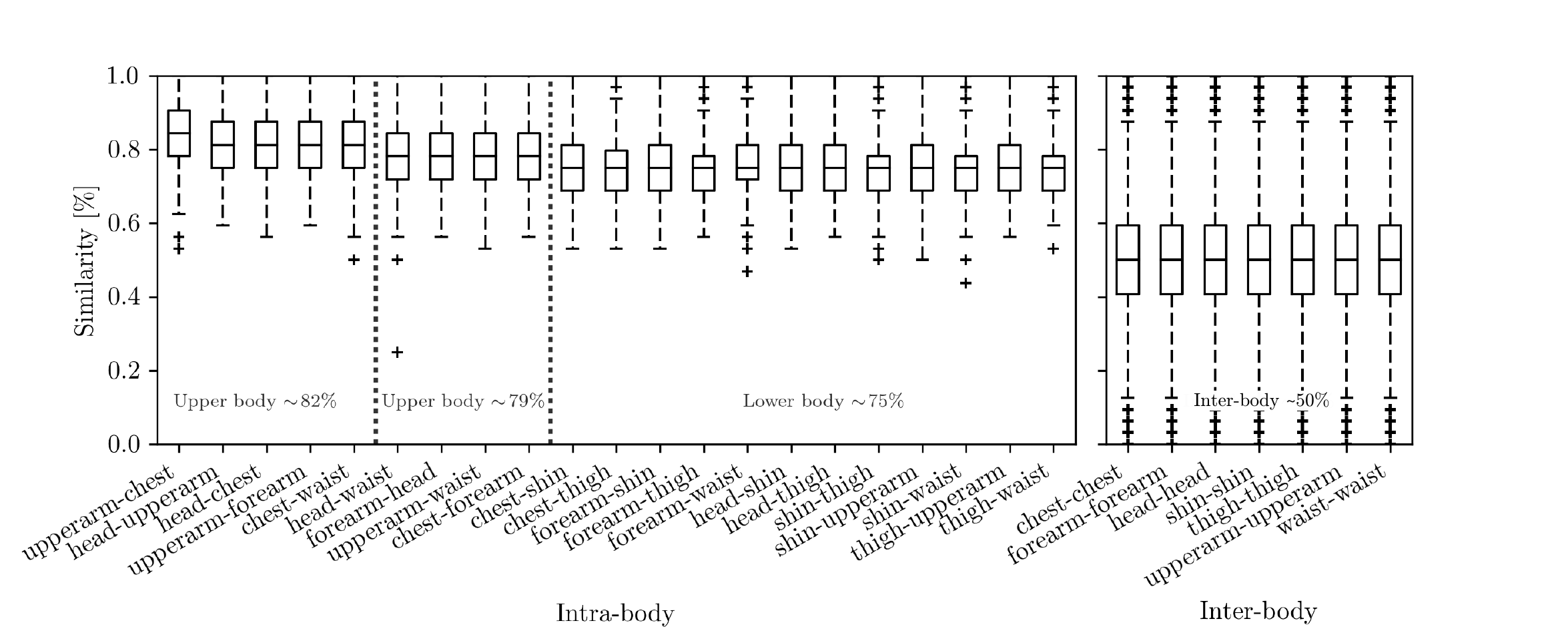}
        \caption{BANDANA}
        \label{fig:similarityBandana}
    \end{subfigure}
    \begin{subfigure}[t]{0.5\textwidth}
        \includegraphics[width=\textwidth, trim={.35cm 0 2.3cm 1cm},clip]{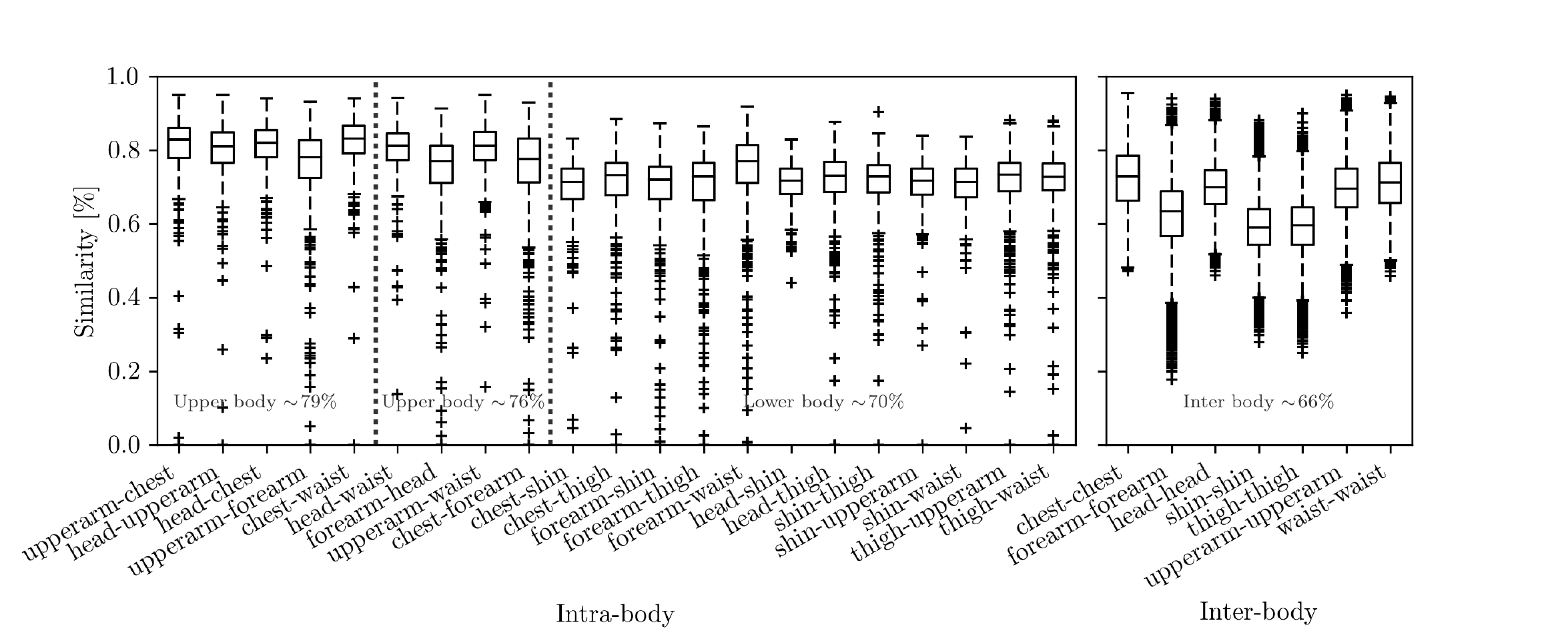}
        \caption{IPI}
        \label{fig:similarityIPI}
    \end{subfigure}

    \caption{Comparison of intra-body against inter-body similarity for the evaluated quantization schemes. Each value in the \emph{intra-body} boxplot is defined by the similarity of two \emph{different} sensor locations on the same subject (all possible combinations within each subject).
    For \emph{inter-body}, each boxplot defines a different sensor location. Only \emph{different} subjects are tested against each other with the \emph{same} sensor location.}
    \label{fig:similarity}
\end{figure*}
\subsection{SAPHE}
\label{sec:similaritySAPHE}
In the SAPHE~\cite{groza2012saphe} protocol, after generating and exchanging the hash $H(r_A)$ ($H(r_B)$) of the random seed $r_A$ ($r_B$) to compute threshold values $\overline{t}_A$ ($\overline{t}_B$), as points in an Acceleration-time coordinate system $\mathbb{K}$, devices derive acceleration sequences $\overline{v}_A$ ($\overline{v}_B$) in $\mathbb{K}$.
Challenges $c_A$ ($c_B$) that describe whether $\overline{t}_A$ ( $\overline{t}_B)$ exceed $\overline{v}_A$ ($\overline{v}_B$) are exchanged together with $r_A$ ($r_B$). 
The protocol does not disclose information on the acceleration during this communication.

We remark though, that the authors propose a second version which leaks information on the acceleration since, in addition, a distance ordering $\overline{o}_A$ ($\overline{o}_B$) between 
$\overline{t}_A$ ($\overline{t}_B$) and $\overline{v}_A$ ($\overline{v}_B$) is exchanged. 
The purpose of this distance ordering is to guard against a specific attack on the hash function (described in~\cite{groza2012saphe}).
However, an adversary could exploit that the threshold points $\overline{t}_A$ ($\overline{t}_B$) with small distance to $\overline{v}_A$ ($\overline{v}_B$) are good estimates of actual acceleration samples from $\overline{v}_A$ ($\overline{v}_B$). 
In addition, those threshold points $\overline{t}_A$ ($\overline{t}_B$) with large distance to $\overline{v}_A$ ($\overline{v}_B$) leak information on the probability of the resulting bit (0 or 1 for larger or smaller threshold). 

As depicted in Figure~\ref{fig:similaritySAPHE}, although affected by outliers, SAPHE's generated key pairs match with high probability of 85\% (lower body) to 86,87\% (upper body) on average on devices worn on the same body (intra-body).
The inter-body case matches on average with 55\% i.e. is 10\% higher than a random guess.
Conclusively, SAPHE is able to generate keys that fulfill the requirement of a clear boundary between intra- and inter-body similarity.

\subsection{Walkie-Talkie}
\label{sec:similarityWalkieTalkie}
The Walkie-Talkie protocol{~\cite{Xu_2016_WalkieTalkie}} is able to extract up to 1 key bit per acceleration sample.

Acceleration samples are interpreted as 0 or 1 conditioned on whether their acceleration is below or above a guard band, while samples that fall inside are ignored (cf. Figure~\ref{fig:quantizationWalkieTalkie}).
Walkie-Talkie has also been utilized in~\cite{revadigar2017smartwearables} to lock a fuzzy vault containing a random key.
The quantization is able to achieve higher bit rates by exploiting multiple thresholds~\cite{xu2017gaitkey}. 
We further discuss the impact of multiple thresholds in Section~\ref{sec:improving-gait-key}.

To mitigate hardware originated differences in acceleration strength, devices exchange and agree on samples in the acceleration sequence that shall constitute the key (\emph{reconciliation}). 
The resulting sequence is thought to be biased towards alternating groups of 1-bits and 0-bits, which is addressed by applying an XOR between consecutive 30 bit long windows. 
We comment on this concern in~\ref{sec:RelWork:Attacks}.

The protocol achieves 60-70\% upper body bit-similarities and 55-65\% for the lower body (cf. Figure~\ref{fig:similarityWalkieTalkie}).
This performance suggests further processing to provide reliable pairing among devices at different body location.
Walkie-Talkie uses Independent Component Analysis as a preprocessing step in order to remove the arm swing (Figure~\ref{fig:protocols}). 

In our implementation, the transformation to the body coordinate system was applied following Mohssen et al.~\cite{mohssen2014human}.
Walkie-Talkie was then executed using the best performing parameters as mentioned in~\cite{Xu_2016_WalkieTalkie}, such as an $\alpha$ of $0.8$ and non-overlapping windows of size $10$.
We did not resample the input data as Walkie-Talkie applies a low pass filter with a cutoff frequency of 10Hz during the preprocessing.
Independent Component Analysis was applied on the complete recording beforehand.
We decided to only exclude arm swing components where they were clearly distinguishable. 

\subsection{BANDANA}
\label{sec:bandanaSimilarity}
In BANDANA, key sequences are generated as a function of the difference between mean and instantaneous acceleration~\cite{schuermann2018jagger}. 
The approach of comparing to the mean at a particular body location serves as a normalisation procedure.
The offset to the mean has a better correlation across various body locations than comparing absolute acceleration values.
Furthermore, \cite{hoang2015gait} argues that this approach might positively impact the distribution of bits in the key sequences towards uniformity as gait patterns are compared to their mean.
To further amplify similarity of sequences generated at different body locations, bits with low difference between mean and instantaneous gait are disregarded. 

The similarity between keys generated at different positions on the body is depicted in Figure~\ref{fig:similarityBandana} for the BANDANA protocol.
The protocol achieves similarity results above 75\% for all location-pairs and is able to render the chances of the adversary (inter-body) to random guess. 
The protocol employs fuzzy cryptography in order to mitigate the remaining 25\% of difference in the key sequences.
We observe, however, a high variance for the inter-body case, which is due to a non-uniform distribution of the key sequences in the key space (cf. Section~\ref{sec:randomness} and Section~\ref{sec:SecurityAnalysis}). 
In Section~\ref{sec:improving}, we discuss how this problem can be addressed with a revised quantization approach.

\subsection{IPI}
\label{sec:IPISimilarity}
The Inter-Pulse-Interval (IPI) protocol~\cite{sun2017secure} exploits the random offset by which individual steps deviate from the mean gait cycle in time domain (cf. Figure~\ref{fig:quantizationIPI}). 
The number of secret bits that can be extracted from the gait signal then depends on the sampling frequency as gait cycle estimation is more accurate with higher sampling rate.
The authors report a standard deviation of 40.8 milliseconds for the IPI. 

Figure~\ref{fig:similarityIPI} shows the similarity achieved for IPI between keys generated from devices located at different positions on the body. 
The similarity in the intra-body case is good. 
IPI also employs fuzzy cryptography to correct remaining bit-errors in the keys generated for devices across the same body. 
However, the figure also shows that the protocol does not prevent a remote adversary from paring with on-body devices, since inter-body similarities are as high as in the intra-body case. 
This is due to limited variation in the generated bit sequences. 
Inter-pulse intervals resemble a normal distribution centered around its mean.
This variation around the mean is similar across subjects and the resolution employed is 4 bits only so that naturally similarity across generated bit sequences is high (cf. Section~\ref{sec:SecurityAnalysis}). 
\begin{figure}
    \begin{subfigure}[t]{0.5\columnwidth}
        \includegraphics[width=\columnwidth]{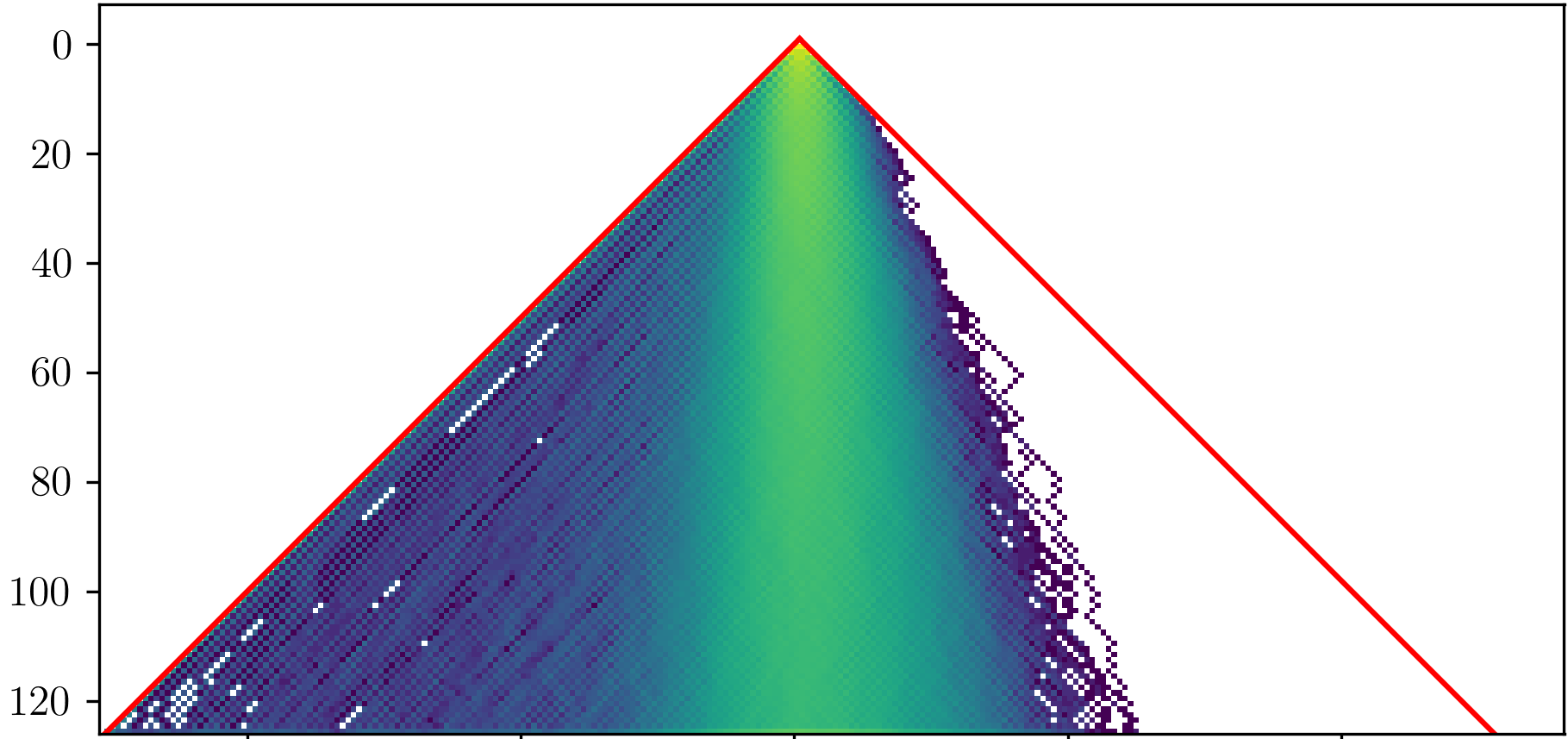}
        \caption{SAPHE}
        \label{fig:heatmapSAPHE}
    \end{subfigure}
    \begin{subfigure}[t]{0.5\columnwidth}
        \includegraphics[width=\columnwidth]{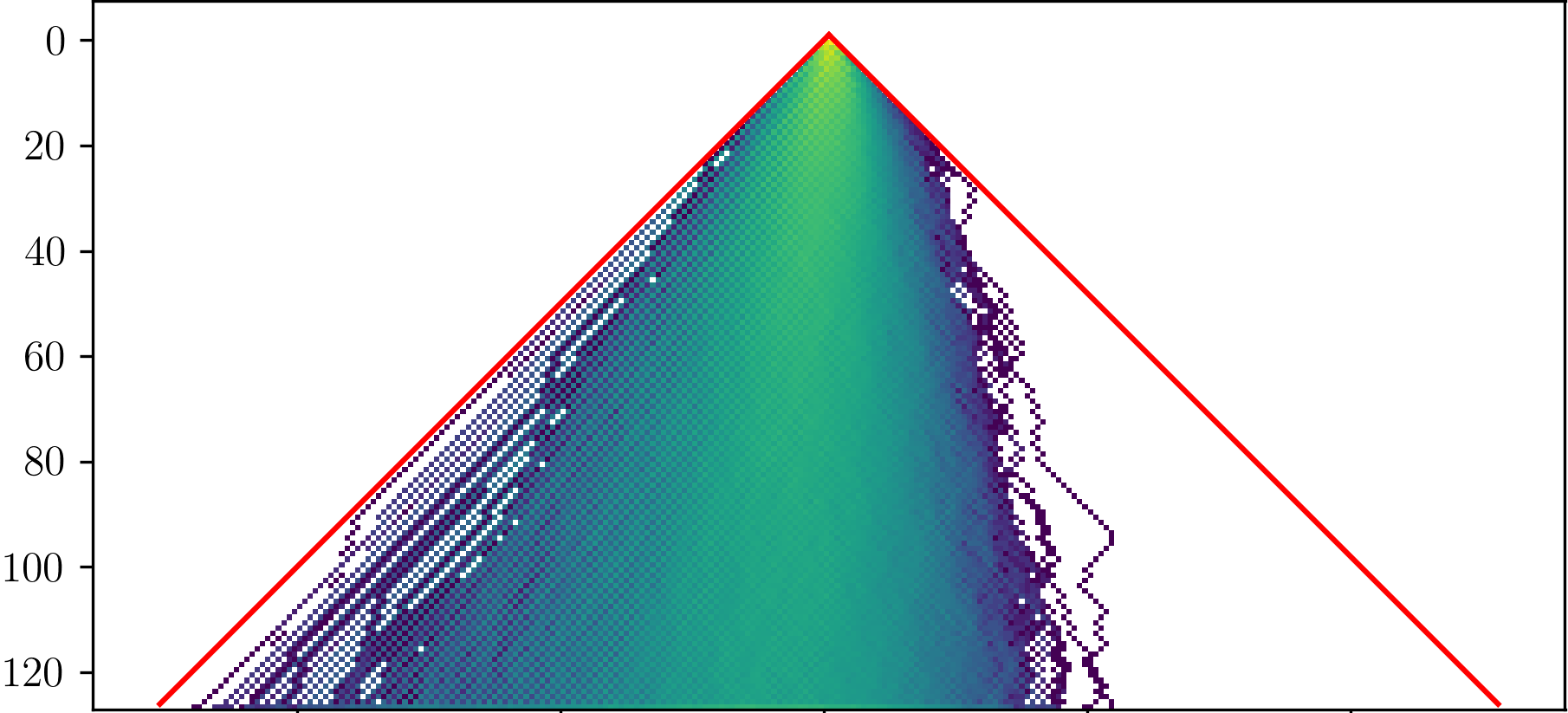}
        \caption{Walkie-Talkie}
        \label{fig:heatmapWalkieTalkie}
    \end{subfigure}
    
    \vspace*{.2cm}
    
    \begin{subfigure}[t]{0.5\columnwidth}
        \includegraphics[width=\columnwidth]{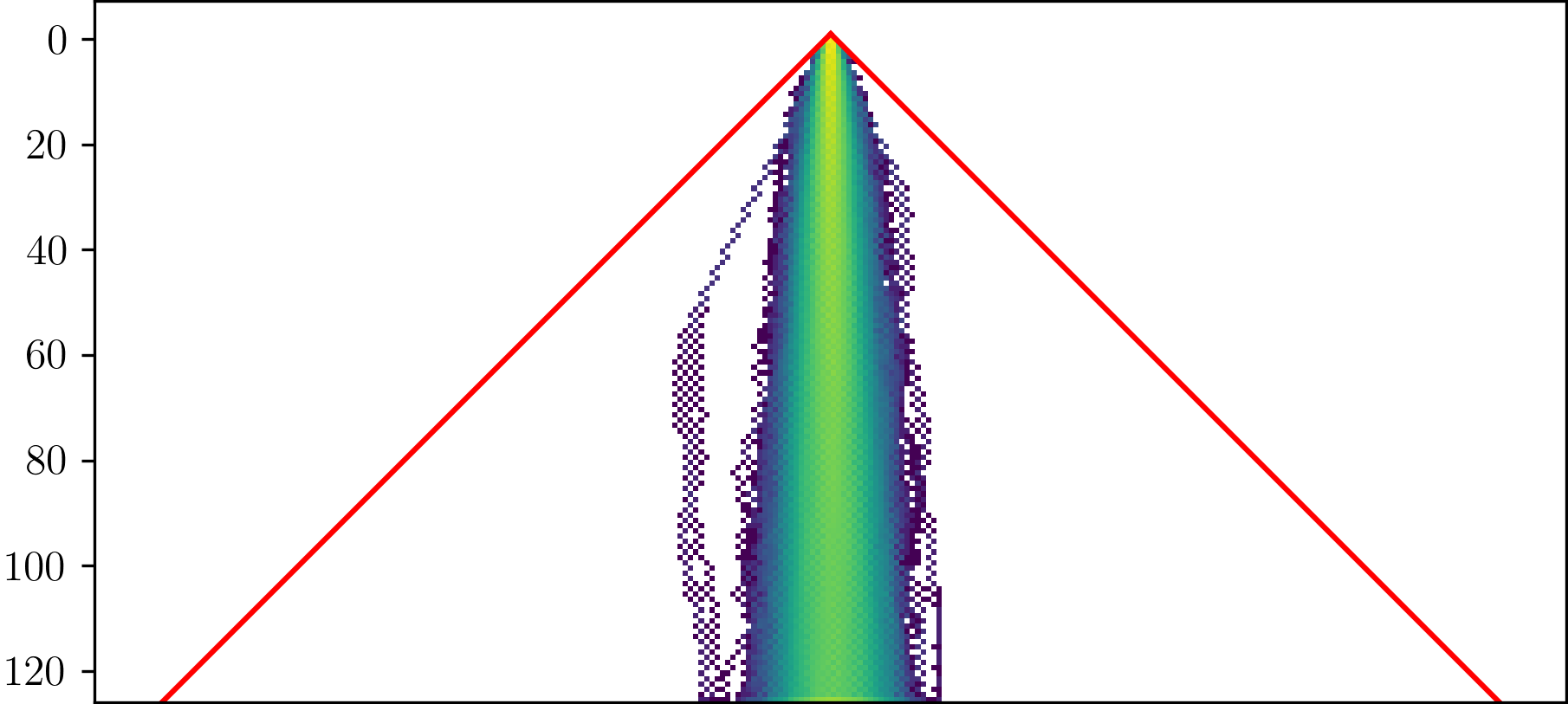}
        \caption{BANDANA}
        \label{fig:heatmapBandana}
    \end{subfigure}
    \begin{subfigure}[t]{0.5\columnwidth}
        \includegraphics[width=\columnwidth]{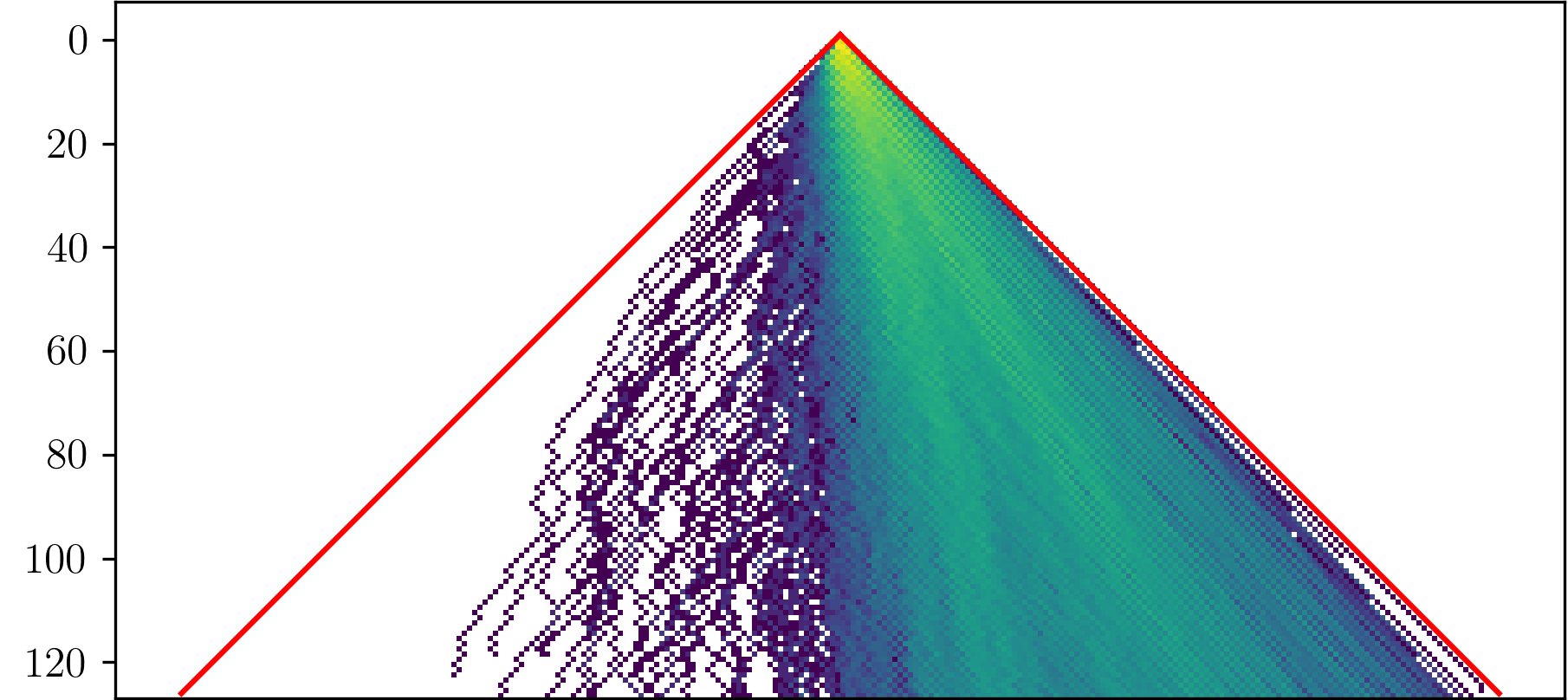}
        \caption{IPI}
        \label{fig:heatmapIPI}
    \end{subfigure}
    
    \caption{Heatmaps of random walks for 128 bit keys generated by the evaluated quantization schemes ($0\rightarrow \text{left}; 1\rightarrow \text{right}$). The red lines depict the boundaries for any possible random walk.}
    \label{fig:heatmap}
\end{figure}

\section{Randomness of Keys}
\label{sec:randomness}
In this section we investigate whether these keys are \emph{sufficiently unpredictable} to withstand a computationally unconstrained adversary.
For this, we analyze the randomness of keys and the results from the DieHarder and ENT Pseudorandom Number Sequence Tests.

\subsection{Bit Distribution}
To describe the randomness of keys, we compare their structure with random walks on a Galton board ~\cite{wangstochastic}.

\begin{figure}
    \begin{subfigure}[t]{0.5\columnwidth}
        \includegraphics[width=\columnwidth,height=2.5cm,trim={0 .65cm 0 0}]{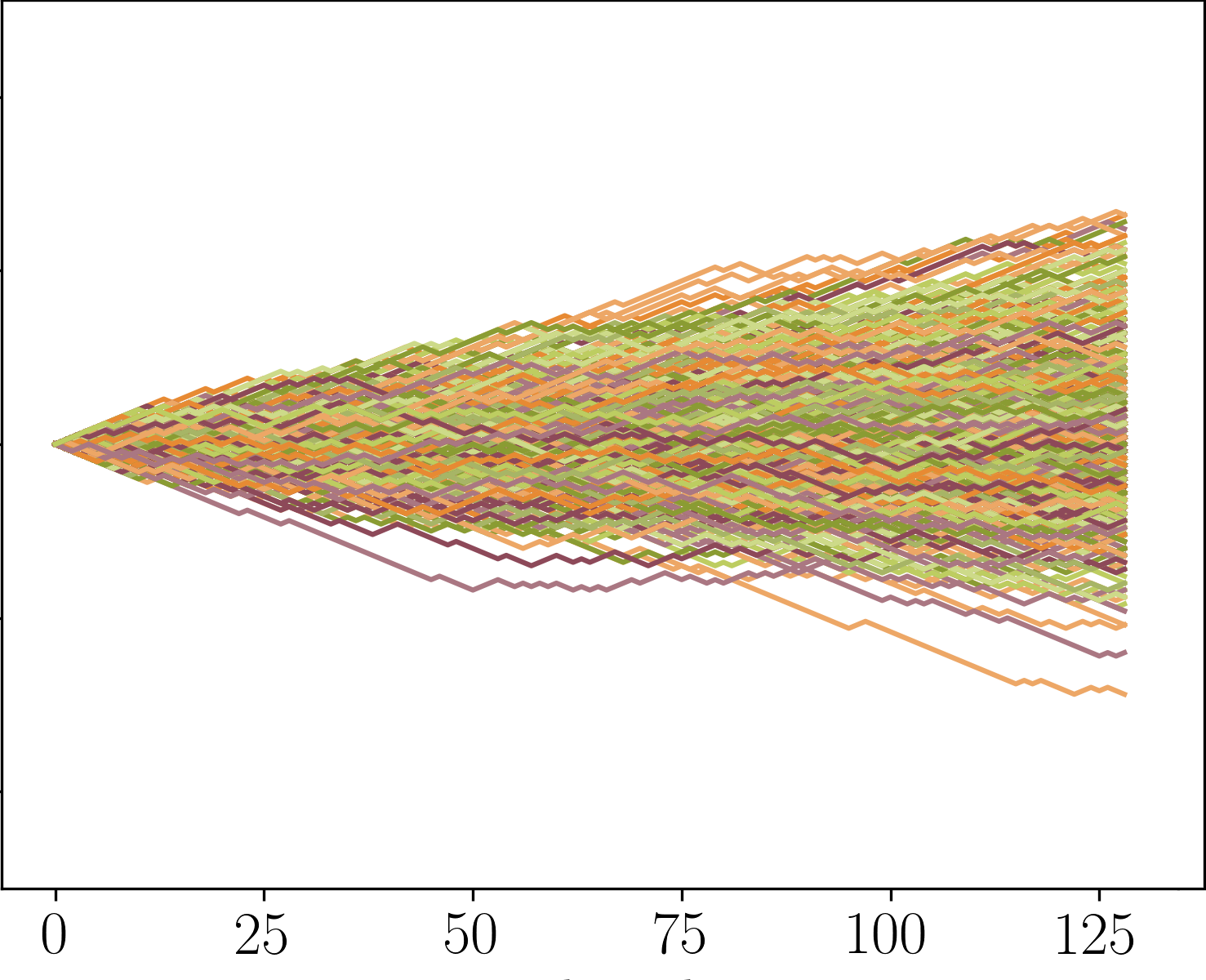}
        \caption{SAPHE}
        \label{fig:SapheStrings}
    \end{subfigure}
    \begin{subfigure}[t]{0.5\columnwidth}
        \includegraphics[width=\columnwidth,height=2.5cm,trim={0 .65cm 0 0}]{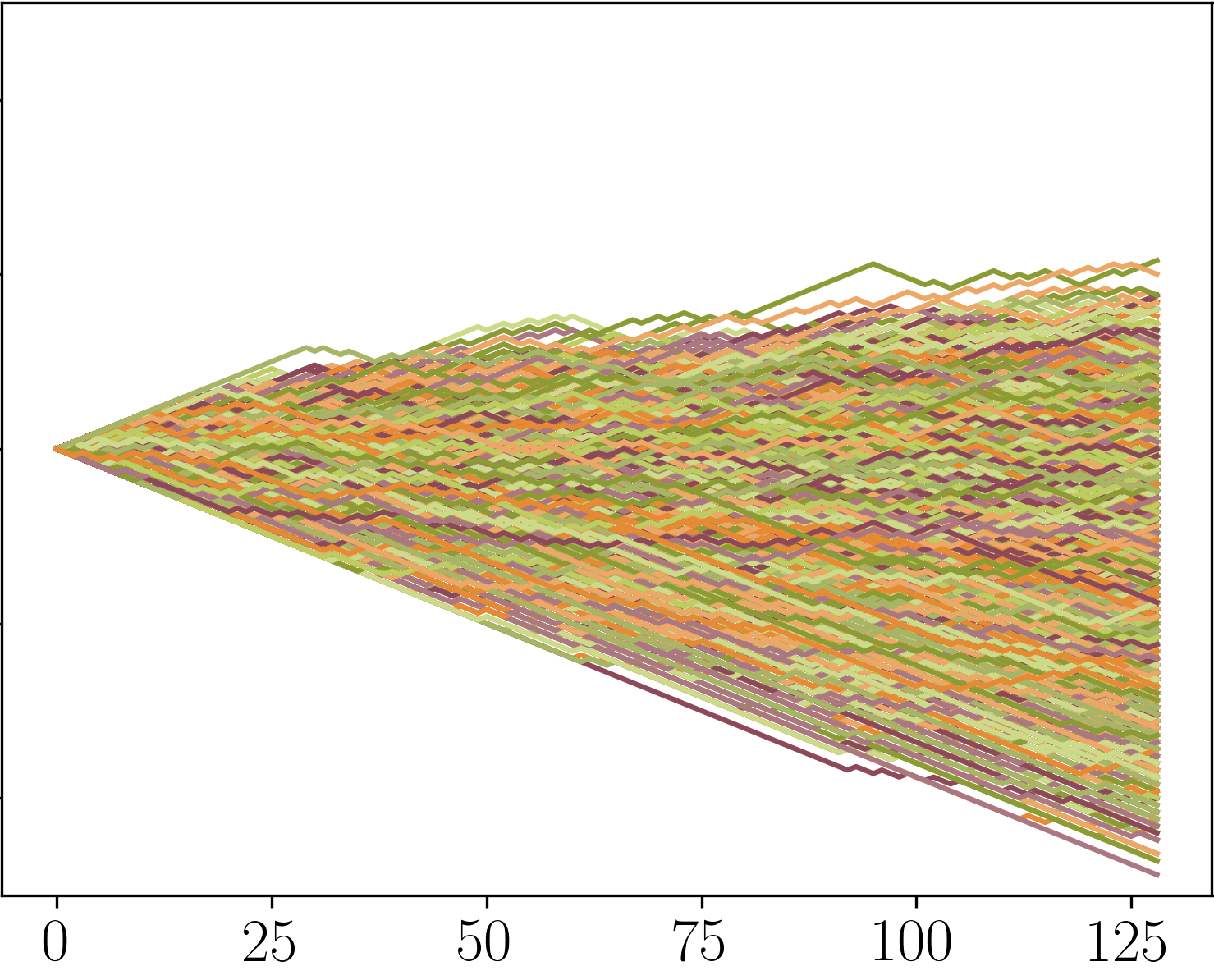}
        \caption{Walkie-Talkie}
        \label{fig:WktkStrings}
    \end{subfigure}
    
    \vspace*{.5cm}
    
    \begin{subfigure}[t]{0.5\columnwidth}
        \includegraphics[width=\columnwidth,height=2.5cm,trim={0 .65cm 0 .7cm}]{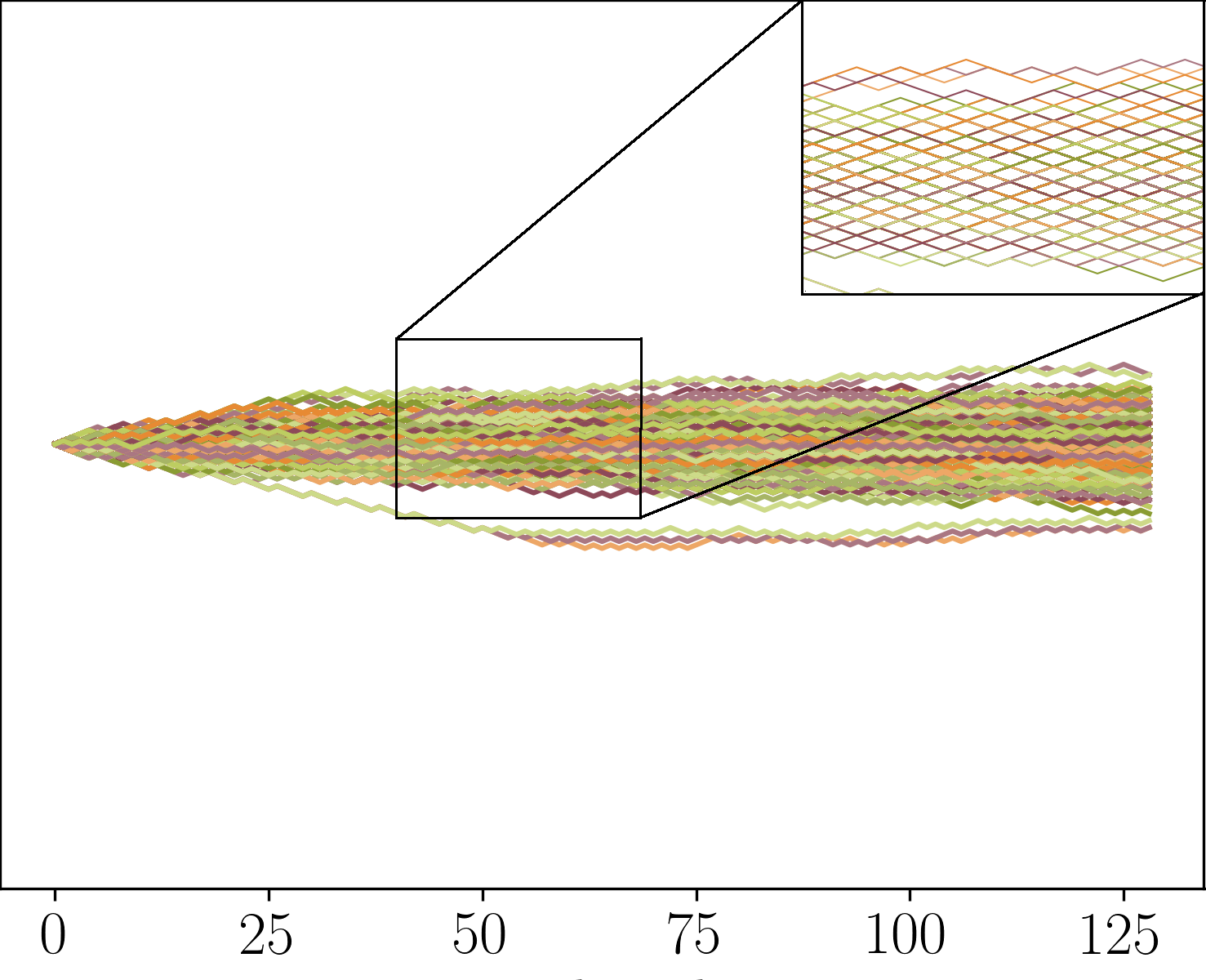}
        \caption{BANDANA}
        \label{fig:BandanaStrings}
    \end{subfigure}
    \begin{subfigure}[t]{0.5\columnwidth}
        \includegraphics[width=\columnwidth,height=2.5cm,trim={0 .65cm 0 .7cm}]{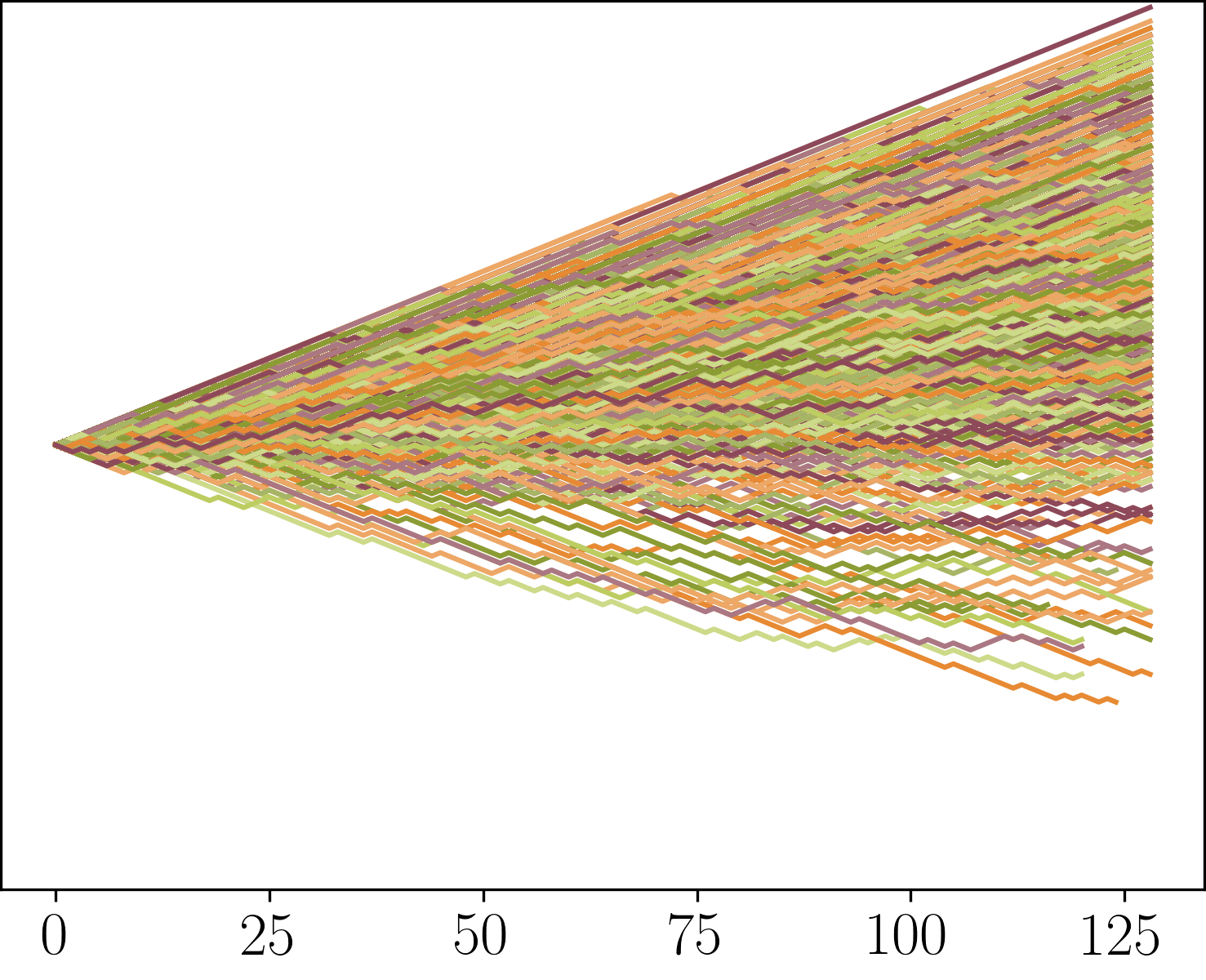}
        \caption{IPI}
        \label{fig:IPIStrings}
    \end{subfigure}
    \caption{Cumulative plot of random walks for 128 bit keys generated by the evaluated quantization schemes ($0\rightarrow \text{bottom}; 1\rightarrow \text{top}$)}
    \label{fig:strings}
\end{figure}

\begin{figure}
    \begin{subfigure}[t]{0.5\columnwidth}
         \includegraphics[width=\columnwidth,height=2.5cm,trim={.5cm 1.65cm .5cm .7cm}]{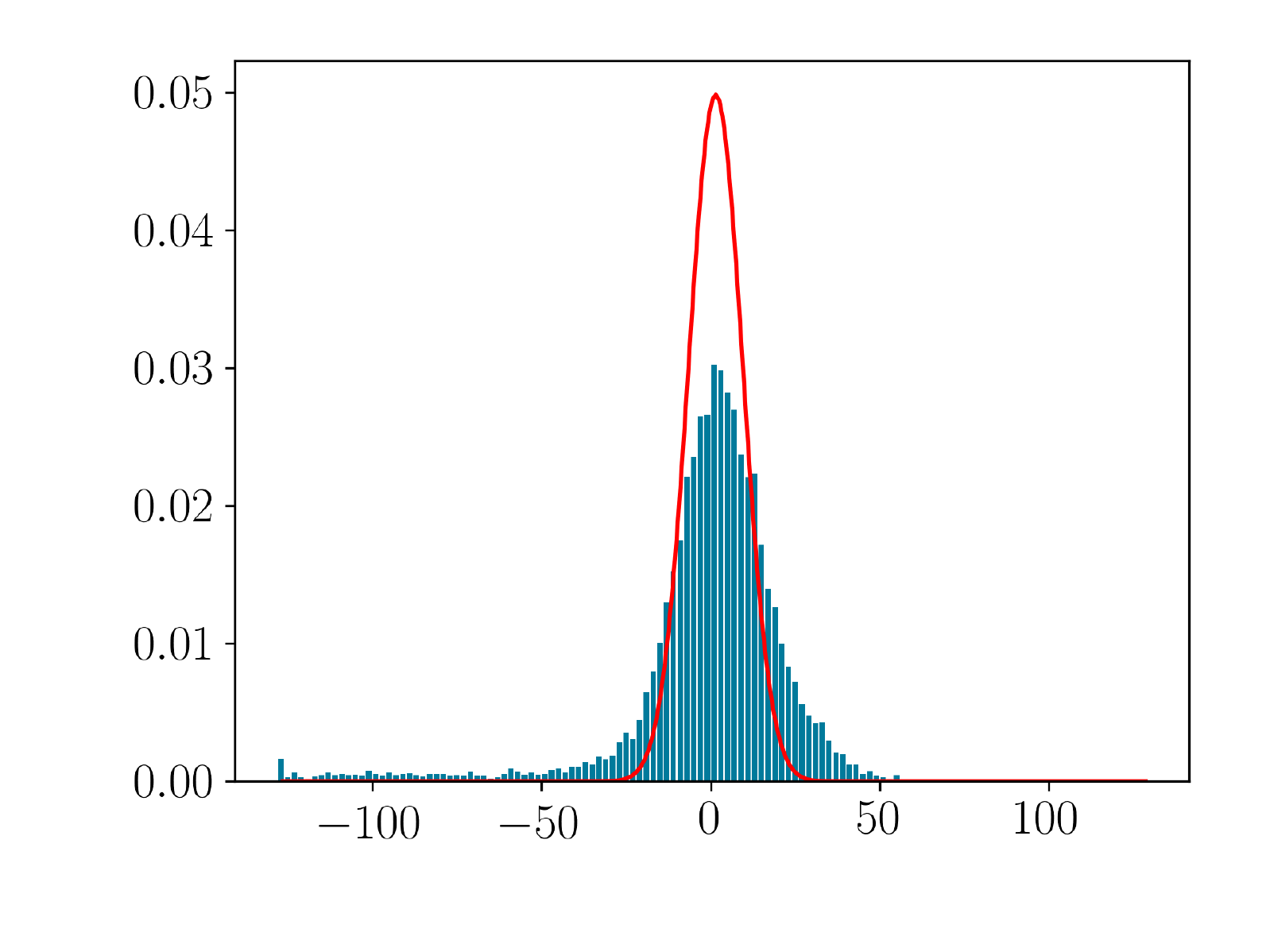}
        \caption{SAPHE}
        \label{fig:SapheDistribution}
    \end{subfigure}
    \begin{subfigure}[t]{0.5\columnwidth}
         \includegraphics[width=\columnwidth,height=2.5cm,trim={.5cm 1.65cm .5cm .7cm}]{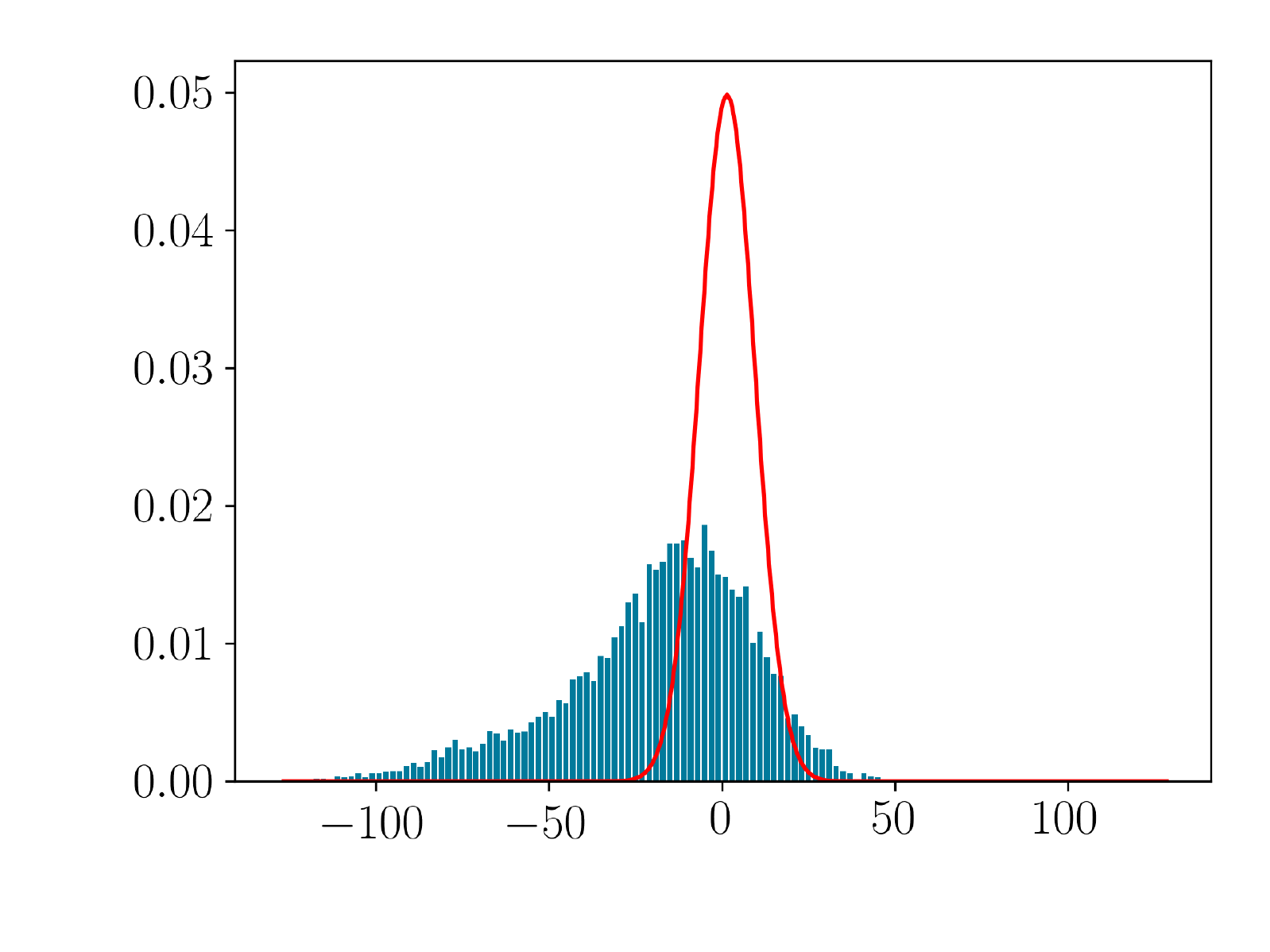}
        \caption{Walkie-Talkie}
        \label{fig:WktkDistribution}
    \end{subfigure}
    
    \vspace*{.5cm}
    
    \begin{subfigure}[t]{0.5\columnwidth}
         \includegraphics[width=\columnwidth,height=2.5cm,trim={.5cm 1.65cm .5cm 1.5cm}]{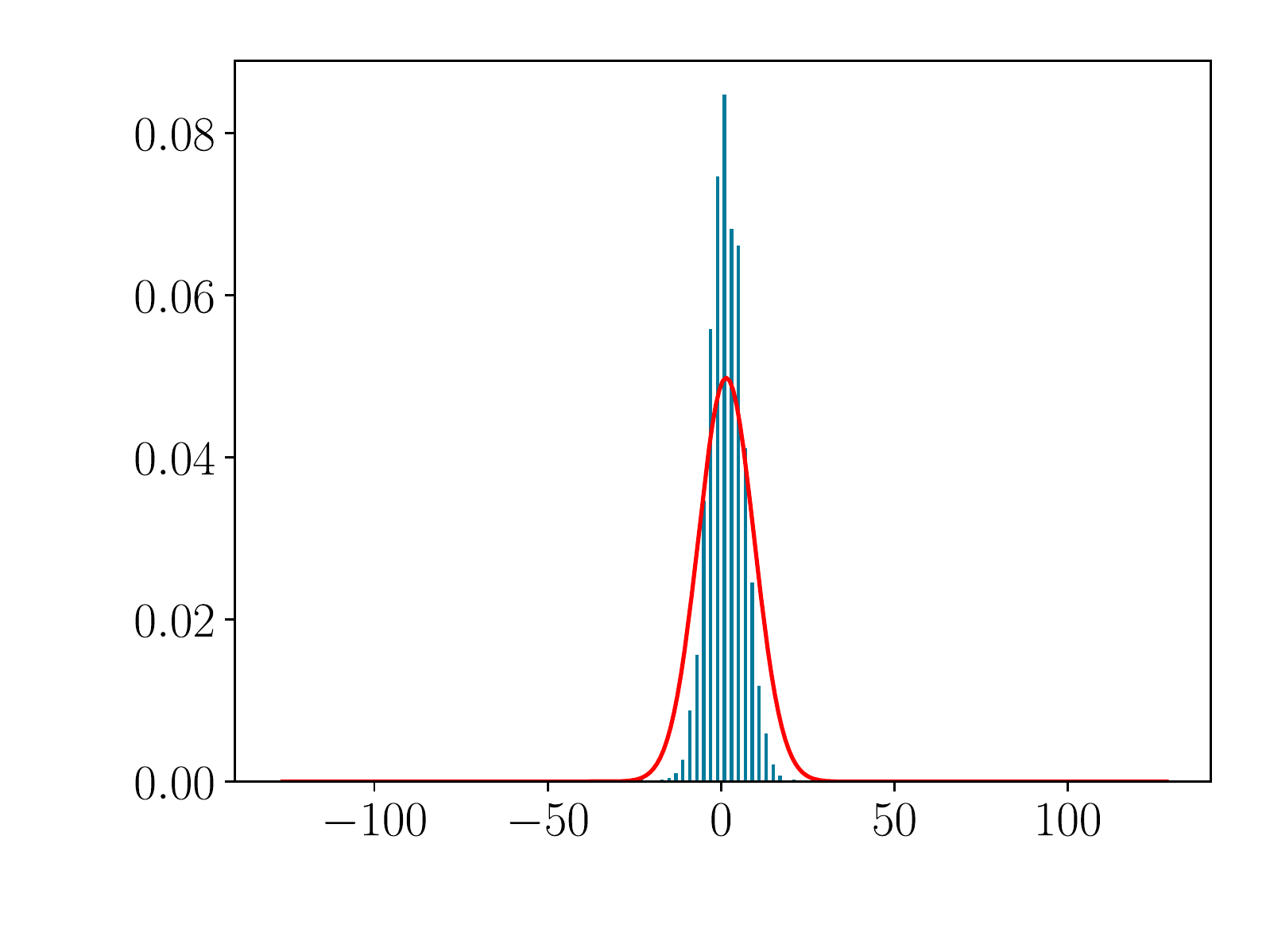}
        \caption{BANDANA}
        \label{fig:BandanaDistribution}
    \end{subfigure}
    \begin{subfigure}[t]{0.5\columnwidth}
         \includegraphics[width=\columnwidth,height=2.5cm,trim={.5cm 1.65cm .5cm 1.5cm}]{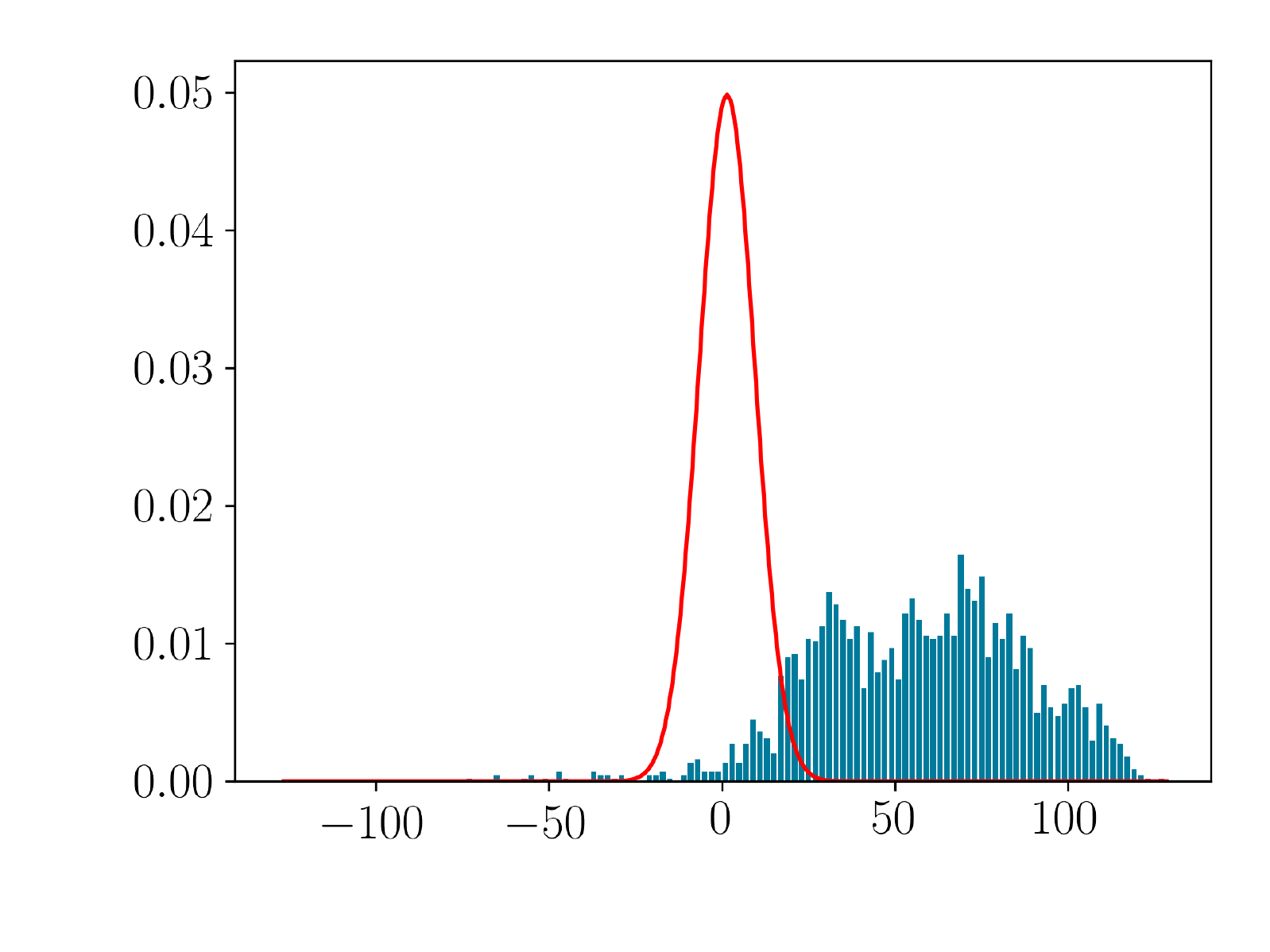}
        \caption{IPI}
        \label{fig:IPIDistribution}
    \end{subfigure}
    
    \caption{Cumulative sums distribution for 128 bit keys (distribution in the last rows in Figure~\ref{fig:heatmap}). 
    Expected biomial distribution in red.}
    \label{fig:distribution}
\end{figure}

Plotting a sufficient amount of these sequences will eventually
show a binomial distribution.
Figure~\ref{fig:heatmap} shows heatmaps of random walks corresponding to the sequences generated by different quantization approaches.
In addition, Figure~\ref{fig:strings} depicts each individual random walk such that specific patterns are observable.
Based on the last row of each heatmap, Figure~\ref{fig:distribution} depicts the cumulative sums distribution.

Key lengths {of 128 bits are chosen for this study,} which means that the acceleration sequence to generate a key varies between the different approaches. 

SAPHE shows a close-to symmetric distribution centered around the mean. 

\begin{figure}
    \centering
    \includegraphics[width=\columnwidth]{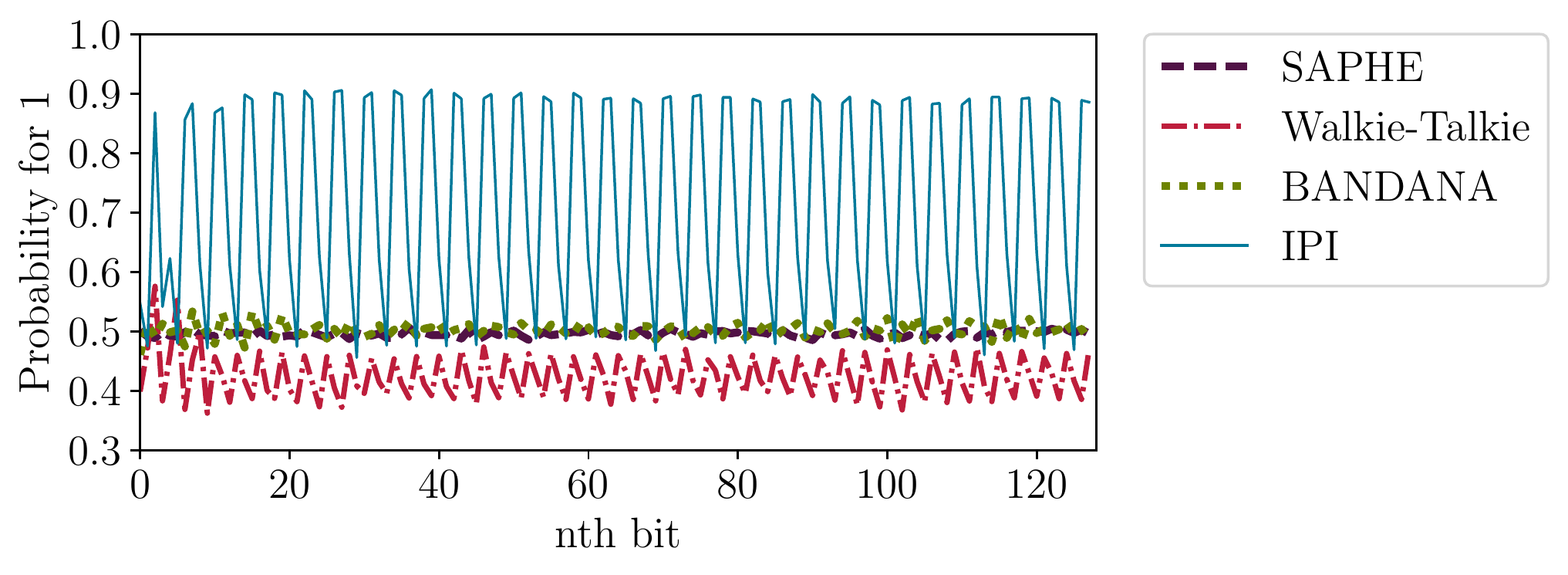}
    \caption{Markov property: Probability of assigning 1 for the $n$th position in 128 bit keys}
    \label{fig:markov}
\end{figure}

The cumulative sums distribution is properly centered but shows deviations to include more `0's for a specific set of keys (cf. Figure\ref{fig:SapheDistribution}). 
We explain this with the characteristic of acceleration readings in our data, which do not necessarily have zero-mean.
{With regard to the binomial distribution (depicted in red), SAPHE's key distribution is slightly stretched.}
Thus, while SAPHE shows good behaviour regarding similarity and usage of space in the Galton board, it carries some characteristics of the input into the output data.
Still, this does not pave the way for a successful attack.
Assuming each bit position to be a state in a Markov chain, Figure~\ref{fig:markov} shows the resulting transition probabilities{, aggregated over all sequences}.
SAPHE shows a good Markov property (cf. Figure~\ref{fig:markov}).

The heatmap and distribution of Walkie-Talkie are depicted in Figure~\ref{fig:heatmapWalkieTalkie} and Figure~\ref{fig:WktkDistribution}.
The individual sequences do not show a bias (cf. Figure~\ref{fig:WktkStrings}).
Walkie-Talkie, however, shows periodicity in the Markov property (cf. Figure~\ref{fig:markov}).
The BANDANA approach features symmetric behaviour but with low variance (cf. Figure~\ref{fig:heatmapBandana}, \ref{fig:BandanaDistribution}).
We can observe from Figure~\ref{fig:BandanaStrings}, that this weakness occurs since bit sequences consist of repetitive `zig-zag' patterns.
We discuss this problem in Section~\ref{sec:SecurityAnalysis} and propose an improved quantization to mitigate it in Section~\ref{sec:improving}.
BANDANA shows a similar Markov property as SAPHE (cf. Figure~\ref{fig:markov}).
Finally, IPI shows good variance but a bias towards including more ones than zeros due to low variation in the quadruples generated as discussed above.
IPI clearly deviates from a binomial distribution  (cf. Figure~\ref{fig:IPIDistribution}).
We observed that consecutive 4-bit chunks repeat with a probability of 60\%.
This clearly shows in IPI's Markov property in Figure~\ref{fig:markov}.
Summarizing, while SAPHE and Walkie-Talkie exhibit reasonable randomness, BANDANA and IPI show biases in the generated keys.

\begin{figure*}
    \begin{subfigure}[t]{0.49\textwidth}
        \includegraphics[width=\textwidth, height=2.6cm]{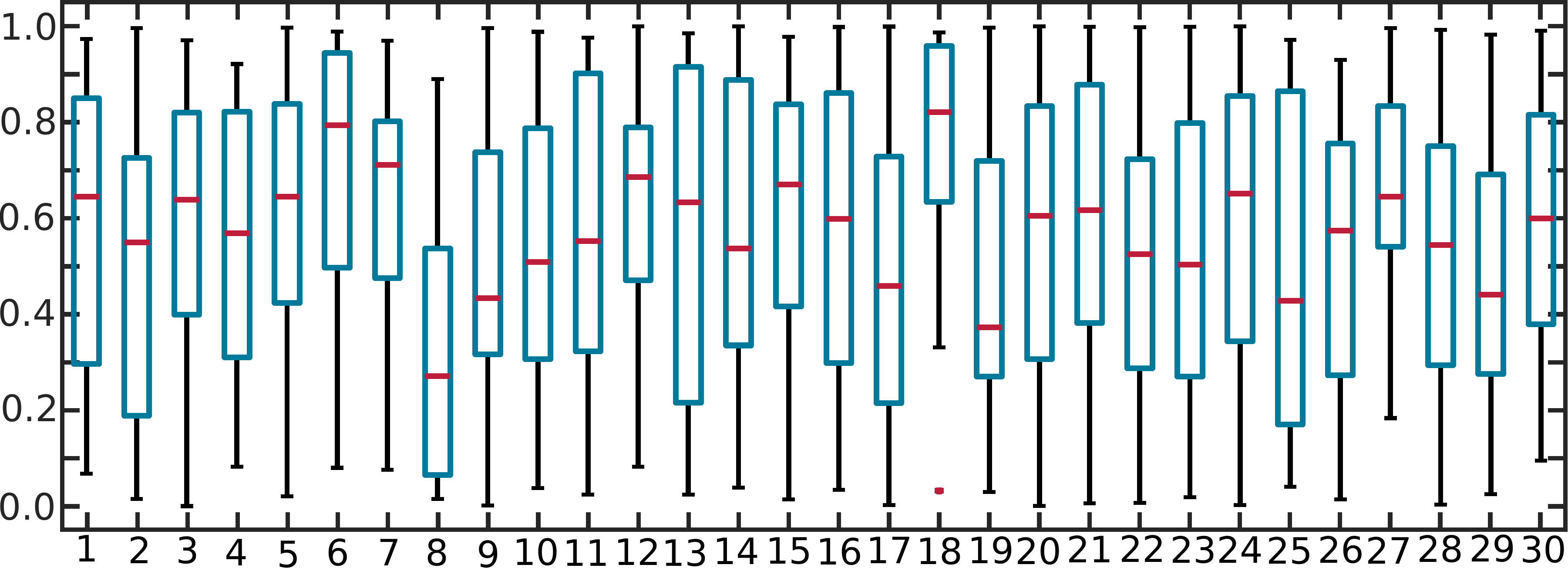}
        \caption{SAPHE}
        \label{fig:statisticsSAPHE}
    \end{subfigure}
    \quad
    \begin{subfigure}[t]{0.49\textwidth}
        \includegraphics[width=\textwidth, height=2.6cm]{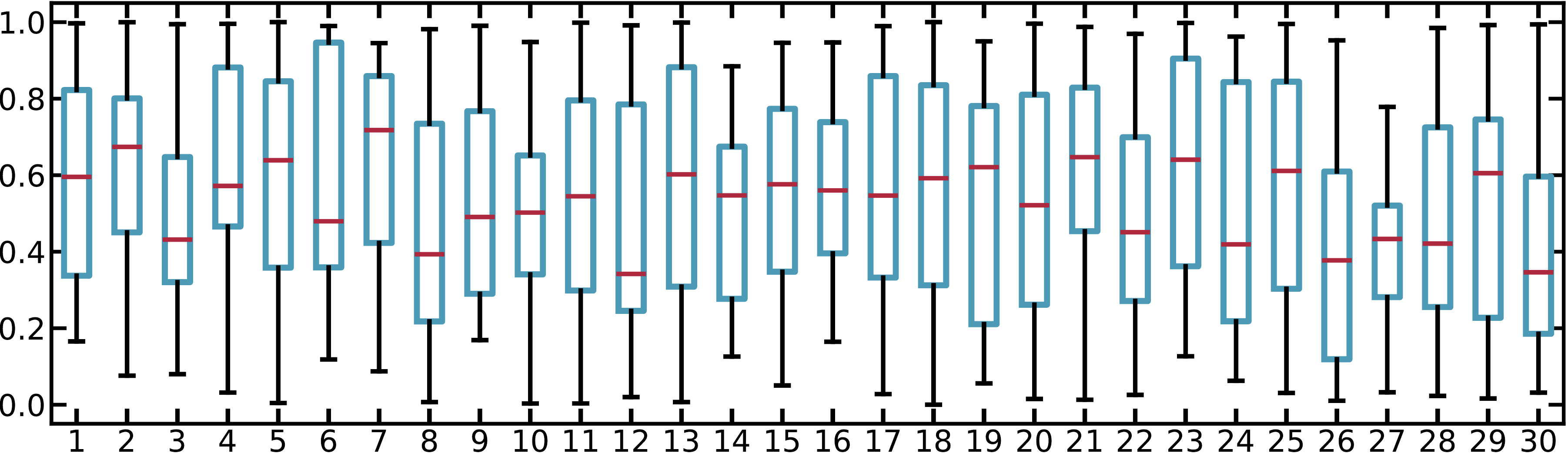}
        \caption{Walkie-Talkie}
        \label{fig:statisticsWalkieTalkie}
    \end{subfigure}
    
    \vspace{.2cm}
    
    \begin{subfigure}[t]{0.49\textwidth}
        \includegraphics[width=\textwidth, height=2.6cm]{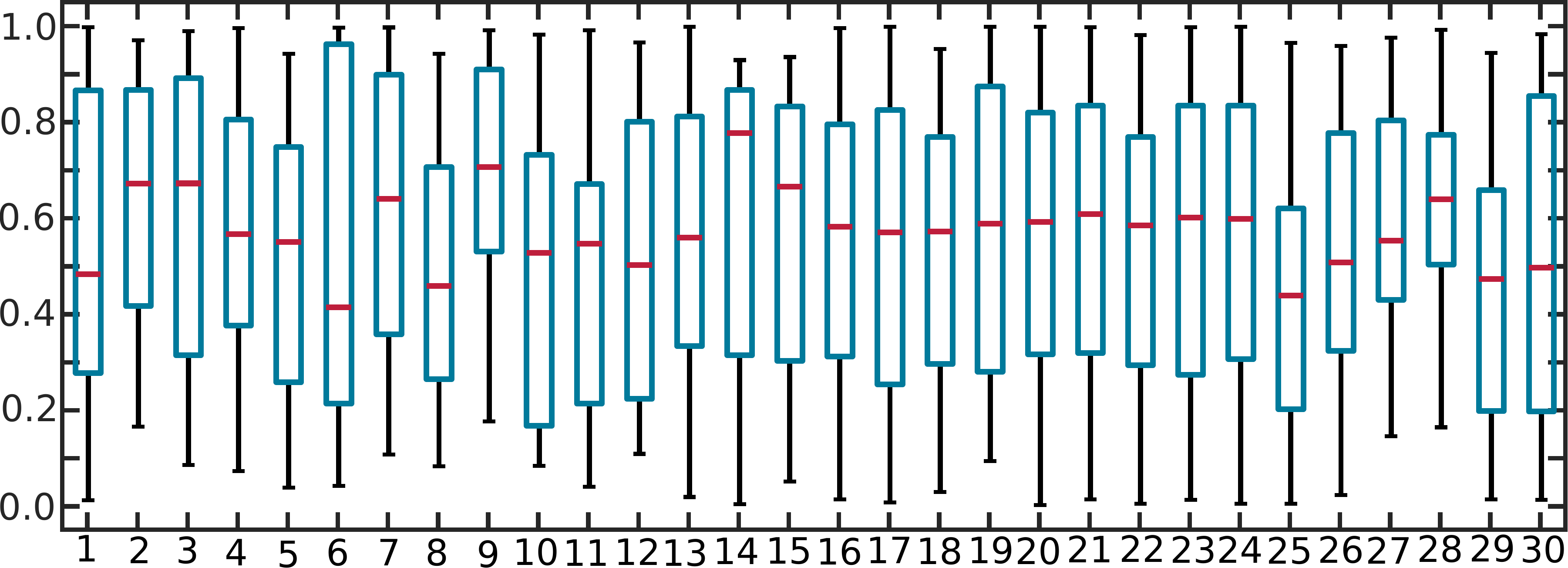}
        \caption{BANDANA}
        \label{fig:statisticsBandana}
    \end{subfigure}
    \quad
    \begin{subfigure}[t]{0.49\textwidth}
        \includegraphics[width=\textwidth, height=2.6cm]{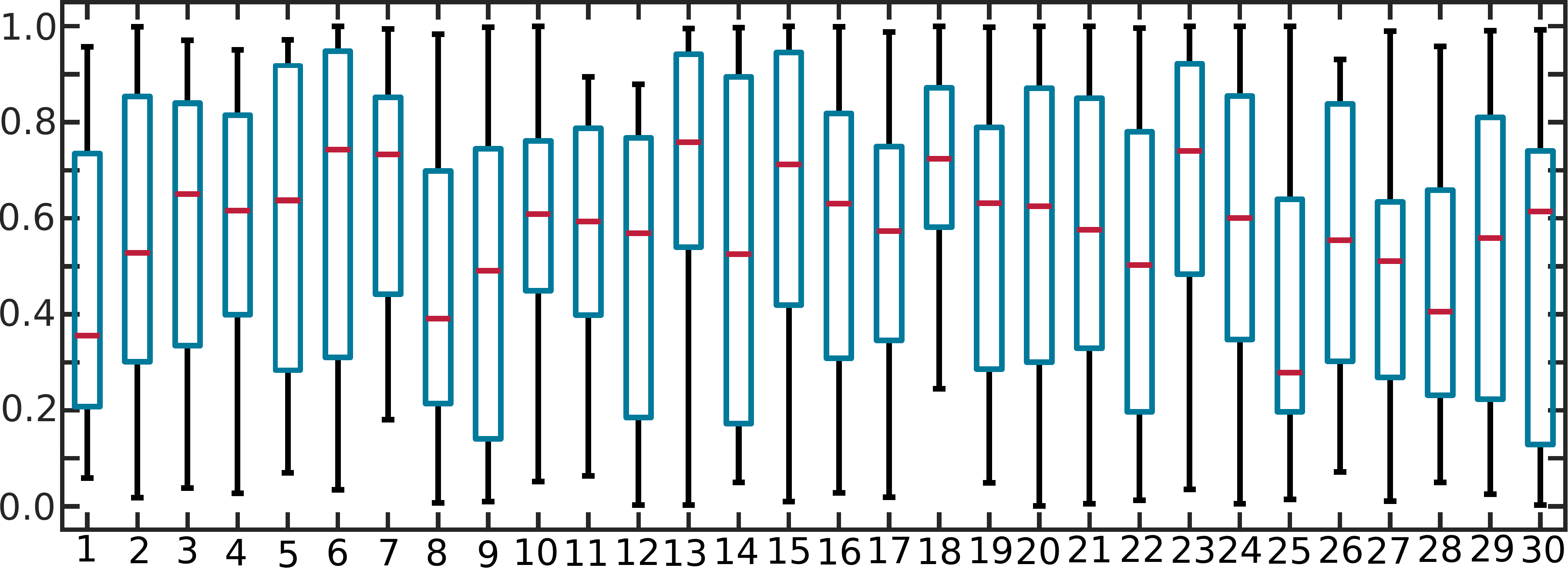}
        \caption{IPI}
        \label{fig:statisticsIPI}
    \end{subfigure}

    \caption{Distribution of p-values achieved for keys after 20 runs of the DieHarder set of statistical tests. Tests are: (1) birthdays (2) operm5 (3) rank32x32 (4) rank6x8 (5) bitstream (6) opso (7) oqso (8) dna (9) count-1s-str (10) count-1s-byt (11) parking (12) 2D circle (13) 3D sphere (14) squeeze (15) runs (16) craps (17) marsaglia (18) sts monobit (19) sts runs (20) sts serial [1-16] (21) rgb bitdistr. [1-12] (22) rgb min dist. [2-5] (23) rgb perm. [2-5] (24) rgb lagged sum [0-32] (25) rgb kstest (26) dab bytedistr. (27) dab dct (28) dab filltree (29) dab filltree 2 (30) dab monobit 2}
    \label{fig:statistics}
\end{figure*}

\subsection{Statistical Tests}
To test the evaluated quantization schemes against bias in the produced random sequences, we ran the DieHarder statistical tests for each scheme.
Figure~\ref{fig:statistics} depicts the p-values computed from 20 runs of the DieHarder tests.

In SAPHE, the {\em dna} and {\em sts monobit} tests appear to be outliers. 
The {\em dna} test considers biases in the occurence of 10 letter words from an alphabet of 4 letters, 
determined by two designated bits in the sequence of random integers being tested. 
The {\em sts monobit} test counts the 1 bits in a long string of random entries and compares this to the expected number. 
Similar to SAPHE, Walkie-Talkie also shows a weakness in the {\em dna} test. 
In addition, the {\em rgb Kolmogorov-Smirnov} test falls out slightly and the {\em 2D sphere} test features some outliers. 
The {\em kolmogorov-Smirnov} test applies a {\em Kuiper KS} test~\cite{Statistical_Kuiper_1962} and the {\em 2D circle} test finds the minimum distance between pairs of randomly selected points to evaluate their randomness. 
BANDANA shows the most stable distribution of p-values. 
A slight bias might be associated with the {\em squeeze} test, which employs a {\em chi-square} test for cell frequencies on the number of multiplication with random integers that are required to reduce $2^{31}$ to $1$.
IPI shows potential weaknesses towards the {\em birthdays} test, the {\em Overlapping Quadruples Sparce Occupancy (oqso)} test, the {\em 3D sphere} test as well as the {\em rgb permutation} and {\em rgb Kolmogorov Smirnov} test.
The {\em rgb permutation} test counts the order of permutations of random numbers. 
{\em Birthdays} test determines the number of matching intervals from 512 `birthdays' drawn from a 24-bit `year' while the {\em oqso} test, similar to the {\em dna} test, considers 4-letter words from an alphabet of 32 letters. 

Additionally, we ran the {\em Ent Pseudorandom Number Sequence} Test\footnote{http://www.fourmilab.ch/random/}.
The information density of bit sequences is computed together with reduction through optimal compression, chi square distribution, arithmetic mean of data bytes as well as serial correlation coefficient (cf. Table~\ref{tab:entropy}). 
We caution that these results are only showing the interdependence of single bits.
Evaluating chunk instead of single bit interdependence, such as 4-bit chunks for BANDANA due to its 4 bit per gait cycle or 30-bit chunks for Walkie-Talkie's privacy amplification, heavily influences the test results.

\setlength{\tabcolsep}{2pt}
\ctable[
    caption = {Results for keys generated by the evaluated protocols after running the ENT Pseudorandom Number Sequence Test Program.},
    label = tab:entropy,
    pos = tbp,
    width=\columnwidth,
    doinside=\scriptsize
]{Xrrrr}{%
}{                          \FL
& \textbf{SAPHE} & \textbf{Walkie-Talkie} & \textbf{BANDANA} & \textbf{IPI} \ML
Sequence size (bit) &1444864 & 3040848  &113792&456104 \NN
Entropy (bits per bit) &0.9999 & 0.9855 &0.9999&0.8929 \NN
Optimum compression rate &0\% &1 \%&0\%&10\% \NN
Chi square distribution &6.91 &61013.17&0.3586&65969.75\NN
Arithmetic mean (random=.5)&0.501094 &0.429175&0.5&0.690156\NN
Monte Carlo Pi value (error) & 3.122155 & 3.331471 &3.642194&2.056830\NN
Serial correlation coefficient & 0.008204& 0.055243 &-0.644796 &-0.002701\NN
(uncorrelated=0.0)& & & & \LL
}

\begin{figure}
    \centering
    \includegraphics[width=.7\columnwidth]{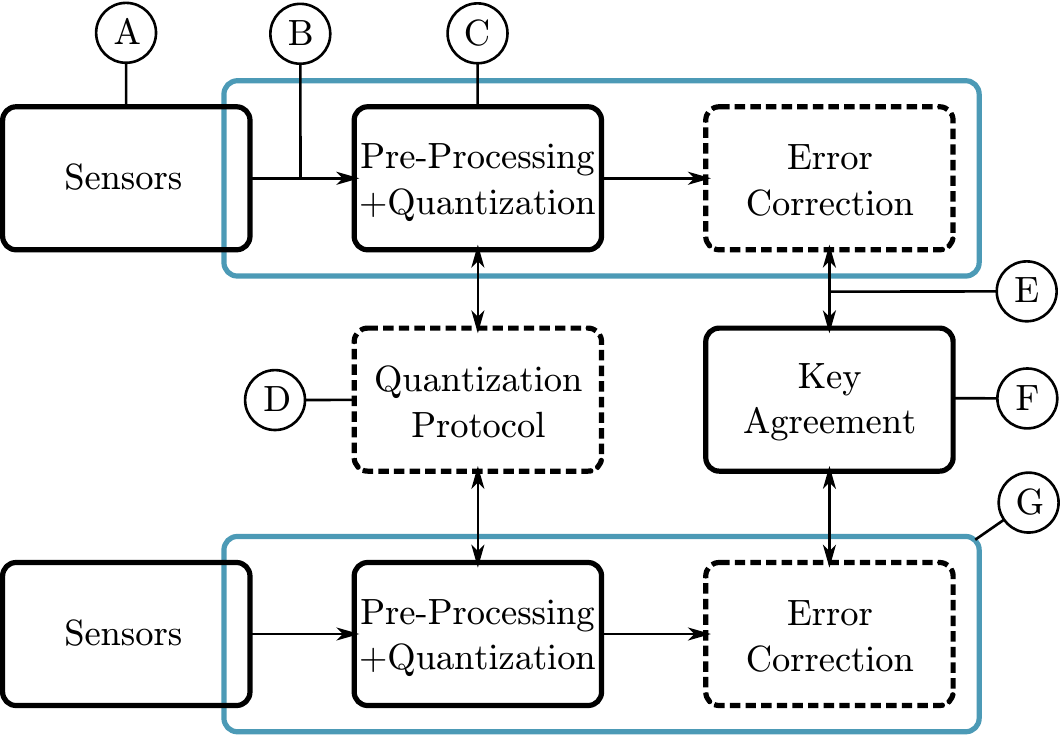}
    
    \caption{Conceptual view of gait-based pairing protocols with attack vectors (blue line depicts device boundary, dashed parts are optional)}
    \label{fig:attacks}
\end{figure}

\section{Security Analysis}
\label{sec:SecurityAnalysis}

As shown in the conceptional view in Figure~\ref{fig:attacks}, the pairing schemes follow a general design. 
Devices measure data, quantize it to bit strings after pre-processing, apply potentially error correction, and agree on a key.

Protection against MitM attacks is achieved only if all parts of a system are resilient. 
Our analysis follows the conceptual aproach proposed in~\cite{muaaz2013analysis,bolle2013guide}.
The discussed attacks are assigned to attack vectors A-G labeled in Figure~\ref{fig:attacks}.

An attack surface is exposed by the sensors (A).
A device owner could be forced to behave in a certain way, e.g., by an adversary controlling stride speed with a treadmill.
It could {further} be possible to bypass data acquisition (B) and reuse data from the past.
With a biased quantization, a naïve brute force attack would become feasible (C).
Some protocols employ communication before the actual key agreement (SAPHE: random seed and distance ordering, Walkie-Talkie: reconciliation, BANDANA: exchange reliability indices) which might potentially leak information (D).
After error correction (e.g. in BANDANA and IPI), the key agreement is executed between both participants.
Here, the risk of a Man-in-the-Middle~(MitM) (E) or impersonation attack (G) must be considered.
Finally, the key agreement could be weak or based on false assumptions, especially if it is not based on established standards (F).
We do not discuss attack vector B as it assumes {a} compromised {device}, which falls outside the focus of this work. 

\subsection{One-Shot Success Probability (E, G)}
\label{sec:attackTypes}
Without requiring additional knowledge about the victim's gait, an attacker may want to exhaust the keyspace $\mathcal{C}$ of all keys $\boldsymbol{k}$ 
to execute a MitM~(E) or impersonation attack~(G).
However, in all discussed protocols, after each single try, a completely new authentication process (new $\boldsymbol{k}$ independent from the previous one) 
is started.
Thus, it is impossible to exhaust $\mathcal{C}$, making this a one-shot attack.
For comparison between protocols, we 
{assume} the same length of 16 bit for $\boldsymbol{k}$.
The length of sequences sampled for a target key $\boldsymbol{k}$ of 16 bit may vary depending on the quantization scheme.

Note that 16 bit provide sufficient entropy since we suggest to implement a PAKE protocol as in \cite{schuermann2018jagger}, which prevents offline attacks and can thus provide a sufficiently large security margin even with short key lengths $K$.

\subsubsection{Candidate Key Protocol Variants}
The candidate key protocol is, for instance, realized in SAPHE~\cite{groza2012saphe}, which resolves its original vunlnerability against MitM attacks.
In particular, first, random challenges are chosen, as depicted in Figure~\ref{fig:quantizationSAPHE} and committed by sharing their hashes.
Afterwards, the acceleration sequence is challenged with respect to these random thresholds where an acceleration point with value lower (higher) than a threshold is interpreted as 0 (1). 
The success probability for a single randomly drawn key $\boldsymbol{k}$ in SAPHE is \begin{equation}
\frac{1}{2^{16}}\approx1.52588\cdot10^{-5}
\end{equation}

\subsubsection{Walkie-Talkie Protocol}
The bits generated in the Walkie-Talkie protocol feature a high bit rate of 15--55 bits per second as reported in~\cite{Xu_2016_WalkieTalkie} (Figure 12(e)).
However, high agreement rates are reached only for $\alpha>0.8$ (Figure 12(d) and 12(f) in~\cite{Xu_2016_WalkieTalkie}), which corresponds to 15--25 bits per second.
A 16 bit binary key can therefore be generated in approximately 1 second and the success probability of an adversary for a single randomly drawn $\boldsymbol{k}$ is then again $\frac{1}{2^{16}}\approx1.52588\cdot10^{-5}$.

\subsubsection{BANDANA Protocol}
In the BANDANA protocol, $M=48$ bit sequences are generated in about ${\SI{12}{\second}}$.
From each sequence, $16$ bit are disregarded for reliability amplification. 
From the remaining $32$ bit fingerprints, up to $8$ bit are corrected by BCH codes, resulting in $|\boldsymbol{k}|=16$ bit keys. 
The success probability of a single randomly drawn fingerprint is then (cf. Section~\ref{sec:fuzzy})
\begin{equation}
\sum_{k=0}^8 \left(\begin{array}{c}32\\k\end{array}\right)/2^{32}
=\frac{\sum_{k=0}^8 \left(\frac{32!}{(32-k)! \cdot k!} \right)}{2^{32}}
\approx  0.0035
\end{equation}

\subsubsection{IPI Protocol}
In the IPI protocol, dependent on the sampling frequency, 2 to 20 secure bits are extracted from each gait cycle (cf. Table I in~\cite{sun2017secure}).
Depending on the sample rate of the accelerometer, the generation of 32 bits in the IPI protocol might therefore require from 2 to 16 seconds.
Since the protocol also employs fuzzy cryptography for error correction, the same success probability as in the BANDANA protocol of 0.0035 applies for a single randomly drawn fingerprint. 

\subsection{Quantization-Specific Attacks (C, D)}
An attacker with insight to {a} quantization scheme might be able to exploit this knowledge in order to boost her one-shot success probability.
We discuss our observations in the Walkie-Talkie, BANDANA and IPI protocols.
For SAPHE, we did not identify any quantization-specific weakness.

\subsubsection{Walkie-Talkie Protocol}
As discussed in \ref{sec:similarityWalkieTalkie}, Walkie-Talkie is thought to be biased towards generating alternating sequences of 1-bits and 0-bits, which should be mitigated
by applying a \emph{privacy amplification}.
We note that if an adversary were able to reconstruct the pattern-prone sequence before the amplification step, she would also be able to compute the \emph{amplification}-step.

\begin{figure}
    \centering
    \begin{subfigure}[t]{0.575\columnwidth}
        \includegraphics[width=\columnwidth, trim={.35cm .5cm 7.3cm 1cm},clip]{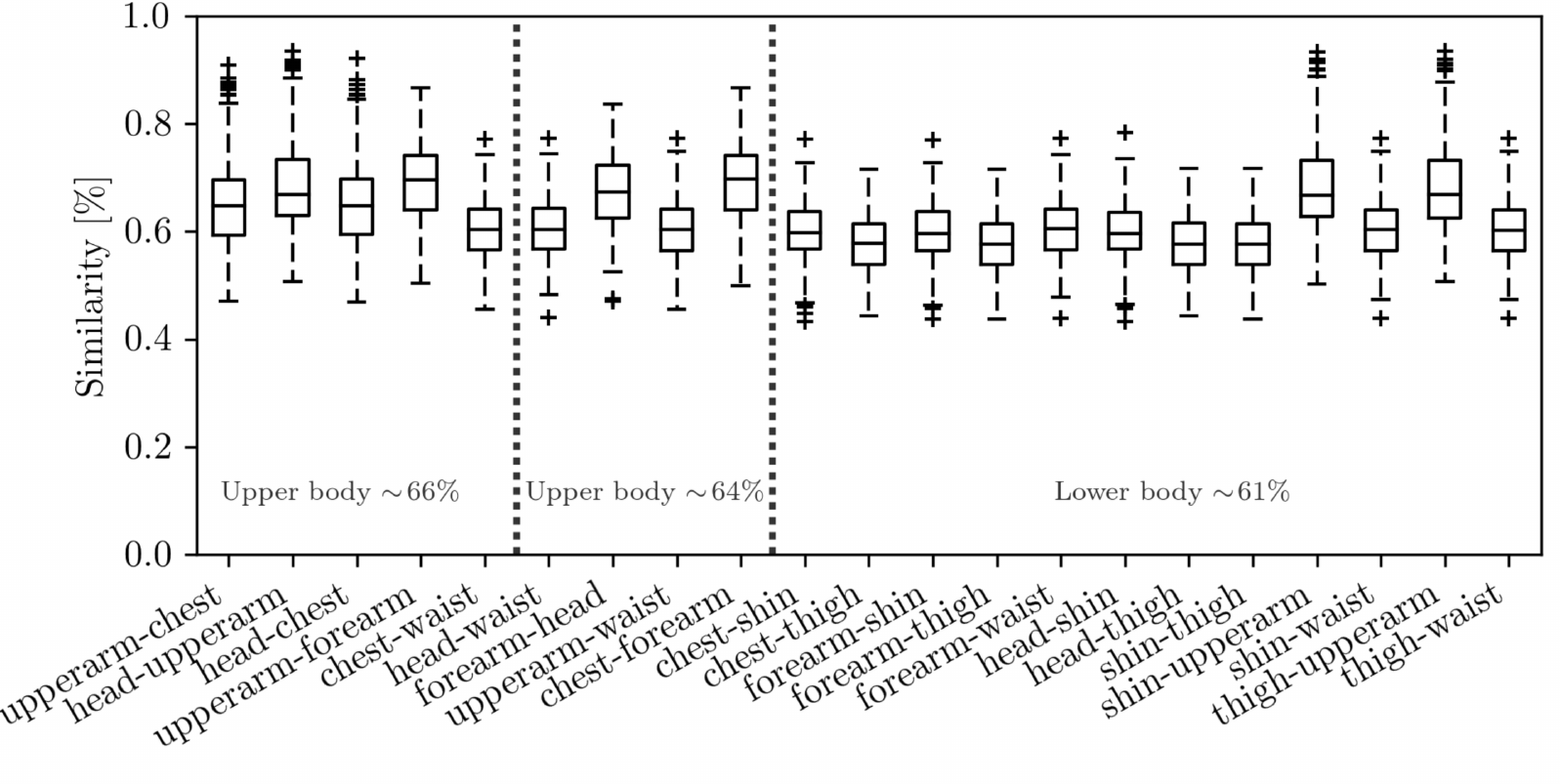}
        \caption{Walkie-Talkie: Overheared bit positions during reconciliation are used to generate bit sequences with similarities higher than 50\%}
        \label{fig:wktkAttack}
    \end{subfigure}
    \begin{subfigure}[t]{0.385\columnwidth}
        \includegraphics[width=\columnwidth]{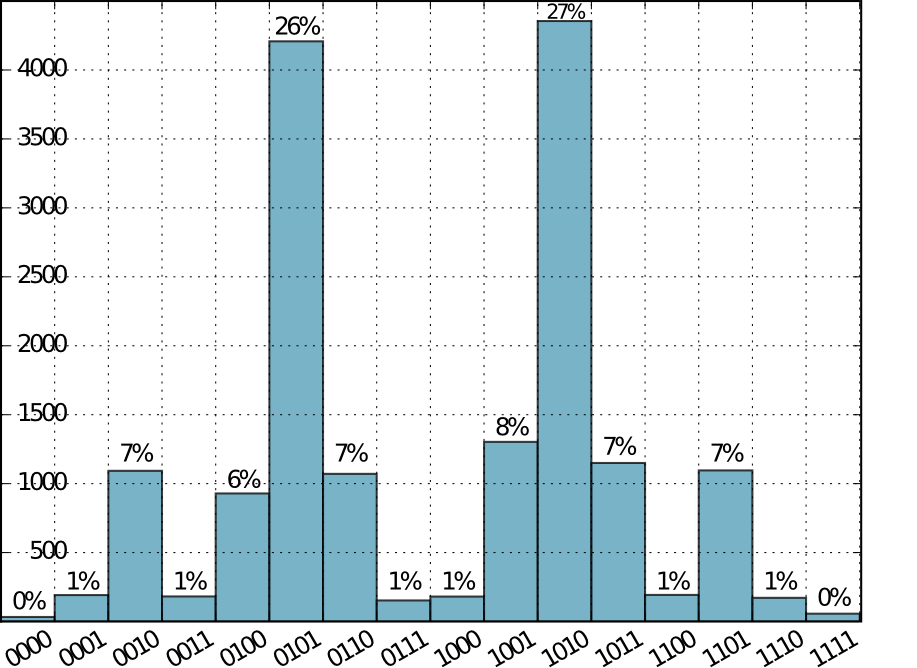}
        \caption{BANDANA: Non-uniform distribution of 4-bit chunks per gait cycles}
        \label{fig:bandanaProblem}
    \end{subfigure}
    \caption{Increasing one-shot success probability due to bias in sequences}
\end{figure}

Figure~\ref{fig:wktkAttack} shows key similarities achieved by this attack when guessed sequences are compared to actual acceleration-based sequences. However, this only works for large window sizes.
For small window sizes such as 10, the consecutive runs of indices become very short. Even worse, they might change signs when running over window borders, due to the newly computed guard band. 
Thus, the concerns about obvious patterns in the generated sequence are effectively mitigated by a small window size.

\subsubsection{BANDANA Protocol}
As indicated in Section~\ref{sec:bandanaSimilarity}, we found that the random success probability for the BANDANA protocol exceeds random guess.
Indeed, as observed in Section~\ref{sec:randomness} (Figure~\ref{fig:heatmapBandana}), the variance in generated sequences is low and, in particular, sequences follow specific patterns (cf. Figure~\ref{fig:BandanaStrings}).
As depicted in Figure~\ref{fig:bandanaProblem}, we found as the reason for this weakness that in the 4-bit chunks, which are generated per gait cycle (and before throwing away bits for reliability amplification), sequences of alternating binary value are significantly more frequent than others. 
In particular, sequences 1111 or 0000, where the instantaneous acceleration constantly exceeds or deceeds the mean acceleration, are seldom. 
Consequently, the distribution of key sequences in the key space is not uniform, and an adversary could utlilize this knowledge to launch an attack (C).
We propose an approach to mitigate this problem in Section~\ref{sec:improving}.

\subsubsection{IPI Protocol}
As discussed in Section~\ref{sec:IPISimilarity}, the IPI protocol suffers from measurement noise in accurately capturing the inter pulse interval due to the limited sampling rate of accelerometers. 
Especially for lower sampling rates, this significantly restricts the size of the key space. 
For instance, with 50 Hz (500Hz) sampling rate, one sample is taken every 20 milliseconds (every 2 ms). 
Since devices are not synchronized, this translates to an unavoidable inaccuracy of up to 10ms (1ms) for the sampled gait on devices (cf. Figure~\ref{fig:quantizationIPI}).
This measurement noise, compared with only 40.8ms standard deviation for the IPI results in a small keyspace and, since gray codes are employed (modulo 16; $q=4$), not all bits in the generated quadruples change. 
In particular, we investigated the variation in 4 bit chunks generated by the IPI protocol on the walking data from~\cite{sztyler2016onbody}. 
In about 63\% of the consecutive 4 bit chunks, all bits are identical. 
Furthermore, in 24\% of all cases, just one bit changed, with 11\% 2 bits changed and with only 0.02\%, 3 bits were different.
An adversary with approximate information on the IPI can therefore boost her guessing success probability significantly beyond chance.

\subsection{Benefits and Pitfalls in using Error Correction}\label{sec:fuzzy}
In biometric authentication systems, noise of the biometric information is an intrinsic property (here: measurement noise in acceleration sensors). 
Fuzzy cryptography has been proposed in order to employ error correcting codes to mitigate such noise. 
Error correcting codes encode messages from a messagespace $m \in \mathcal{M}$ into codewords of the (larger) codespace $c \in \mathcal{C}$ introducing redundancies.
This process allows to correct errors introduced to $c$ by decoding it back to $m$.
In fuzzy cryptography, the biometric information or fingerprints contain noise or errors that can be corrected after mapping into $\mathcal{C}$. 
The redundancy introduced in the encoding process, however, dictates that an adversary also does not have to guess all bits in the fingerprint correctly, but can be sloppy. 
For instance, assume a key length of $K$ and an error correcting code able to correct a fraction of $u$ bits from the total fingerprint length $N$. 
This means that the success probability of a single randomly drawn fingerprint is not $2^N$, but instead only
\begin{equation}
\sum_{k=0}^u \left(\begin{array}{c}N\\k\end{array}\right)/2^{N}
=\frac{\sum_{k=0}^u \left(\frac{N!}{(N-k)! \cdot k!} \right)}{2^{N}}
\end{equation}
since up to {$u$} errors are allowed at arbitrary position in the fingerprint sequence. 
Careful choice of the parameters is therefore demanded to limit the advantage gained by an adversary through the use of fuzzy cryptography.

From the protocols we investigated, BANDANA~\cite{schuermann2018jagger} and the IPI-protocol~\cite{sun2017secure} employ BCH codes for error correction.
\cite{revadigar2017smartwearables} integrates the fuzzy vault design that operates on order-invariant tuples generated by Walkie-Talkie.
The Gait-Key~\cite{xu2017gaitkey} variant, which is further discussed in Section~\ref{sec:improving-gait-key}, implements a scheme by Yan et al.~\cite{yang2016muscle}.
\color{black}

\subsection{Gait Mimicry (A)}
As recently discussed in~\cite{muaaz_2017}, it is unlikely that an attacker would be able to mimic natural gait of a victim to a degree where gait sequences {are} 
sufficiently similar to break gait-based authentication or pairing schemes.
In particular, the authors employed professional actors to mimic the gait of victims with similar physical properties (age, weight, height, shoe size, upper leg length) and showed that after guided training and instructions, all actors failed to mimic the observed gait of victims. 
In a second test, by walking next to a victim one out of five attackers was able though to achieve sufficient similarity in the gait acceleration sequence. 
In particular, the authors assumed that the victim instinctively adapted her walking speed to the common step pattern with the adversary.
This was, however, not further investigated. 

\subsection{Impersonation via Video Recording (G)}
\label{sec:videoattacks}
Cameras are omnipresent in these days, for instance as CCTV systems, personal camcorders, or mobile phones. 
The quality of captured videos is sufficient to discriminate subtle movements. 
An adversary with camera-support might therefore be able to extract pairing keys from recorded video (G).
In this section, we investigate the threat of video-based side-channel attacks. 
In particular, we consider how accurate acceleration sequences describing gait can be estimated by tracking movement of body parts from video. 

\begin{figure}
    \centering
    \includegraphics[width=\columnwidth]{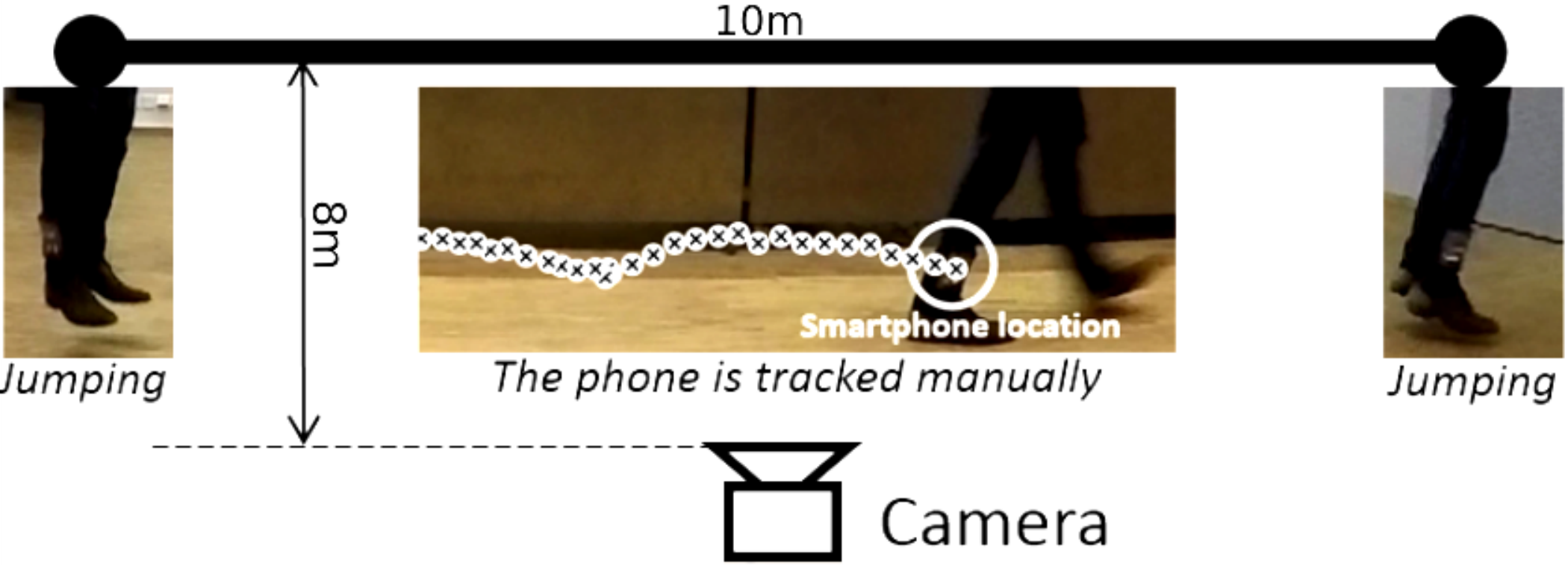}
    \caption{Experimental setup for video-based attack on gait-based pairing}
    \label{fig:videosetting}
\end{figure}

For our experiment, we captured movement of a subject both by a wearable inertial measurement unit (smartphone) and with a high-speed camera.
The smartphone was attached to one leg.
Five subjects (4 male; height: 1.63-1.95m; $\mu=1.76$m) walked in a straight line in approximately 8m distance to the camera (1080p resolution; 90fps) mounted on a tripod (cf. Figure~\ref{fig:videosetting}).
Acceleration data was sampled at 50Hz.
For synchronization between video and inertial sensor, a single jump both at the beginning and at the end framed the walking segment. 
Each subject conducted the experiment twice.
We utilized Tracker\footnote{http://physlets.org/tracker/} to manually track the location of the smartphone on the recorded video.
Although human pose estimation~\cite{Mehta2017} is able to estimate leg movements, we achieved higher accuracy by manually marking the location of the smartphone on the video frames.

\begin{figure}
    \centering
    \includegraphics[width=\columnwidth]{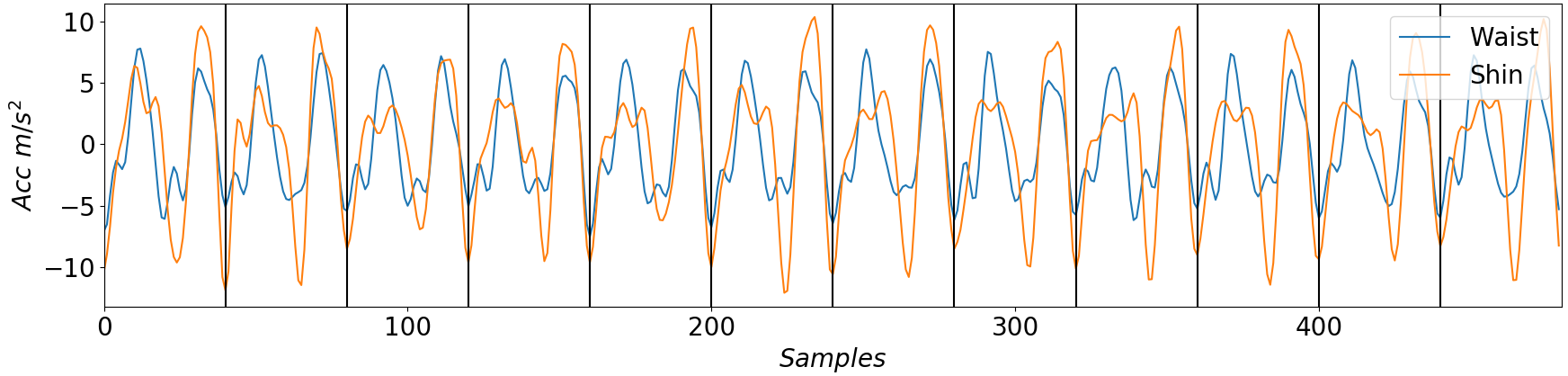}
    \caption{ Gait cycles extracted from shin and waist. 
    }
    \label{fig:videogaitcycles}
\end{figure}

For gait-based on-body pairing, the attacker is free to estimate gait according to the most easy to attack body location, since the protocols are inherently designed to pair acceleration sequences from arbitrary body location pairs.
The Spearman's coefficient (1: perfect monotonically increasing relationship; 0: non monotonic relationship; -1: perfect monotonically decreasing relationship)~\cite{Spearman1904} for gait sequences extracted at waist and shin in the dataset~\cite{sztyler2016onbody} is $0.44$, which reflects their moderate increasing monotonic association. 
For instance, correlation between gaits extracted from these locations can be observed in Figure~\ref{fig:videogaitcycles}. 

\begin{figure}
    \begin{subfigure}[t]{0.48\columnwidth}
        \includegraphics[width=\columnwidth]{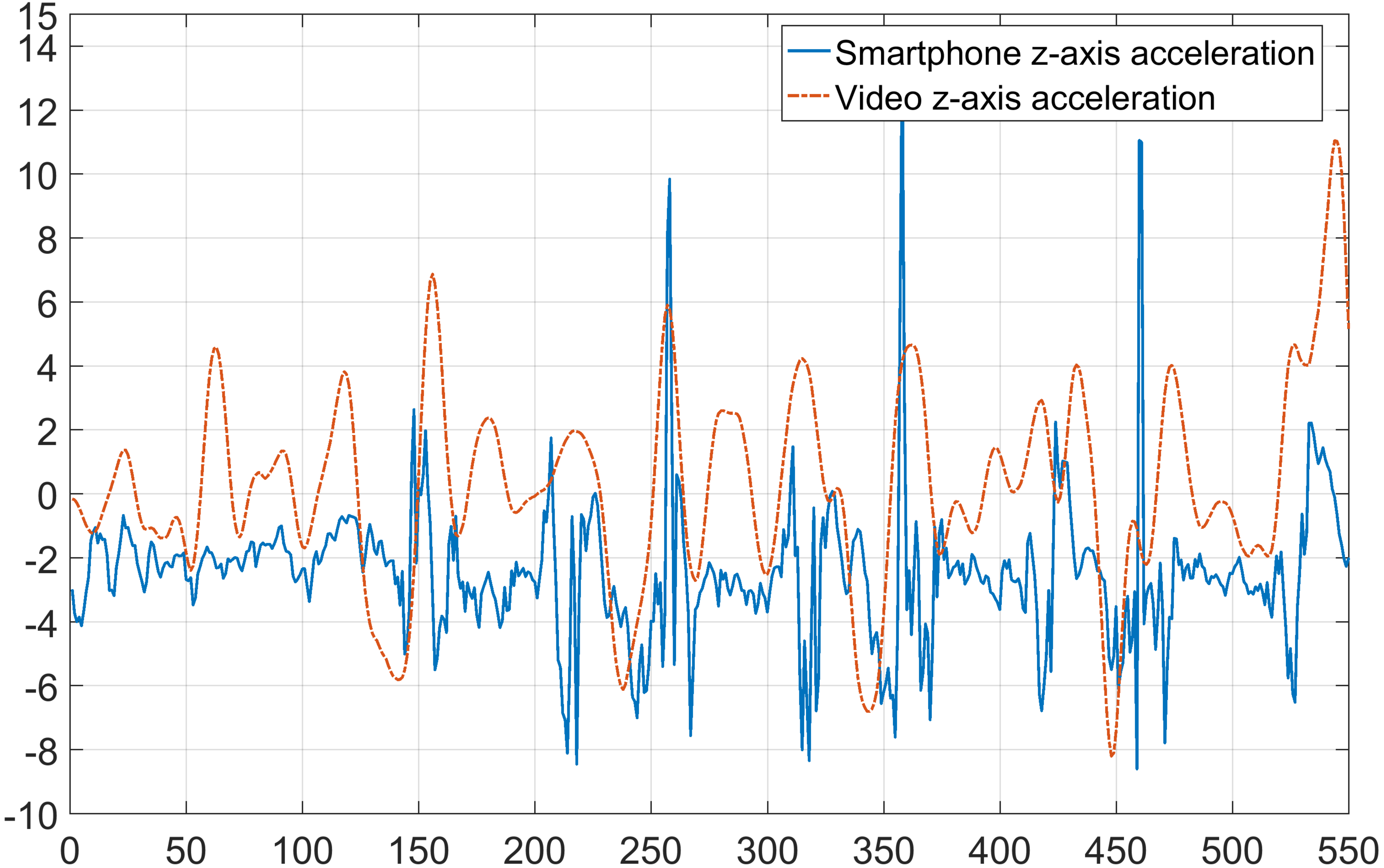}
        \caption{Alignment of acceleration sequences from smartphone and video}
        \label{fig:alignment}
    \end{subfigure}
    \quad
    \begin{subfigure}[t]{0.48\columnwidth}
        \includegraphics[width=\columnwidth]{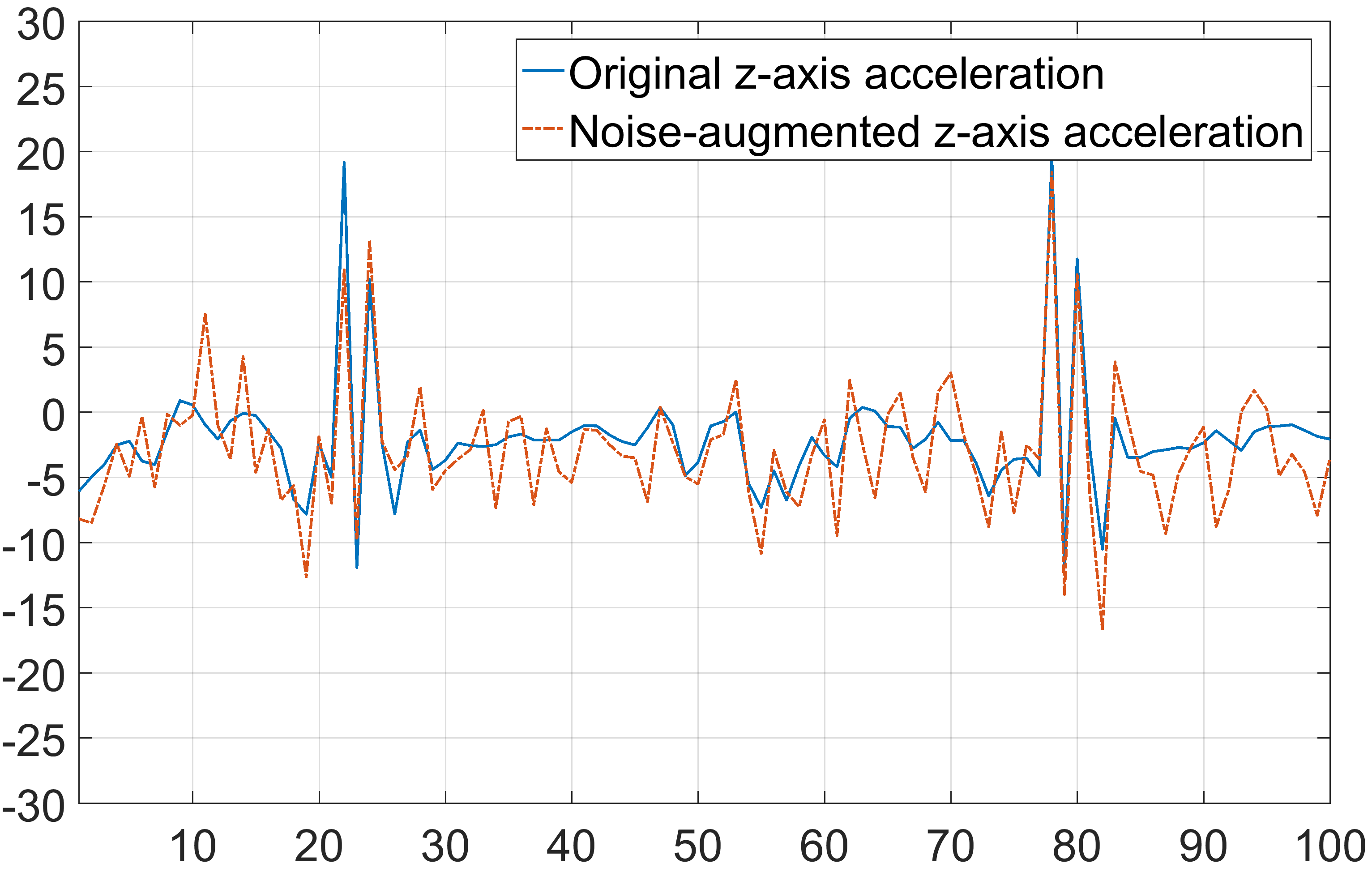}
        \caption{Acceleration sequence augmented with low noise level ($\mathcal{N}(\frac{\mu_v}{2},\,\frac{\sigma_v^{2}}{4})$)}
        \label{fig:videonoise}
    \end{subfigure}
    
    \begin{subfigure}[t]{0.48\columnwidth}
        \includegraphics[width=\columnwidth]{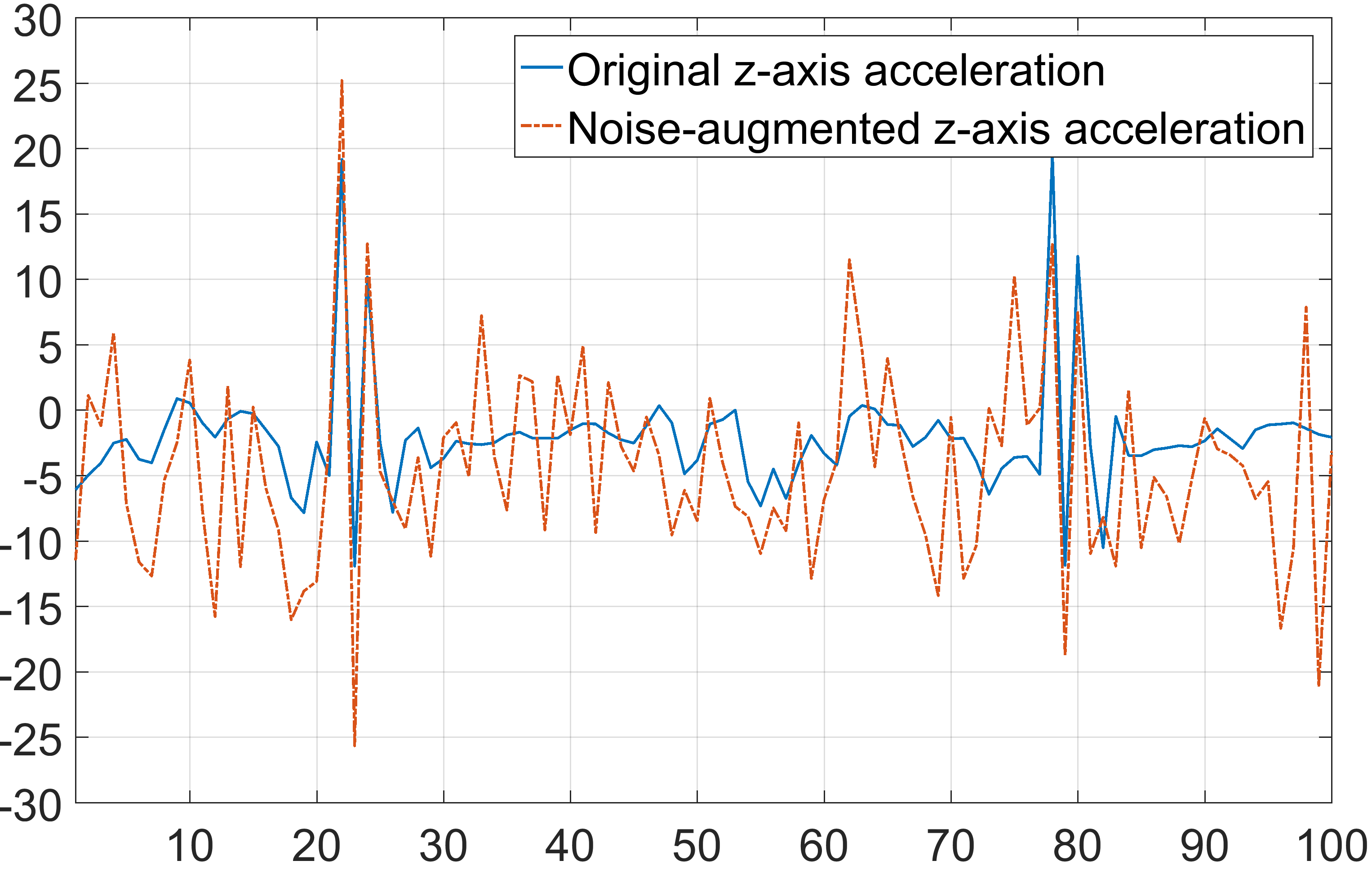}
        \caption{Acceleration sequence augmented with video noise level ($\mathcal{N}(\mu_v,\,\sigma_v^{2})$)}
        \label{fig:lownoise}
    \end{subfigure}
    \quad
    \begin{subfigure}[t]{0.48\columnwidth}
        \includegraphics[width=\columnwidth]{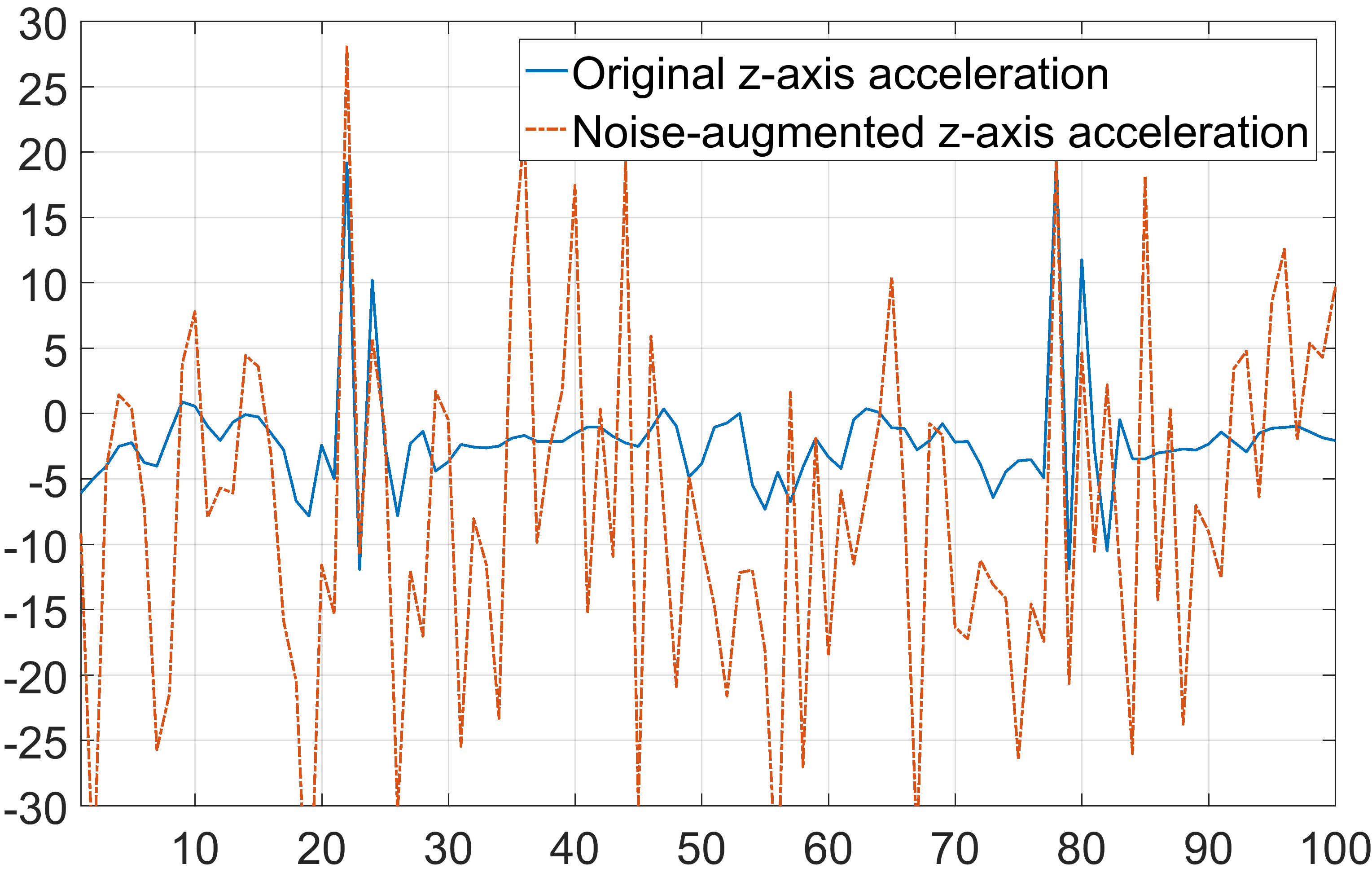}
        \caption{Acceleration sequence augmented with high noise level ($\mathcal{N}(2\cdot\mu_v,\,4\cdot \sigma_v^{2})$)}
        \label{fig:highnoise}
    \end{subfigure}
    \caption{Acceleration signals featuring different noise levels}
    \label{fig:noiseaugmentation}
\end{figure}

From the tracked trajectory we estimated the acceleration of the smartphone.
We calculated the velocity in horizontal and vertical direction before computing the acceleration.
The obtained result is smoothed by a Gaussian filter to reduce annotation noise.
This estimated acceleration sequence is then re-sampled to match the 50Hz sampling rate of the inertial sensor.
Note that we estimated movement orthogonal to ground since any rotation is implicitly corrected by the pairing scheme (Figure~\ref{fig:alignment}). 

To estimate the pairing performance and noise from video-extracted acceleration, in the dataset~\cite{sztyler2016onbody}, we estimated the mean $\mu_v = 2.0921$\footnote{From the amplitude estimation error due to inaccurate distance measurement between camera and walking subject.} and standard deviation $\sigma_v = 6.0210$ of disparity values between optimally synchronized\footnote{We refined the synchronization between the estimated and recorded acceleration sequences by shifting both sequences until a minimum root mean squared error is achieved} gait acceleration sequences (estimated and recorded) in our experiment.
These values were then used as parameters for noise distributions, which we added to the walking data recorded by the dataset in~\cite{sztyler2016onbody}.
We generated Gaussian, Laplacian, and uniformly distributed noise\footnote{$p_n(n) = \frac{1}{\sqrt{\pi\sigma^2}}\text{e}^{\frac{(n-\mu)^2}{-\sigma^2}}$; $p_n(n) = \frac{1}{\sqrt{2\sigma}}\text{e}^{\frac{\sqrt{2}|n-\mu|}{-\sigma}}$; $p_n(n) = \frac{1}{2 \sqrt{3} \sigma}$}.

We then generated noisy acceleration signals with $\mathcal{N}(\mu_v,\,\sigma_v^{2})$ (noise observed from video-based acceleration estimation), $\mathcal{N}(\frac{\mu_v}{2},\,\frac{\sigma_v^{2}}{4})$ (low noise) and $\mathcal{N}(2\cdot\mu_v,\,4\cdot \sigma_v^{2})$ (high noise) as illustrated in Figure~\ref{fig:noiseaugmentation} for Gaussian additive noise.
Other noise models are treated similar. 

\begin{figure}
    \begin{subfigure}[t]{0.49\columnwidth}
        \includegraphics[width=\columnwidth, height=2.5cm]{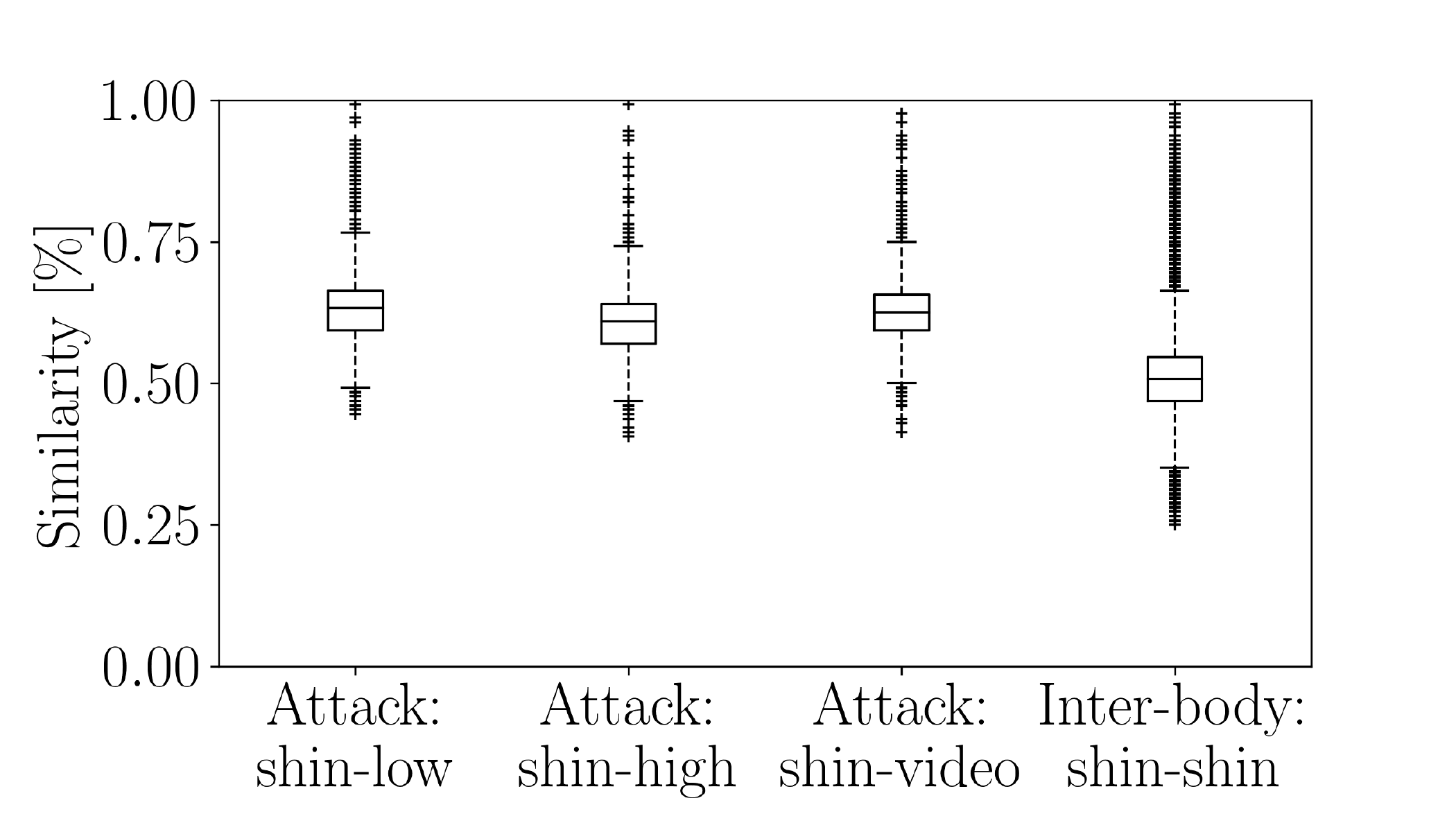}
        \caption{SAPHE}
        \label{fig:comparisonvideo-saphe}
    \end{subfigure}
    \begin{subfigure}[t]{0.49\columnwidth}
        \includegraphics[width=\columnwidth, height=2.5cm]{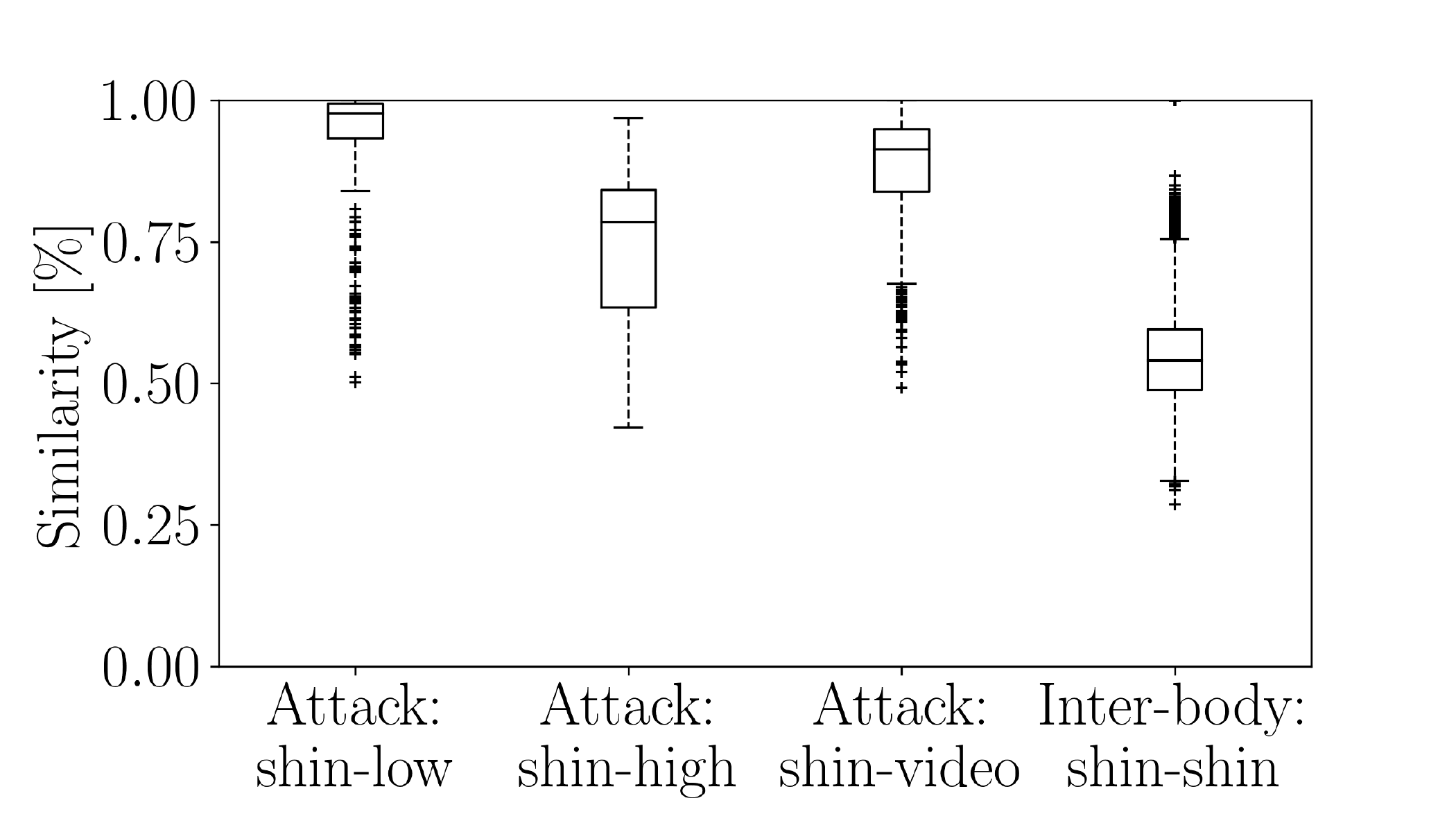}
        \caption{Walkie-Talkie}
        \label{fig:comparisonvideo-walkietalkie}
    \end{subfigure}
    
    \begin{subfigure}[t]{0.49\columnwidth}
        \includegraphics[width=\columnwidth, height=2.5cm, trim={0cm 0cm 0cm 0cm},clip]{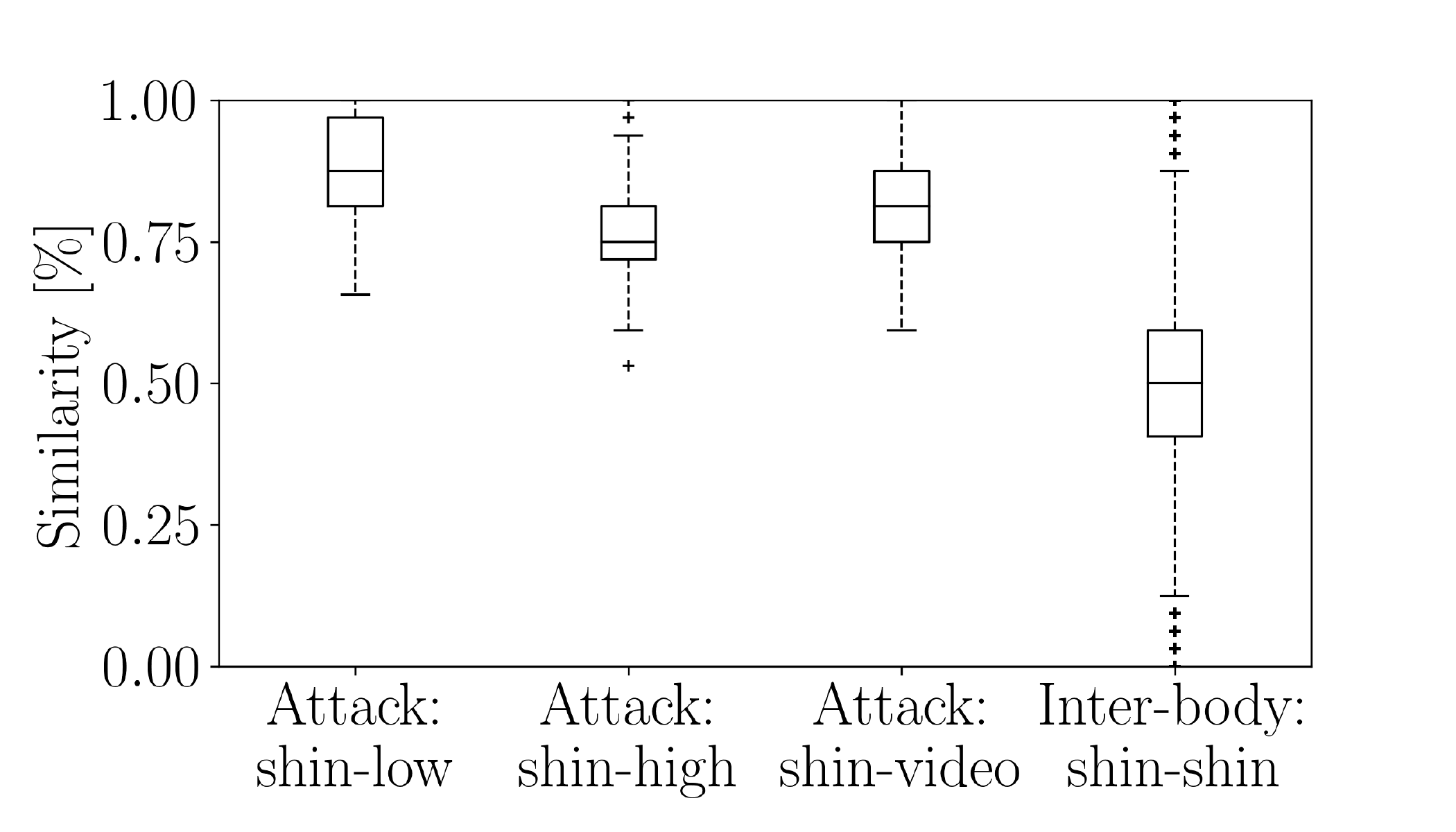}
        \caption{BANDANA}
        \label{fig:comparisonvideo-bandana}
    \end{subfigure}
    \begin{subfigure}[t]{0.49\columnwidth}
        \includegraphics[width=\columnwidth, height=2.5cm]{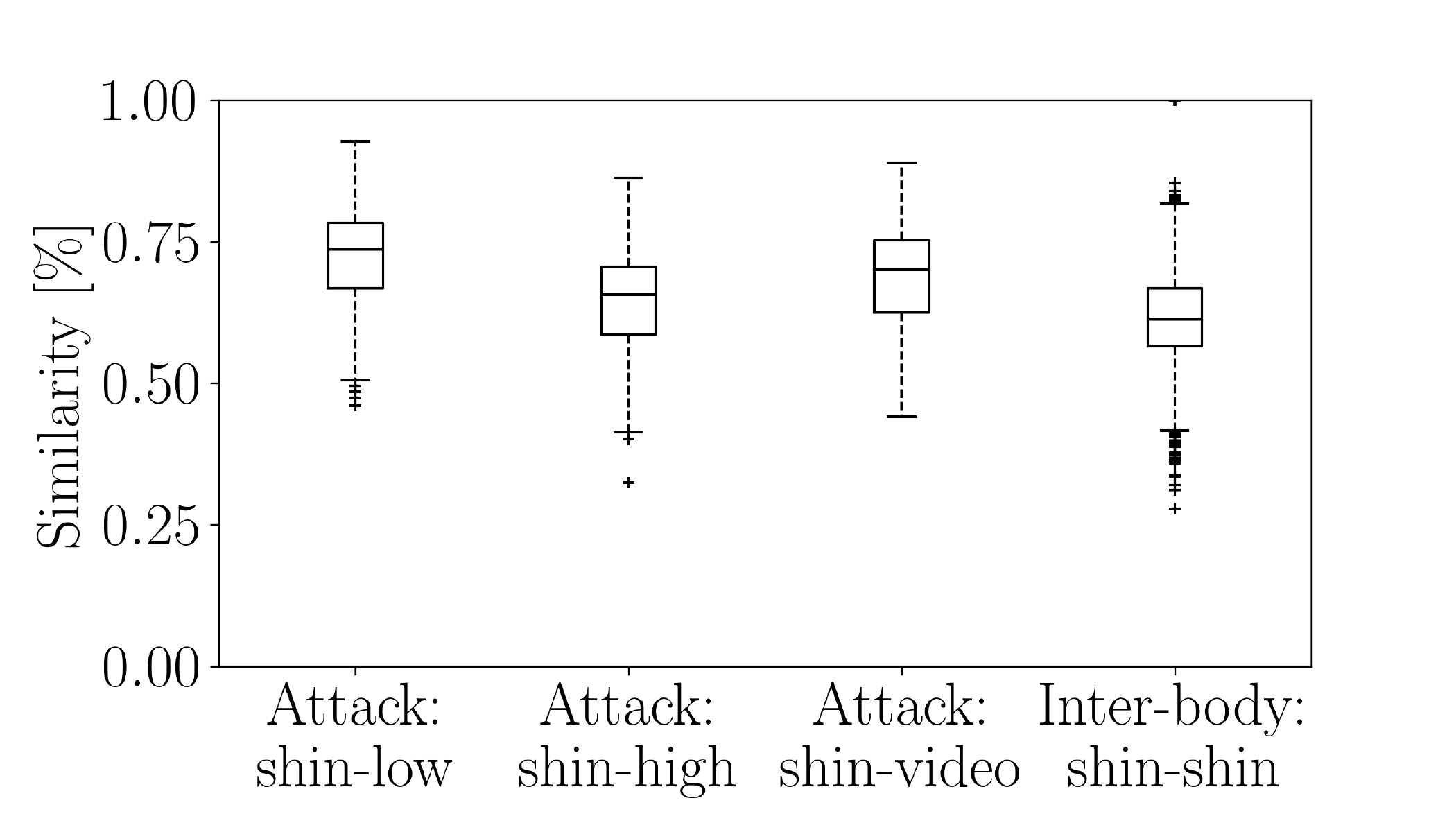}
        \caption{IPI}
        \label{fig:comparisonvideo-ipi}
    \end{subfigure}
    \caption{Attacks using Video-based impersonation: Similarity of gait-fingerprints with different noise levels over four pairing schemes
    }
    \label{fig:comparisonvideo}
\end{figure}

Figure~\ref{fig:comparisonvideo} details the similarity for intra-body, inter-body, and video-based acceleration sequences with three noise levels.
We assessed the effectiveness of video-based attacks on the four quantization schemes.
Video-based acceleration is able to generate fingerprints which are sufficiently close to the actually recorded acceleration sequence, so that this attack can break the gait-based pairing protocol for all three noise distributions considered.
Walkie-Talkie~\cite{Xu_2016_WalkieTalkie} is the most vulnerable protocol under the video-based attacks.
On the other hand, SAPHE~\cite{groza2012saphe} is the most secure protocol against video-based attacks (cf. Figure~\ref{fig:comparisonvideo}).

\section{Protocol Variants}
\label{sec:improving}
In this section, we discuss improvements to SAPHE and BANDANA as well as a variant of Walkie-Talkie, exploiting n-ary quantization for higher bit-rate.

\subsection{SAPHE}
From the pairing schemes considered, SAPHE is the most promising as it introduces randomness instead of relying solely on gait-implicit randomness.
As a potential improvement to our current implementation with a range of $1g$, we propose to implement a dynamic range.
This would prevent outlier threshold values independent of the acceleration.
Due to SAPHE's quantization, an attack, where a simple sinusoidal acceleration signal is artificially generated in alignment with the heel-strike, might {then} lead to a good estimate of the key.
We propose to choose the threshold values as close to the acceleration reading as possible while still not revealing the actual unique gait features.
This could be achieved by filtering out the dominant gait frequencies.
Finally, instead of using hashed heuristic trees~\cite{groza2012saphe}, we propose the usage of extensively studied cryptographic building blocks, such as fuzzy cryptography and a Password Authenticated Key Exchange.

\subsection{Walkie-Talkie}
\label{sec:improving-gait-key}
In \cite{xu2017gaitkey} Xu et al. present an evolved version of Walkie-Talkie, called Gait-Key. 
In contrast to Walkie-Talkie, multiple guard bands lead to several \emph{quantization levels} and multiple bits. 

We implemented this protocol and used four-ary quantization with $\alpha=0.9$ as recommended in~\cite{xu2017gaitkey}. 
As a window size for quantization we chose 50 samples.
We applied reconciliation and privacy amplification as in Walkie-Talkie. 
\begin{figure}[!t]
    \includegraphics[width=\columnwidth]{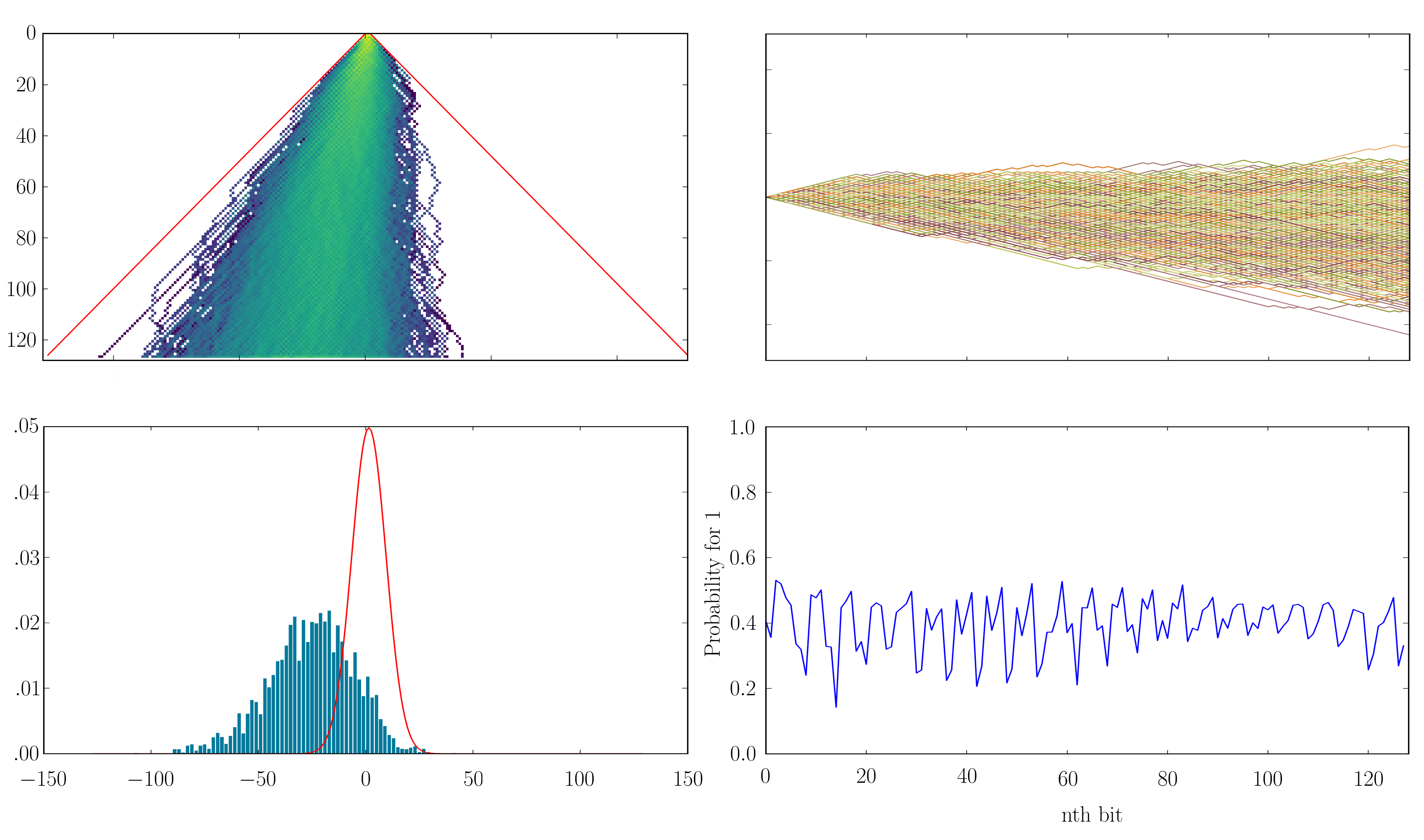}
     \caption{Evaluations for Gait Key show an offset towards more zeros.}
     \label{fig:gaitkey}
\end{figure}
Figure~\ref{fig:gaitkey} shows the randomness evaluation for Gait-Key. 
Similar to Walkie-Talkie, the key distribution is slightly shifted towards including more zeros. 
Remarkably, the Markov property shows periodic behaviour.
The reasons for this are twofold: 
First, the quantization scheme calls for normal or equal distribution of acceleration samples. 
Biased distributions along a certain axis lead to unequal occurences of 1's and 0's. 
Due to the privacy amplification, exploiting XOR, this results in a larger amount of 0's in the final key (equal bits are mapped to 0). 
Second, slicing the acceleration space into several areas and assigning these with multiple bits per sample implies  
that consecutive samples generate bit sequences from identical or neighbouring areas.
Hence, n-ary quantization achieves an improved, higher bit-rate but also introduces interdependence between bits while one guard band delivers the best performance. 

\begin{figure*}[!t]
    \begin{subfigure}[t]{0.5\textwidth}
        \includegraphics[width=\textwidth,trim={.3cm .2cm 2.3cm 1cm},clip]{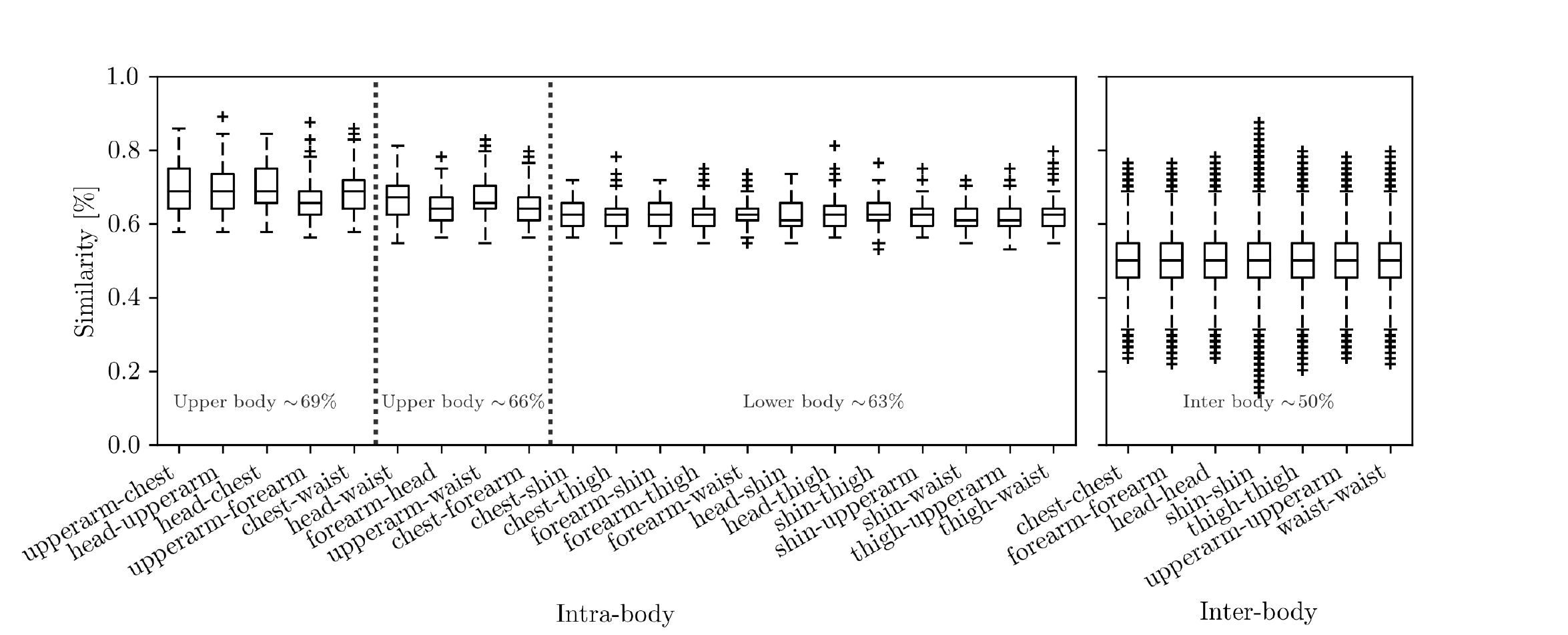}
        \caption{Mapping approach}
        \label{fig:bandanaCorrectedMapped}
    \end{subfigure}
    \begin{subfigure}[t]{0.5\textwidth}
        \includegraphics[width=\textwidth,trim={.3cm .2cm 2.3cm 1cm},clip]{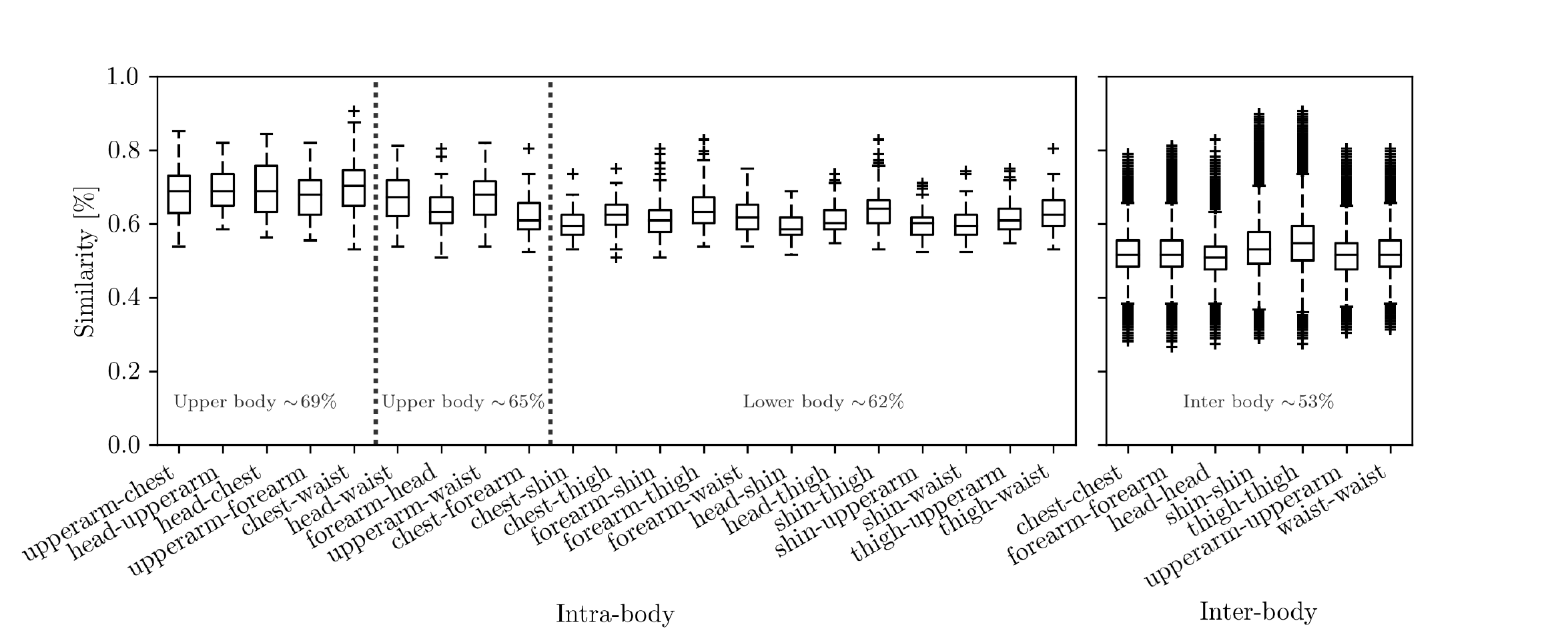}
        \caption{Normalization approach}
        \label{fig:bandanaCorrectedNorm}
    \end{subfigure}
    \caption{BANDANA improvements: Comparison of intra-body against inter-body similarity for our proposed improvements}%
    \label{fig:bandanarevised}
    \vspace*{-.5cm}
\end{figure*}

\begin{figure}[!t]
    \begin{subfigure}[t]{0.48\columnwidth}
        \includegraphics[width=\columnwidth]{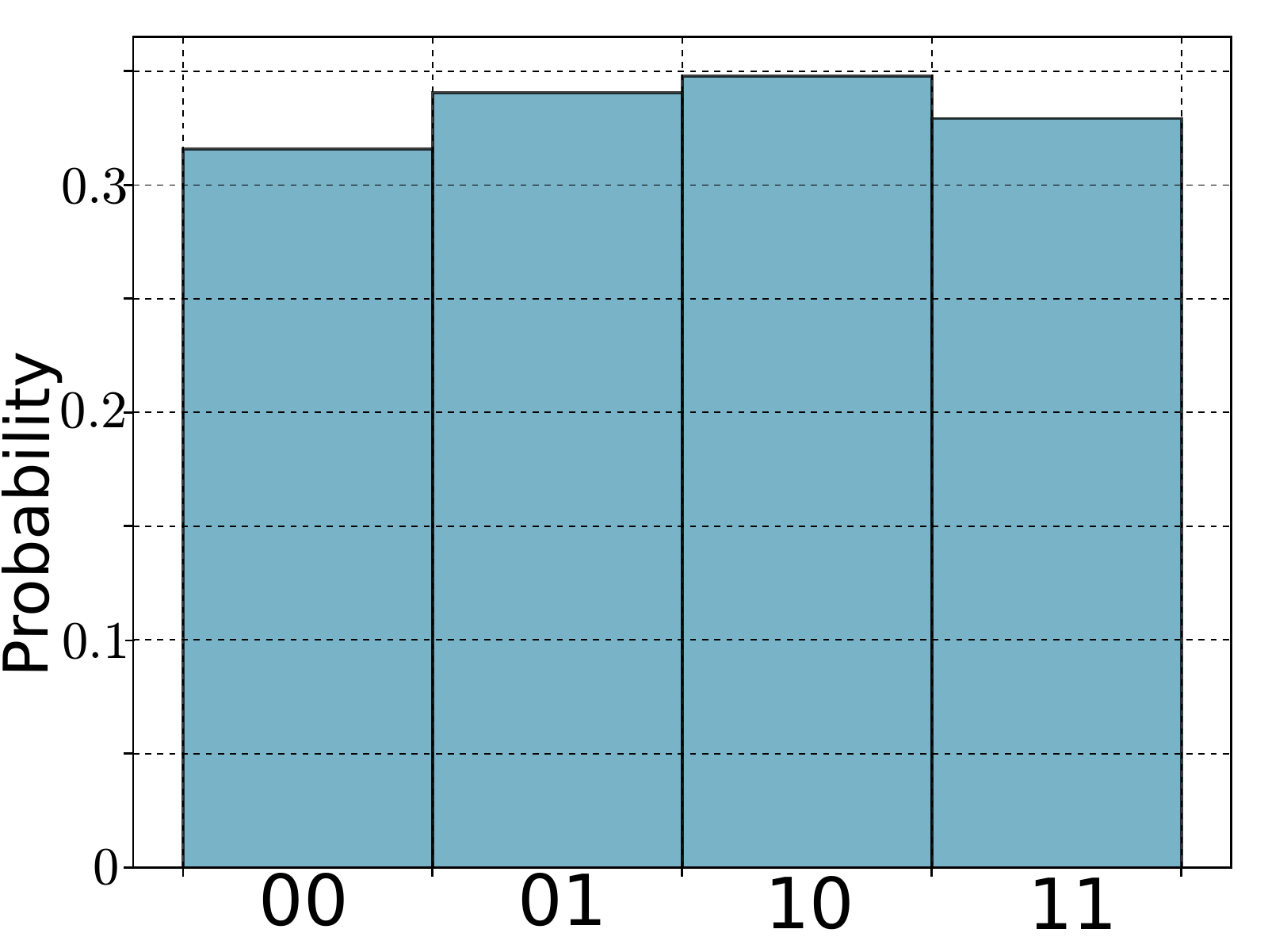}
        \caption{Mapping approach --  2 bits/bin}
        \label{fig:histogramMapping2bits}
    \end{subfigure}
    \quad
    \begin{subfigure}[t]{0.48\columnwidth}
        \includegraphics[width=\columnwidth]{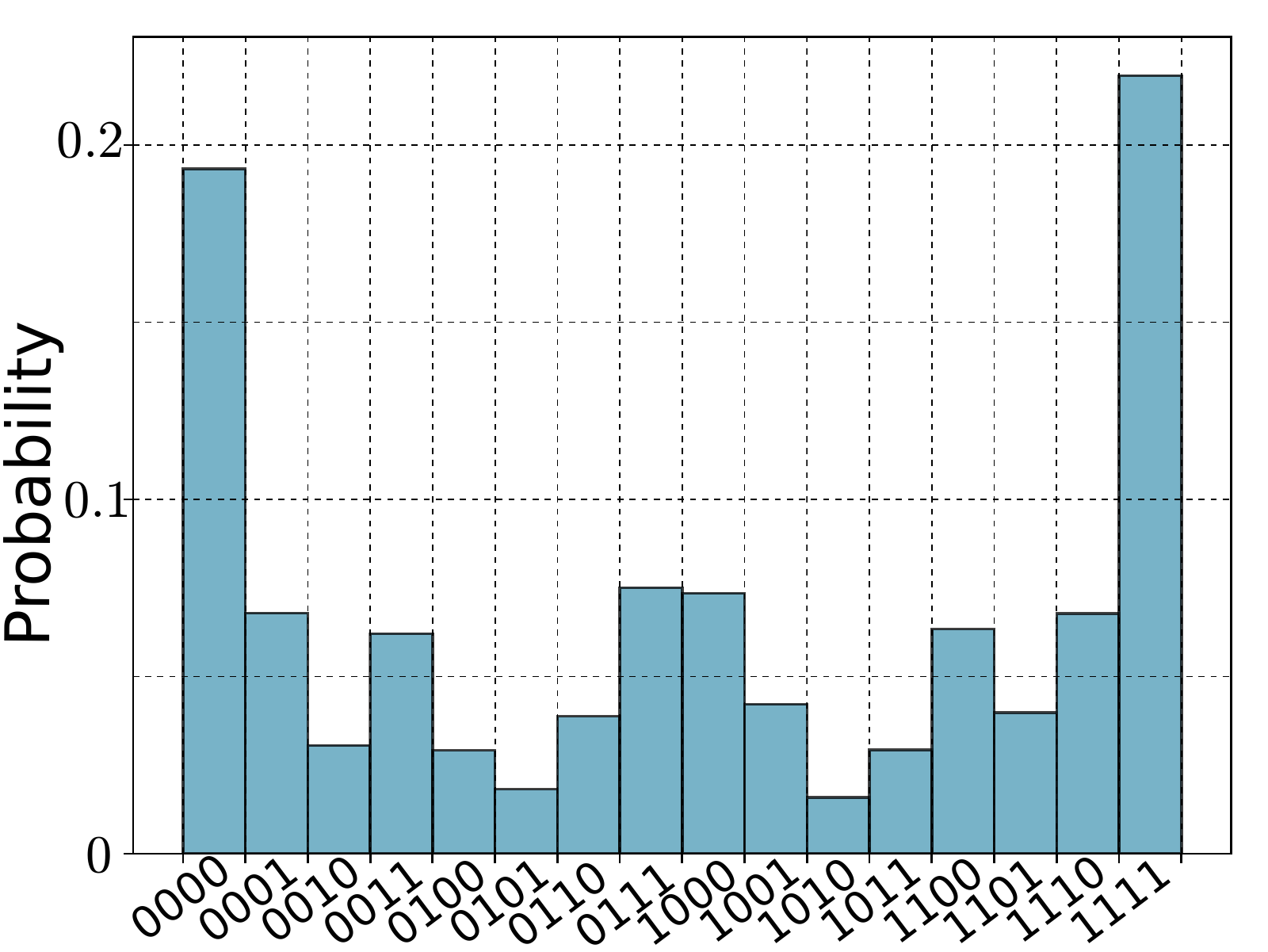}
        \caption{Mapping approach -- 4 bits/bin}
        \label{fig:histogramMapping4bits}
    \end{subfigure}
    \vspace{.2cm}
    \begin{subfigure}[t]{0.48\columnwidth}
        \includegraphics[width=\columnwidth]{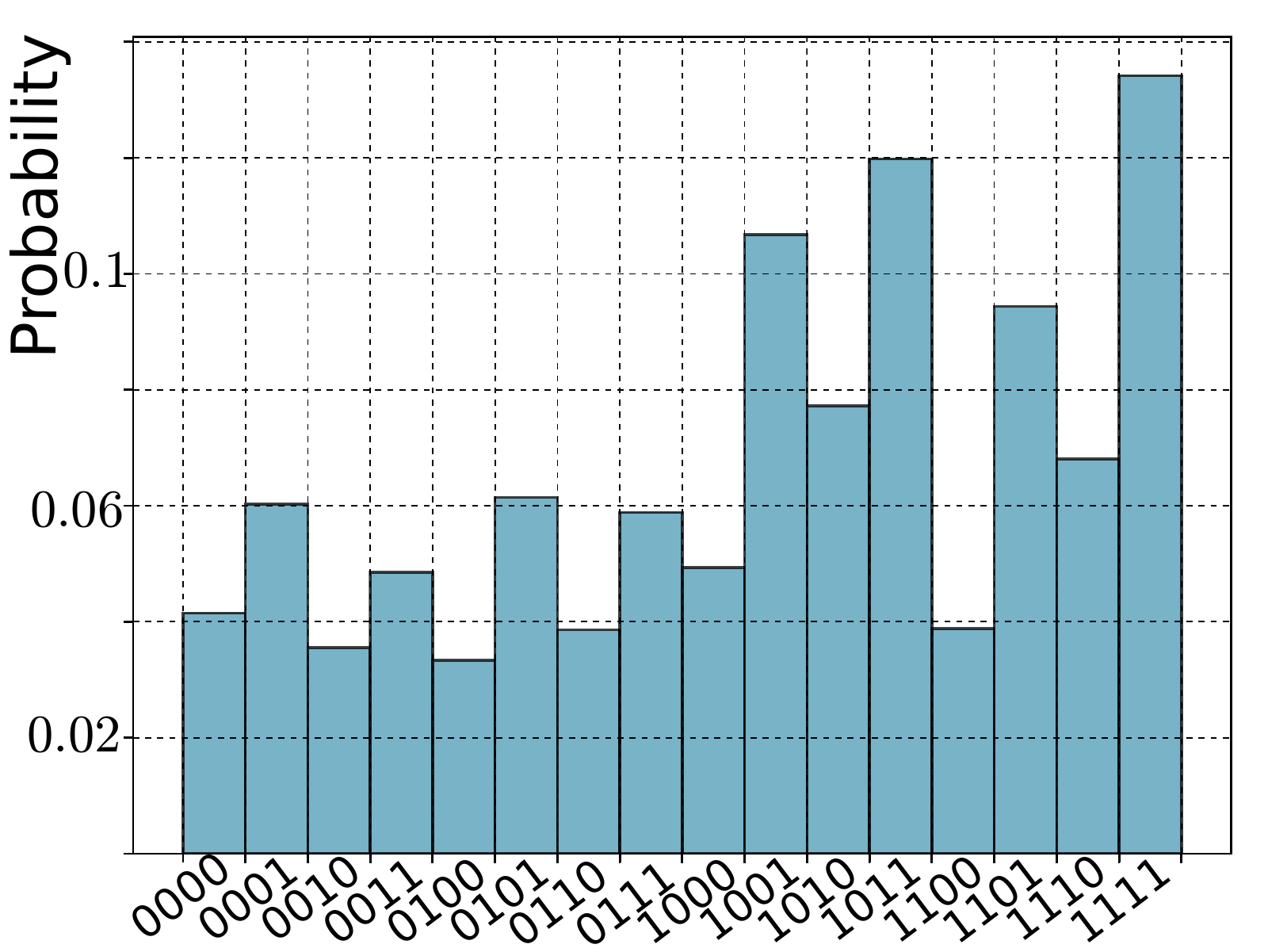}
        \caption{Normalization approach -- normalizing acceleration amplitudes}
        \label{fig:histogramNormalized}
    \end{subfigure}
    \quad
    \begin{subfigure}[t]{0.48\columnwidth}
        \includegraphics[width=\columnwidth]{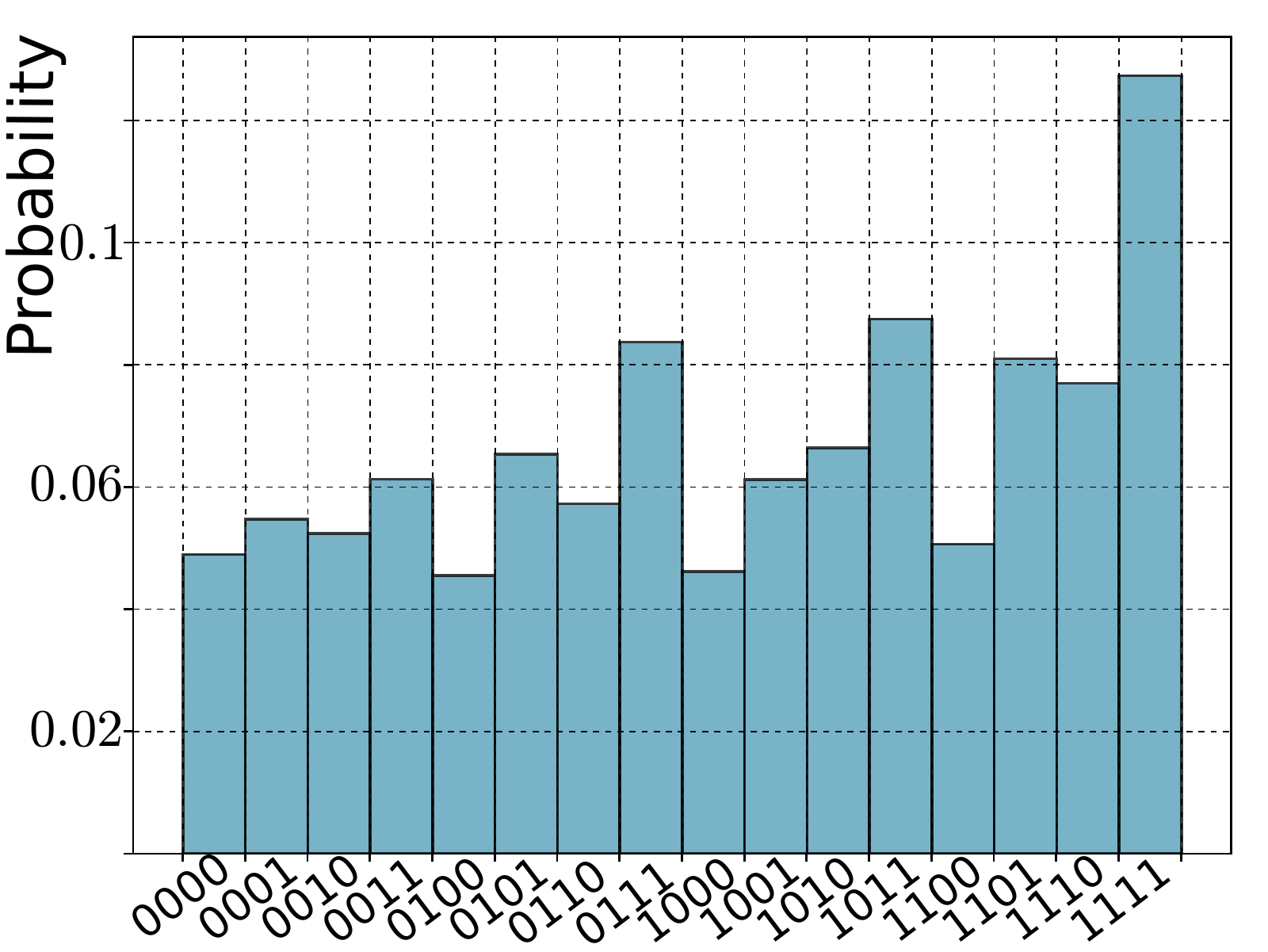}
        \caption{Normalization approach -- additional disregarding of patterns according to inverse occurrence probabilities}
        \label{fig:histogramNormalizedThrowAway}
    \end{subfigure}
    \caption{BANDANA improvements: Histograms generated from different improved versions of BANDANA}
\end{figure}

\subsection{BANDANA}
The quantization approach of BANDANA is biased towards specific patterns which are generated significantly more often than others (cf. Section~\ref{sec:quantization}). 
A straightforward solution is to disregard these 4-bit patterns with probability inverse to their occurrence frequency.
However, due to the significant distortion of the histogram (cf. Figure~\ref{fig:bandanaProblem}), this is not feasible. 
Since some patterns occur with a frequency of 1\% or less, close to all frequent patterns would have to be discarded to arrive at a balanced random distribution.

\begin{figure}[!t]
\begin{subfigure}[t]{0.5\columnwidth}
        \includegraphics[width=\columnwidth,trim={1.2cm 2.8cm 1.3cm 2.6cm},clip]{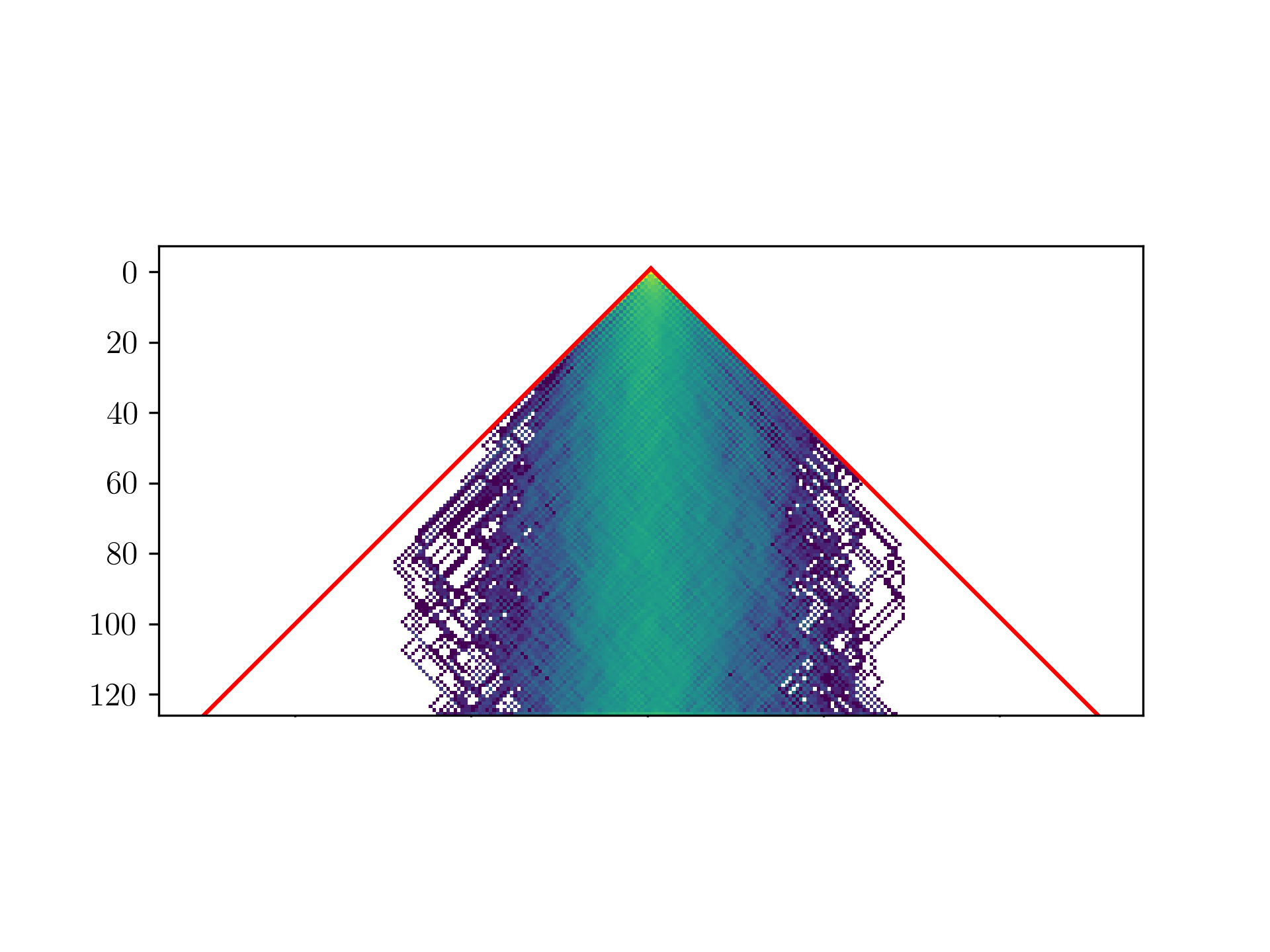}
        \caption{Mapping approach}
        \label{fig:heatmapBandanaImproved01}
    \end{subfigure}
    \begin{subfigure}[t]{0.5\columnwidth}
        \includegraphics[width=\columnwidth,trim={1.2cm 2.8cm 1.3cm 2.6cm},clip]{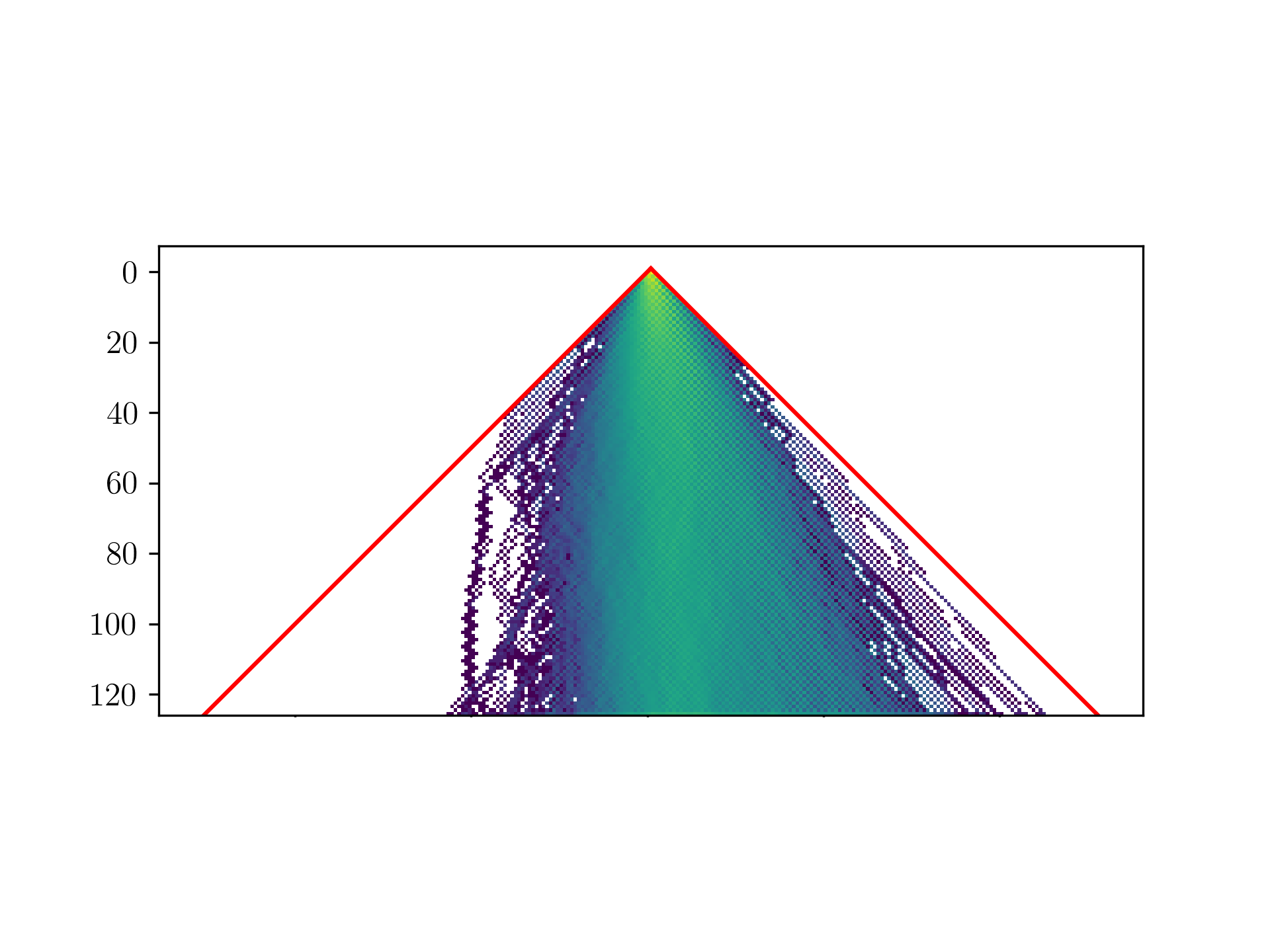}
        \caption{Normalization approach}
        \label{fig:heatmapBandanaImproved02}
    \end{subfigure}
    \caption{BANDANA improvements: Heatmaps of random walks for 128 bit keys generated by improved versions}
\end{figure}

Instead, we map each pair of consecutive bits in the generated key sequence to a single bit ($01,11\rightarrow1$, $10,00\rightarrow0$)\footnote{This does not leak information since 01 and 10 (11 and 00) are equally probable due to symmetry in the histogram in Figure~\ref{fig:bandanaProblem}}.
Figure~\ref{fig:histogramMapping2bits} and~\ref{fig:heatmapBandanaImproved01} show the distribution of bit sequences after the mapping as well as the heatmap for fingerprints generated with the modified protocol. 

The weakness described in Section~\ref{sec:quantization} could be mitigated, however, due to the strong unbalancedness, some bias still remains even after the mapping as depicted in the histogram in Figure~\ref{fig:histogramMapping4bits}.
A further mapping can reduce this bias, however, this process also increases the time required to generate a particular key sequences as well as the similarity for intra-body pairings (cf Figure~\ref{fig:bandanaCorrectedMapped}).

Another solution is to modify the comparison of gait sequences. 
The mean gait features an average amplitude with respect to the instantaneous gait sequences. Also, the acceleration peaks of the instantaneous gait fall with about equal probability to the left or right of the mean gait sequence. 
Consequently, the quantization, exploiting the difference between mean and instantaneous gait generates 0101 and 1010 patterns more often than other patterns. 
We suggest to normalize both mean and instantaneous gait  prior to comparing them for gait generation. 
The heatmap and histogram for bit sequences generated with this modified versions are depicted in figures~\ref{fig:histogramNormalized} and~\ref{fig:heatmapBandanaImproved02}. 

The distribution is improved. 
Unfortunately, a bias towards including more `1'-s is introduced. 
However, since this bias is less severe than in the original BANDANA protocol, the effect can be {damped} by disregarding patterns with probability inverse to their observed occurrence frequency (cf. Figure~\ref{fig:histogramNormalizedThrowAway}).
We observe in Figure~\ref{fig:bandanaCorrectedNorm} that the similarity for intra-body pairing is slightly reduced. 

\section{Conclusion}
\label{sec:conclusion}
We analyzed four acceleration-based pairing schemes. 
We have compared their quantization and discussed quantization-specific attacks. 
Their on-body pairing performance, statistical properties and entropy of generated key sequences were investigated based on walking data from 15 subjects and devices at 7 on-body locations.

Although not originally designed for the purpose of gait-based pairings, the SAPHE protocol achieved best results. 
We modified it towards gait-pairing by executing filtering and re-orientation before the pairing process.
Still, room for improvement remains as shown in the randomness analysis.

The Walkie-Talkie protocol, which is able to generate the highest number of key bits achieves exact matching keys only across upper body locations and with low confidence. 
Together with SAPHE, it has the lowest one-shot success probability. 
This is, however, put into different perspective by a design flaw in the protocol.
Even a naive adversary is able to boost her success probability to 0.125 by analysing the communication during the pairing process. 

A similar quantization mechanism is utilized in the Gait-Key implementation, which suffers from lack of randomness introduced by an n-ary quantization. 
The BANDANA protocol produces high similarity for different and also remote locations on the same body. 
However, the keys show a bias towards specific patterns.
This problem originates from the quantization utilized and we proposed alternative mechanisms that address these issues.

Finally, the IPI protocol is also able to achieve high similarity across keys generated at different location on the same body.
Our investigation revealed that the protocol suffers from a low variance in the generated binary patterns, so that similarity is also high for random gait sequences. 

We further analyzed the threat of a video attack on gait authentication and pairing and found that a sophisticated attacker with video support and real-time gait estimation is able to break the studied gait-based pairing approaches. 

\section*{Acknowledgment}
We appreciate partial funding in the frame of an EIT Digital HII Active
project, as well as from the Academy of Finland and from the German Academic
Exchange Service (DAAD).

\ifCLASSOPTIONcaptionsoff
  \newpage
\fi

\bibliographystyle{IEEEtran}

\vspace{-1cm}
\begin{IEEEbiography}[{\includegraphics[width=1in,height=1.25in,clip,keepaspectratio]{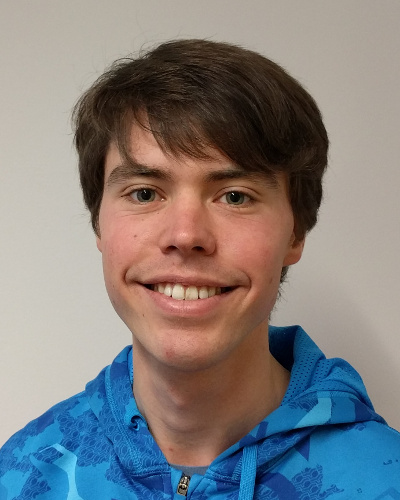}}]{Arne Brüsch}
received his B.Sc. in 2015 and his M.Sc. in 2017, both in computer science. Currently, he works as a research assistant at the Institute of Systems Security at TU Braunschweig. 
His research interests include quantization schemes for key generation in usable security as well as contextual security in general.  
\end{IEEEbiography}
\begin{IEEEbiography}[{\includegraphics[width=1in,height=1.25in,clip,keepaspectratio]{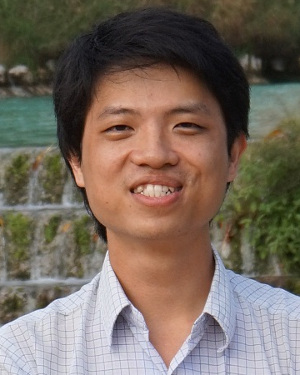}}]{Ngu Nguyen}
is a doctoral student at Ambient Intelligence Group, Aalto University.
He completed his bachelor's and master's degree at University of Science, Ho Chi Minh City, Vietnam.
His research focuses on usable security and distributed machine learning.
\end{IEEEbiography}
\begin{IEEEbiography}[{\includegraphics[width=1in,height=1.25in,clip,keepaspectratio]{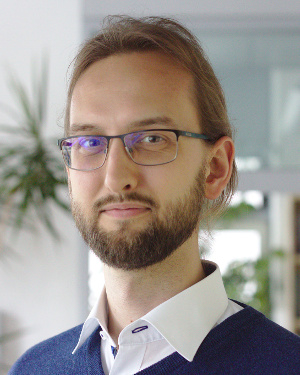}}]{Dominik Schürmann}
received the M.Sc. and Ph.D. degrees in 2014 and 2018 respectively, from TU Braunschweig.
After working as a research fellow at the Institute of Operating Systems and Computer Networks at TU Braunschweig, he co-founded the company Confidential Technologies GmbH.
His research interests include interaction-free security based on physical context and usable security in general.
\end{IEEEbiography}
\begin{IEEEbiography}[{\includegraphics[width=1in,height=1.25in,clip,keepaspectratio]{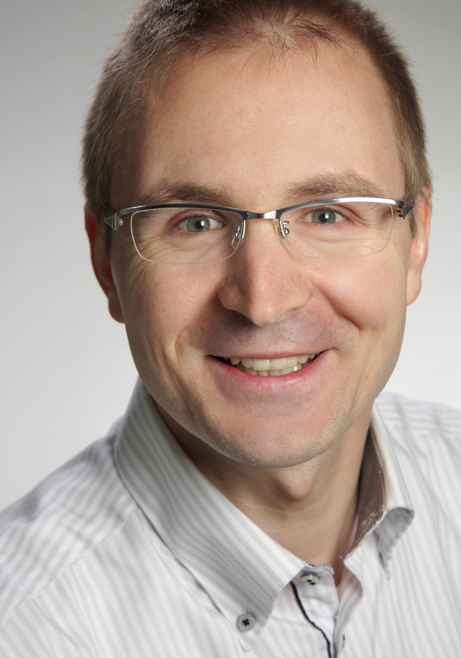}}]{Stephan Sigg}
received his M.Sc. degree in computer science from TU Dortmund, in 2004. and his Ph.D. degree from Kassel University, in 2008. 
Since 2015 he is an assistant professor at Aalto University, Finland. 
He has served as a TPC member of many conferences including IEEE PerCom, Ubicomp, etc.
His research interests include Pervasive Computing, activity recognition, usable security and optimization of algorithms in mobile distributed systems. 
\end{IEEEbiography}
\begin{IEEEbiography}[{\includegraphics[width=1in,height=1.25in,clip,keepaspectratio]{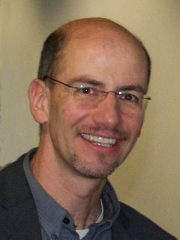}}]{Lars Wolf} received his Ph.D. in 1995.
In 1999 he was an associated professor at the computer science department of Universität Karlsruhe (TH).
Since spring 2002 Lars Wolf is full professor for computer science at the TU Braunschweig where he is head of the Institute of Operating Systems and Computer Networks.
His current research interests include wireless networking in general, sensor networks, vehicular networks, delay-tolerant networks, and mobile systems.
\end{IEEEbiography}

\end{document}